\documentclass{svjour3}
\smartqed
\usepackage{graphicx,epsfig,natbib}
\journalname{Space Science Reviews}

\newcommand{\kms}{\mbox{{\rm\thinspace km\thinspace s$^{-1}$}}}
\newcommand{\ms}{\mbox{{\rm\thinspace m\thinspace s$^{-1}$}}}

\newcommand{\mss}{\mbox{{\rm\thinspace m\thinspace s$^{-2}$}}}
\newcommand{\km}{\mbox{\rm\thinspace km}}

\newcommand{\hei}{He\,{\sc i}}
\newcommand{\heii}{He\,{\sc ii}}

\newcommand{\caii}{Ca\,{\sc ii}}

\newcommand{\arcsec}{$^{\prime\prime}$}

\newcommand{\ha}{${\rm H\alpha}$}
\newcommand{\hb}{${\rm H\beta}$}
\newcommand{\degr}{\hbox{$^{\circ}$}}

\newcommand{\aap}{{Astron. Astrophys.}}
\newcommand{\aaps}{{Astron. Astrophys. Suppl. Ser.}}
\newcommand{\araa}{{An. Rev. Astron. Astroph.}}

\newcommand{\apj}{{Astrophys. J.}}
\newcommand{\apjl}{{Astrophys. J. Lett.}}

\newcommand{\solphys}{{Solar Phys.}}
\newcommand{\mnras}{{Month. Not. Roy. Astr. Soc.}}
\newcommand{\pasj}{{Pub. Astron. Soc. Japan}}
\newcommand{\aj}{{Astron. Journal}}
\newcommand{\nat}{{Nature}}
\newcommand{\jqsrt}{{Journal Quant. Spectr. Rad. Trans.}}
\newcommand{\na}{{New Astron.}}
\newcommand{\ssr}{{Space Sci. Rev.}}
\newcommand{\azh}{{Astronomicheskij Zhurnal}}
\newcommand{\memsai}{{Memorie della Societa Astronomica Italiana}}
\begin{document}

\title{Solar fine-scale structures. I. Spicules and other small-scale,
jet-like events at the chromospheric level: observations and physical parameters}

\titlerunning{Solar fine-scale structures}

\author{G. Tsiropoula$^1$, K. Tziotziou$^1$,\\
I. Kontogiannis$^1$, M. S. Madjarska$^{2*}$,\\
J.G. Doyle$^2$, Y. Suematsu$^3$}

\institute{$^1$Institute for Space Applications and Remote Sensing,
              National Observatory of Athens, Lofos Koufos, 15236 P. Penteli, Greece\\
           $^2$Armagh Observatory, College Hill, Armagh BT61 9DG, UK\\
           $^3$Hinode Science Center, National Observatory of Japan,
           2-21-1 Osawa, Mitaka, Tokyo 181-8588, Japan\\
           $^*$now at: UCL-Mullard Space Science Laboratory, Holmbury St Mary, Dorking, Surrey, RH5 6NT, UK\\
           Corresponding author: G. Tsiropoula (e-mail: georgia@noa.gr)\\}

\date{Received: date / Accepted: date}

\maketitle

\begin{abstract}
Over the last two decades the uninterrupted, high resolution
observations of the Sun, from the excellent range of telescopes
aboard many spacecraft complemented with observations from
sophisticated ground-based telescopes have opened up a new world
producing significantly more complete information on the physical
conditions of the solar atmosphere than before. The interface
between the lower solar atmosphere where energy is generated by
subsurface convection and the corona comprises the chromosphere,
which is dominated by jet-like, dynamic structures, called mottles
when found in quiet regions, fibrils when found in active regions
and spicules when observed at the solar limb. Recently, space
observations with Hinode have led to the suggestion that there
should exist two different types of spicules called Type I and Type
II which have different properties. Ground-based observations in the
\caii\ H and K filtergrams reveal the existence of long, thin
emission features called straws in observations close to the limb,
and a class of short-lived events called rapid blue-shifted
excursions characterized by large Doppler shifts that appear only in
the blue wing of the \caii\ infrared line. It has been suggested
that the key to understanding how the solar plasma is accelerated
and heated may well be found in the studies of these jet-like,
dynamic events. However, while these structures are observed and
studied for more than 130 years in the visible, but also in the UV
and EUV emission lines and continua, there are still many questions
to be answered. Thus, despite their importance and a multitude of
observations performed and theoretical models proposed, questions
regarding their origin, how they are formed, their physical
parameters, their association with the underlying photospheric
magnetic field, how they appear in the different spectral lines, and
the interrelationship between structures observed in quiet and
active regions on the disk and at the limb, as well as their role in
global processes has not yet received definitive answers. In
addition, how they affect the coronal heating and solar wind need to
be further explored. In this review we present observations and
physical properties of small-scale jet-like chromospheric events
observed in active and quiet regions, on the disk and at the limb
and discuss their interrelationship.


\end{abstract}

\section{Introduction}
During the last 2 -- 3 decades, observations from space combined
with ground-based data, as well as state-of-the-art, post-processing
reconstruction techniques combined with progress in numerical
modelling and theory have increased our understanding of the
physical processes occurring in the solar atmosphere. Apart from
demonstrating that the solar atmosphere is highly dynamic and
inhomogeneous these studies have also provided a deep insight into
all phenomena occurring in its different parts. They have revealed a
number of different types of structures that exist in the different
atmospheric layers for which a range of formation mechanisms have to
be investigated. They have also provided important information about
the physical conditions of the emitting plasma, as well as hints
about the physical nature of related dynamic phenomena which
continuously occur on the Sun over a large range of spatial and
temporal scales.

The solar chromosphere is an intricately structured and dynamic
layer of the Sun. It is traditionally defined as the layer, about
2\,000\km\ thick, lying between the photosphere and the transition
region and has long been known not to be in hydrostatic equilibrium.
Therefore, its nature is that of dynamic fine-scale structures which
uniquely characterize this part of the solar atmosphere. These
appear to be structured by the magnetic fields that extend from
network boundaries and plages and be directly related to its
evolution. The appearance of the underlying atmosphere is, mainly,
due to processes involving the emergence, buffeting and
re-arrangement of the magnetic flux generated by interior dynamo
action. The resulting spatial inhomogeneity produces a fruitful
environment for the emerging fluxes to undergo quite complex changes
as they dynamically evolve and interact with the pre-existing
magnetic fields in the ambient medium. Complex and concentrated,
mainly unipolar, strong magnetic fields produce the so-called active
regions, which are the locations of some jet-like structures, such
as the (traditional) fibrils and dynamic fibrils (DFs). Outside
active regions, what is generally known as the quiet solar
chromosphere, is not really quiet. It is continually subject to a
variety of small-scale events, such as mottles, straws and the,
recently reported, rapid blue excursions, occurring at the
boundaries of cellular patterns that constitute the magnetic
network. The magnetic network is believed to be sustained by
magnetic bipoles, which emerge in the interiors of supergranules,
move apart and are fragmented by granular buffeting. Eventually, the
fragments are driven to the supergarnular boundaries where they
interact with the existing magnetic flux \citep{wang1996,
schrijver1997}. Opposite polarity fluxes cancel and submerge,
whereas like-polarity flux merges to form larger flux
concentrations.

When observing the chromosphere at different wavelengths, from the
EUV to the radio domain its very large range of inhomogeneity and
structuring is revealed. Chromospheric structures can be seen
against the solar disk, by means of monochromatic or narrow-band
filters or spectrometers operating in strong spectral lines, such as
\ha, \caii\ H \& K, and the \caii\ infrared triplet (IR). In
particular, filtergrams recorded in the wings and line center of
\ha\ reveal a wealth of fine-scale structures. The most prominent
small-scale features residing at the network boundaries are
certainly mottles. Mottles are thin jet-like features, better
observed in \ha. \citet{rutten2007} identified in \caii\ H images
taken close to the solar limb as very thin and short-lived bright
structures which occur in ``hedge rows''. He called them ``straws''.
Recently, \citet{langangen2008c} and \citet{rouppe2009} found
sudden, large line shift changes in both \ha\ and \caii\ IR lines on
their blue side. The authors interpreted these events, which occur
at the edges of rosettes (cluster of mottles expanding radially
around a common center), as chromospheric up-flows and called them
``rapid blue-shifted excursions'' (RBEs). In active regions, a
subset of dark structures called fibrils includes relatively short,
thin, jet-like structures called dynamic fibrils (DFs). They are
found in the central area of plages where the magnetic field is more
dense, and more vertically oriented. Another large subset includes
very inclined, almost horizontal, long fibrils which are clearly
more stable than the DFs. The relationship between all these on-disk
structures, as well as their relationship to spicules, thin,
elongated, jet-like structures observed at the solar limb, is under
strong debate. Spicules are the most obvious component of the
chromospheric limb. Although they were discovered almost 130 years
ago, they still remain one of the mysterious phenomena in the solar
atmosphere because, at least until recently, remained very close to
the resolution limits of current solar observations. As a result,
several of their properties have not been well defined and this has
led to a plethora of theoretical models for their interpretation
(for a review see \citet{sterling2000}). Recently, with its high
resolution and high cadence, as well as an uninterrupted view of the
Sun, the Hinode space mission provided a new picture of spicules.
\citet{depontieu2007b}, based on time series observations of the
solar limb taken with Solar Optical Telescope (SOT) aboard Hinode in
the \caii\ H broad filter, suggested that spicules can be grouped in
two categories, called Type I and Type II, which seem to have
different properties and formation mechanisms. The origin of short
duration/high velocity spicules (Type II) has been attributed to
magnetic field reconnection. On the other hand, the origin of Type
I, classic spicules (long duration/low velocities) seems to be
associated with the leakage of p-modes into the upper atmosphere and
subsequent development of shocks that follow the magnetic field
lines.

While spicules are ubiquitous at the limb, there is still a debate
on the issue of their counterparts are on the solar disk.
\citet{grossmann1992} concluded from the different velocity
distributions of mottles and spicules, that mottles are not the disk
counterparts of (classical) spicules. \citet{tsiropoula_sch1997}, on
the other hand, based on the similarity of their physical
parameters, argued for a close relationship between them. Type I
spicules appear to rise up from the limb and fall back again. They
show a similar dynamical behaviour as active region DFs
\citep{depontieu2007a}, as well as a subset of mottles. Type II
spicules, on the other hand, seem to exhibit an upward motion
followed by rapid fading from the Hinode \caii\ H passband, although
more recent work by \citet{zhang2012}, were unable to locate any
Type II spicules. \citet{rouppe2009} suggested that the disk counterpart of
Type II spicules are RBEs based on the similarities of their
properties.

When observed in the line center of \ha, the chromospheric plasma is
clearly seen to be organized along fine-scale magnetic structures
which have different inclinations. Recently,
\citet{kontogiannis2010a} and \citet{kontogiannis2010b} using
high-resolution \ha\ observations together with magnetograms
obtained with the spectropolarimeter on Hinode, revealed through a
potential magnetic field extrapolation another very important aspect
of these structures when observed on the disk. They showed that
magnetic flux tubes following the local inclination of the magnetic
field lines, define the layer of the magnetic canopy, i.e. the
plasma-$\beta \sim 1$ layer, where the plasma-$\beta$ is the ratio
of gas pressure to magnetic pressure. It is at this layer, where
acoustic waves generated by the convective motions, undergo mode
conversion, refraction, reflection and transmission with important
effects to the upper solar atmosphere and thus these structures play
a very important role in wave propagation.

The fact that the solar corona has a temperature more than 1 million
degrees is still a puzzling problem of solar physics, despite the
considerable theoretical and experimental efforts. Since the energy
release in the largest explosive events (flares and microflares)
does not supply enough power to heat the corona, the behaviour of
smaller-scale and less energetic, but much more frequent events is
an essential constituent of the problem \citep{doyle1985,
parker1988}. Both magnetic reconnection and waves related to
spicules are powerful mechanisms for energy release and transfer and
have been considered as heating agents of the solar corona.
\citet{tsiropoula1994} and \citet{tziotziou2003} examining the
temporal evolution of the line-of-sight velocity of mottles found
bi-directional flows and proposed a model according to which
spicules are due to magnetic reconnection. \citet{tsiropoula2004}
showed that if magnetic reconnection is considered as the driving
mechanism of mottles, the material they supply to the solar corona
is in excess of that needed to compensate for coronal mass losses in
the solar wind, while the amount of energy released to heat the
corona depends on several parameters, among which are the number of
the events, their axial velocity and magnetic field, etc., and can
be negligible or substantial. \citet{depontieu2007b}, on the other
hand, showed that the energy flux provided by the Alfv\'en waves
identified in spicules is large enough to supply the energy
necessary to heat the quiet solar corona and drive the solar wind.
Thus the question arises of how these small-scale features
participate and contribute into the larger scale observable
phenomena, and how a more accurate determination of the physical
parameters, which are related to the physical processes, can be
extracted from observations.

This review is the first part of a series of four reviews. The aim
of these reviews is to present an overview of our current
understanding (observations, modelling, physical parameters, etc.)
of solar small-scale, jet-like structures observed both on the disk
and at the limb, in active, as well as in quiet regions. In this
review (Part I) we present an overview of the chromospheric
fine-scale structures. More specifically, we will present their
morphological properties, their derived physical parameters, such as
height, width, lifetime, velocities, etc., as well as methods used
for the derivation of some of them, their dynamical behaviour,  and
comment on their interrelationship. In subsequent papers we will
discuss transition region small-scale events (Part II), and the role
of the various small-scale events in coronal heating (Part III).

\section{The solar photosphere}

\begin{figure}[ht]
\begin{center}
\includegraphics[width=0.5\textwidth]{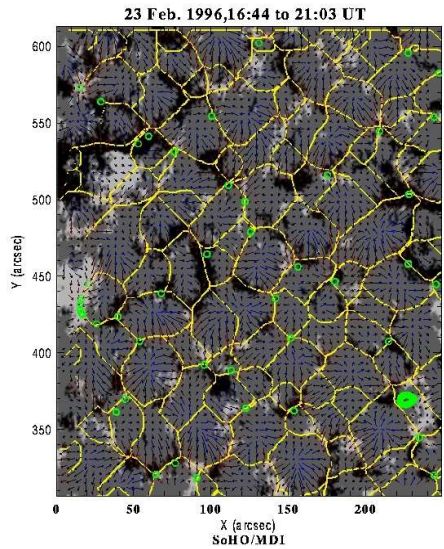}
\end{center}
\caption{A SOHO/MDI magnetogram. The boundaries of supergranule
cells are marked in yellow, and arrows indicate the supergranular
flow pattern (Credits SoHO/MDI). }\label{mdi}
\end{figure}

The solar surface is threaded by a complex network of magnetic
fields. Observations of these fields at the photospheric level
reveal them to be concentrated into regions of mixed (negative and
positive) polarities covering a wide range of values for the
(unsigned) magnetic flux and lifetimes. In {\it quiet areas} the
dynamic interplay between the magnetic field concentrations and the
plasma has been called the Sun's ``magnetic carpet''
\citep{title1998}. In these areas, photospheric motions are
dominated by the flow pattern of large convective cells called
supergranules. These cells range in diameter from roughly
10\,000\km\ to 50\,000\km, with an average diameter of between
13\,000 and 18\,000\km\ \citep{hagenaar1997}. Their flow pattern
takes the form of an up-flow at the cell center, followed by radial
outflow of $\sim$0.5\kms, and down-flow at the cell boundaries
\citep{simon_lei1964}. \citet{wang_zir1989} derived an upper limit of
0.1\kms\ for the magnitude of the supergranule's vertical velocity
(both up-flow and down-flow). The strongest down-flow occurring at
points where two or more cells meet, and as a result, the magnetic
flux tends to build up along the boundaries of supergranular cells
after being swept from the center by the radial outflow. It is now
well established from investigations of high resolution magnetograms
that new bipolar elements emerge continuously inside the cell
interiors and are, subsequently, swept by the supergranular flow at
the periphery of supergranular cells where they gather and form the
network boundaries with kG field strengths \citep{wang1996,
schrijver1997}. This is illustrated well by the SoHO/MDI magnetogram
shown in Figure~\ref{mdi}. The supergranule cell boundaries are
marked in yellow, and it can be seen that the majority of the flux
concentrations are located along these lines, and in particular at
the intersection of multiple cells. These concentrations called
\emph{network concentrations} are typically found at sites of strong
down-flow, at the boundaries of supergranular cells
\citep{martin1988}. Their fluxes are of the order of 10$^{18}$ --
10$^{19}$\,Mx and have typical diameters of 1\,000 -- 10\,000\km\
\citep{parnell2001}. They tend to be larger than \emph{ephemeral
regions} which are clumps of newly emerged flux of varying strengths
with total net flux equal to zero. Ephemeral regions emerge
preferentially near supergranular boundaries, without influencing
the flow pattern. After emergence, they most often fragment into
several concentrations where their flux quickly merges into the
pre-existing network. \citet{schrijver1997} give the average flux of
an ephemeral region to be $\sim$ 1.3$\times$\,10$^{19}$\,Mx.
\citet{martin1990} determined that the network concentrations are
produced from the residuals of other flux concentrations. It is
estimated that 90\% or more of their flux originates from ephemeral
regions, with the remaining 10\% being contributed by
\emph{intranetwork fields}. \emph{Intranetwork (IN) fields} are the
smallest of the three types of small-scale magnetic flux
concentrations. They typically last 2 to 4 hours. Their mean
diameter is about 2\,000\km, while their total flux content ($\sim$
10$^{16}$\,Mx) tends to be close to the detection limit of present
instruments. They emerge within supergranule cells due to the
emergence of small bipolar pairs. They are then swept towards the
cell boundaries by radial flows, where they interact with network
fields \citep{martin1988}. Recently, \citet{centeno2007} using
observations obtained with Solar Optical Telescope
Spectropolarimeter (SOT/SP) on board the Hinode spacecraft presented
clear evidence of an emerging IN magnetic loop showing strong
horizontal magnetic signal and traces of vertical opposite
polarities on each side of it. The lower parts of the loop were
advected into the intergranular lanes, where they aggregate to other
magnetic field concentrations resulting in larger flux elements.
\citet{lites2008} based on the high spatial resolution and
polarimetric sensitivity of the SOT/SP reported the existence of
horizontal magnetic fields in the IN regions whose mean field
strength surpasses considerably that of the vertical component. The
vertical fields are concentrated in the intergranular lanes, whereas
the stronger horizontal fields occur preferentially at the edges of
the bright granules or inside them aside from the vertical fields.
This kind of spatial relation between the horizontal and the
vertical magnetic fields suggests that the IN field consists of
small ${\Omega}$ loops having their vertical roots in the
intergranular lanes and connected by an horizontal field above
granules.

From the previous discussion it becomes clear that the magnetic flux
patches within the magnetic carpet, while diffusing along the
superganular boundaries due to supergranular flows, interact with
one another by the following four main processes: Flux
\emph{emergence} which is the appearance in pairs or in clusters of
new magnetic flux (with equal amounts of flux in both the positive
and negative polarities). \emph{Cancellation} which occurs when
concentrations of opposite polarity come into contact and mutually
lose flux \citep{livi1985, martin1985}. If two or more
concentrations of the same polarity merge together to form a single
larger concentration, this is known as \emph{coalescence}. Finally,
\emph{fragmentation} which is the splitting of a large concentration
of flux into several different smaller concentrations. Using a
series of full-disc 96 minute MDI magnetograms, \citet{hagenaar2001}
found that the time taken for all flux within the quiet Sun
photosphere to be replaced (``the flux replacement timescale'') was
around 14 hours. However, later studies determined this value to be
an order of magnitude smaller, at just 1 -- 2 hours
\citep{hagenaar2008}.

The magnetic network formed by the driving of the magnetic flux to
the boundaries of supergranules is made up of clusters of kilogauss
flux tubes. By definition the smallest observable structure of the
magnetic flux in the photosphere are magnetic elements which have
various field strengths and diameters close to or less than the best
currently achievable spatial resolution ($\sim$ 150\km). Magnetic
elements are very important because they not only structure the
photosphere but also provide a link to the chromosphere and corona
along which energy can be transported from the Sun's interior to the
outer layers. They were first observed as ``magnetic knots''
\citep{Beckers_Sch1968} -- small dark structures equivalent to what
we now call ``micropores'' \citep{Topka97}-- and as ``filigree'',
elongated chains of bright points \citep{Dunn_Zir1973} which have
been resolved into strings of adjacent bright points by
\citet{Mehltretter74}. When observed at disc center they are usually
best seen in the ``G-band filtergrams'', which are obtained with a
wide-band filter centered around 430.5\,nm, named G-band by
Fraunhofer. The first observations of the network at this wavelength
window were carried out by \citet{Muller_Rou1984}. The name ``network
bright points'' (NBPs) was introduced \citep{Stenflo_Har1985,
Muller1985} for the bright grains which are visible in filtergrams
and are arranged in roughly cellular patterns making up the network
boundaries. Since bright points have been presumed to be the visible
manifestation of thin magnetic flux tubes and thus related to the
magnetic elements, extensive work has been initiated and it is now
well established that NBPs are brightness manifestations of the
small, strong field magnetic elements that make up the magnetic
network \citep{Muller_Rou1984, Berger_Tit1996, Berger_Tit2001,
Rouppe2005}.

In {\it stronger field areas}, on the other hand, the increased
number density of magnetic flux tubes gives rise to large
chromospheric areas called plages which consist of many bright
elements (sometimes referred to as facular granules) packed
together. Granules near network bright points and in plage regions
are smaller, display slower temporal evolution and have a lower
contrast than those found in quiet areas \citep{title1992}. The
granular pattern in these areas is referred too as ``abnormal''
\citep{dunn1974}. Flux concentrations take on structure, behave less
passively and seem to affect the granular flow. The larger magnetic
filling factor in plage regions gives rise to large amorphous,
elongated ``ribbons'' that contain high net flux. They are
characterized by large flux concentrations having a brighter edge
and an intensity depression in the center. The majority of them are
not resolved into flux tubes, but instead show a substructure which
is interpreted as an indication of a range of magnetic field
strengths within the structure. Circular manifestations of spread
out ribbon structures have been called ``flowers''. The ``flowers''
and ``ribbons'' are small flux concentrations \citep{berger2004},
while larger flux concentrations are called (micro)-pores and have a
distinct dark center. There is a continuous transition between these
features evolving from one to another depending on the amount of the
magnetic flux that is collected or dispersed. The merging and
splitting of ``flux sheets'', which are concentrations of the
magnetic field due to the granular flow, and the transitions between
the ribbons, flowers, and micropores are described by
\citet{Rouppe2005}. These authors found also that the plasma is
basically at rest in these structures with small concentrations of
weak up-flow sites, while narrow sheets with down-drafts are found
right at the edges of the magnetic field concentrations.

Due to the continual motion of mixed polarity concentrations in the
magnetic carpet and the different flux evolution processes, the
quiet Sun photosphere is highly dynamic. Since magnetic fields from
the magnetic carpet extend up into the solar chromosphere and lower
corona, it is expected that the quiet Sun chromosphere and corona
are also highly dynamic responding to the movement and evolution of
the different flux concentrations. Even more, the spatial
distribution and rates of the different processes of flux evolution
play a major role in determining the distribution of heating events
in the upper solar layers. The driving of foot-points of the
small-scale magnetic fields on the surface of the Sun plays a
fundamental role in sustaining the chromospheric network and leads
to the build up of currents and/or the generation of waves,
mechanisms considered as potential sources of energy for solving the
coronal heating problem. On the other hand, the larger magnetic
filling factor in plage regions gives rise to more extended
structures. As stated by \citet{Rouppe2005}, the merging and
splitting of flux sheets and the continuous transition between
ribbons and micropores, are difficult to reconcile with models that
regard these structures as being composed of discrete flux tubes
that keep their identity over a long period of time. They also
suggested that it should be more appropriate to characterize the
temporal behaviour of the magnetic field in these areas as
fluid-like, where the ``magnetic fluid'' is in continuous
interaction with the field-free plasma, forming the extended
concentrations. Due to the differences between quiet and active
regions at the photospheric level described above, differences
between the different underlying structures related to them should
be expected.

It should be made clear at this point that the aim of this section
is not to give an extensive review on the solar photosphere and its
magnetic structure, but rather to provide the background for the
overlying chromosphere and its structures which is the main topic of
this review. The interested reader can be referred to recent, more
detailed reviews on the solar photosphere, like e.g. those by
\citet{dewijn2009} and \citet{muller2011}.

\begin{figure}[t]
\centering
\includegraphics[width=0.95\textwidth]{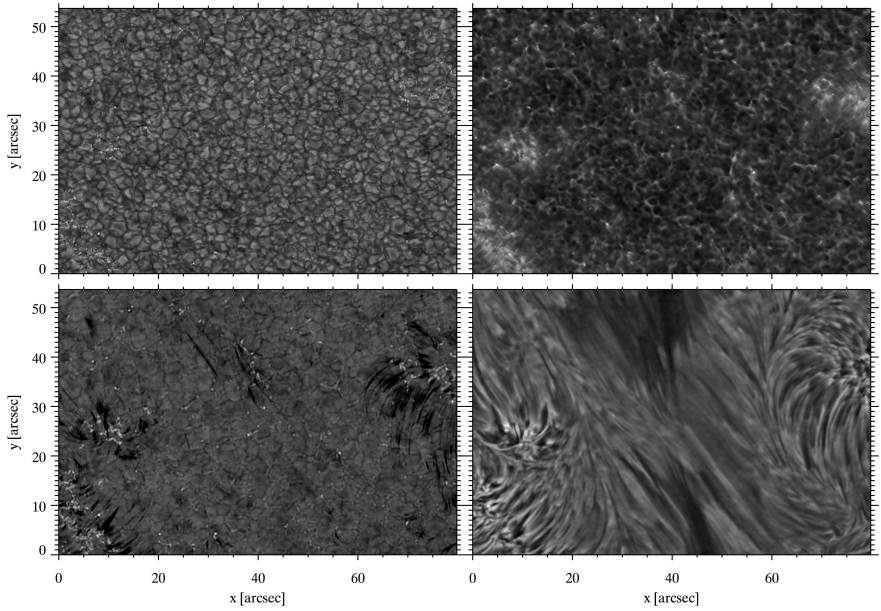}
\caption{Images of a quiet-Sun area containing a quiescent filament
obtained by the Dutch Open Telescope (DOT) in La Palma (Canary
Islands). {\it Top left}: G-band. {\it Top right}: \caii\ H. {\it
Bottom left}: blue \ha\ wing at $\Delta\lambda$ = $-$0.08\,nm from
line center. {\it Bottom right}: \ha\ line center. The left-hand one
samples quiet network, the right-hand one more enhanced network
(figure courtesy of Dr J. Leenaarts, see also \citet{Leenaarts2006})}
\label{mag_elem}
\end{figure}

\section{The solar chromosphere}
The photospheric magnetic fields of different forms and strengths
extend through the chromosphere into the corona providing the link
between the different solar layers. As stated in the previous
section in the photospheric layers the supergranular flows carry the
magnetic fields to the edges of the supergranular cells resulting in
large magnetic flux concentrations. The magnetic flux tubes that
constitute the magnetic network expand upwards and appear as bright
patches that constitute the chromospheric network. A strong spatial
coincidence exists between the photospheric magnetic network
concentrations and the overlying network boundaries in the
chromosphere. In \caii\ H\&K images, which sample the low
chromosphere, NBPs show up with a larger brightness enhancement over
the surrounding area than they do in G-band (cf Fig. 2 of
\citet{Lites1999}), but with lower sharpness due to strong resonance
scattering and possibly also due to increasing flux tube width with
height (Fig.~\ref{mag_elem}, upper row). Also high resolution images
taken in the blue wing of \ha\ were found to display strikingly
intense small-scale brightenings at the locations where magnetic
elements form the NBPs (Fig.~\ref{mag_elem}, lower row, left).
\citet{Leenaarts2006} showed the potential of these \ha\ bright
points as proxies of the magnetic elements.

The solar chromosphere being only about 2\,500\,km thick
($\sim$\,3\arcsec) actually covers the layer where the solar
atmosphere turns from the gas-dominated photosphere/lower
chromosphere into the magnetic field dominated upper
chromosphere/corona (for recent reviews on the solar chromosphere,
as well as to its coupling with the other solar layers the reader
can be referred to \citet{rutten2007}, \citet{rutten2010},
\citet{wedem2009}). Chromospheric plasma, especially when seen in the
center of the \ha\ line, is very clearly organized along fine-scale,
dark, elongated structures that span out from regions of enhanced
magnetic flux. Some of them are connected to neighbouring regions
and others fading in between, while covering the IN cells
(Fig.~\ref{mag_elem}, lower row, right). Such filamentary structures
portray the magnetic field topology and are consistent with the
increasing dominance of the magnetic pressure over the gas pressure
with increasing heights in the atmosphere. This dominance has as a
result the upward expansion of the magnetic fields which fill the
space in the chromosphere and affect its dynamics. The relative
importance of the pressures is given by the ratio of plasma pressure
to magnetic pressure (also referred to as the plasma-$\beta$) and is
related to the formation of the magnetic canopy zone. This zone, in
which $\beta=$ 1 and the sound speed equals the Alfv\'en speed,
separates the solar atmosphere into magnetic and non-magnetic
regions and plays an important role in the refraction and reflection
of the upward propagating waves, as well as in converting acoustic
waves into other modes, such as fast and slow magnetoacoustic
waves\citep[see e.g.][]{rosenthal2002, bogdan2003}).
\citet{kontogiannis2010a} and \citet{kontogiannis2010b} have shown the
important role the dark \ha\ fine-scale magnetic structures play in
the formation of the magnetic canopy (for details see Section
\ref{canopy}). As one goes from the \ha\ line center to the wings,
the picture gradually changes (see lower left panel of
Fig.~\ref{mag_elem}). It is, usually, suggested that in dark
structures observed in the \ha\ line wings the intensity is coming
from lower layers than the core of the line, although it cannot be
excluded that it might come from highly Doppler shifted features
that are located higher in the atmosphere than the features observed
in line center. The dark structures endings in the \ha\ blue wing
image (Fig.~\ref{mag_elem}, lower row, left) -- which are part of
the elongated mottles seen in \ha\ line center- are likely to
emanate from photospheric bright points which are concentrated into
clusters. The relationship between the ends of the dark structures
and the photospheric bright points is presented in Section
\ref{pbp}.

From the above it becomes clear that the photosphere and
chromosphere are coupled via magnetic fields and, furthermore, any
description of the solar atmosphere in terms of average properties
is incomplete without the consideration of the fine-scale structures
and their dynamics.

\section{Fine-scale, jet-like chromospheric structures}
\label{sec:1}

Mottles, spicules, fibrils and dynamic fibrils, straws, rapid blue excursions (RBEs) are
some of the jet-like, fine-scale features that dominate the dynamic and
highly-structured chromosphere. Direct, narrow-band or broad-band filtergrams have
provided the means for obtaining parameters such as lifetime, size, spatial
distribution, inclination, etc., of these structures. On the other hand, spectroscopic
observations in different lines are essential to extend our knowledge on the physical
conditions, particularly the temperature, density and flows. Determination of the
physical conditions in the observed features is also based on the study of the spatial
and temporal variation of the observed radiation intensity in different wavelengths,
i.e. line intensity or contrast profiles. On the basis of observational findings various
theoretical models and numerical simulations have been developed to explain the
mechanism responsible for their formation. However, in spite of the remarkable advances
made by the high resolution observations and by the development of theories and
numerical simulations, their interrelationship, the determination of their physical
parameters, the definition of their formation mechanism, as well as their possible role
in the heating of the solar corona still remain uncertain. Ambiguities are mainly due to
the differences in their appearance when observed in different spectral lines, but also
at different wavelengths within the same line, to the different instruments used for
their observation (which usually have different spatial and temporal resolution), to the
different methods used to infer their physical parameters, especially velocities, to
their relatively short duration and rapid changes and, most important, to the different
names used by different authors to describe the same structure.

Below we present a description of the appearance and the physical
properties of these structures. We follow the traditional
terminology, as well as the recent one which has resulted from high
resolution observations of some ``new'', ``spicular-like"
structures.

\begin{figure}[t]
\centering
\includegraphics[width=0.7\linewidth]{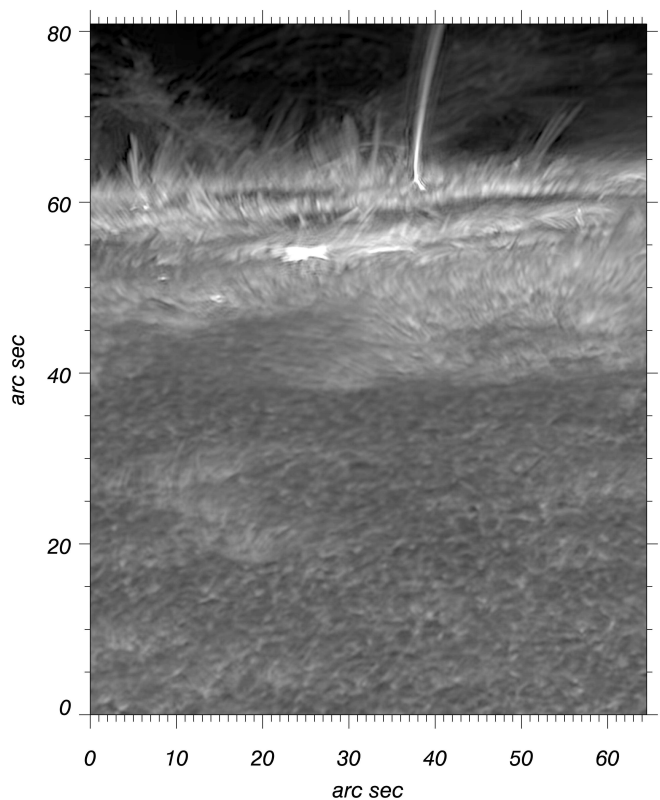}\\
\caption{A sample DOT \caii\ H image obtained on November 4, 2003
showing numerous jet--like structures (spicules, active region
fibrils, superpenumbral fibrils) clearly visible on the limb in
addition to a large surge. The dark elongated structures near the
limb are sunspots. At the bottom of the image thin bright
structures, called straws, are emanating, from the chromospheric
network (which is hardly visible in this image), while around the
active regions several dynamic fibrils and penumbral fibrils are
visible (from \citet{tziotziou2005})} \label{dot}
\end{figure}

\subsection{Off-limb structures}

\subsubsection{Spicules}
The traditional term ``spicule'' refers to the relatively thin,
elongated jet-like structures, seen at the limb of the Sun which
appear as bright jet-like features against the dark background of
the solar corona. The term was first used by \citet{Roberts1945}, to
describe jets which dominate the chromosphere seen at the limb
shooting up to 10\,000\km\ heights. Spicule-like features, however,
were first observed and described by Father Angelo Secchi of the
Vatican Observatory in 1877 \citep{secchi1877}. As first reported by
\citet{lippincott1957}, spicules seem to show a group behavior, the
so-called ``porcupine'' and ``wheat'' field patterns. In the former,
spicules seem to radiate outward from a common centre forming a
18\,Mm wide configuration along the solar limb, whereas in the
latter all spicules seem to have the same orientation over a width
of 140\,Mm. Initially, spicules were associated with quiet Sun
regions at the limb, nowadays, however, it is believed that on-disc
quiet Sun structures called mottles constitute only a fraction of
the observed off-limb spicules and that active region fibrils, as
well as superpenubral fibrils are seen as spicules when these
regions cross the limb (see Fig.~\ref{dot}). Despite their discovery
over 130 years ago, spicules are among the least understood
phenomena of the solar chromosphere. Until recently, this was due to
the lack of high-resolution observations (their diameters of a few
hundred kilometers and lifetimes of a few minutes were close to the
observational limits), as well as to the highly simplified and
poorly constrained models used for their modelling. Several recent
advances not only on the observational side, but also on the
modelling side have provided significant progress on the
understanding of these structures. However, we still lack definitive
answers on several of their physical properties and
interrelationships with other limb and on-disk structures, while
models still face difficulties in describing them accurately.
\citet{beckers1968, Beckers1972} give comprehensive reviews of early
work on mottles and spicules, while a more recent review on the
modelling of these structures is given by \citet{sterling2000}.

\begin{figure}[t]
\begin{center}
\includegraphics[width=0.95\textwidth]{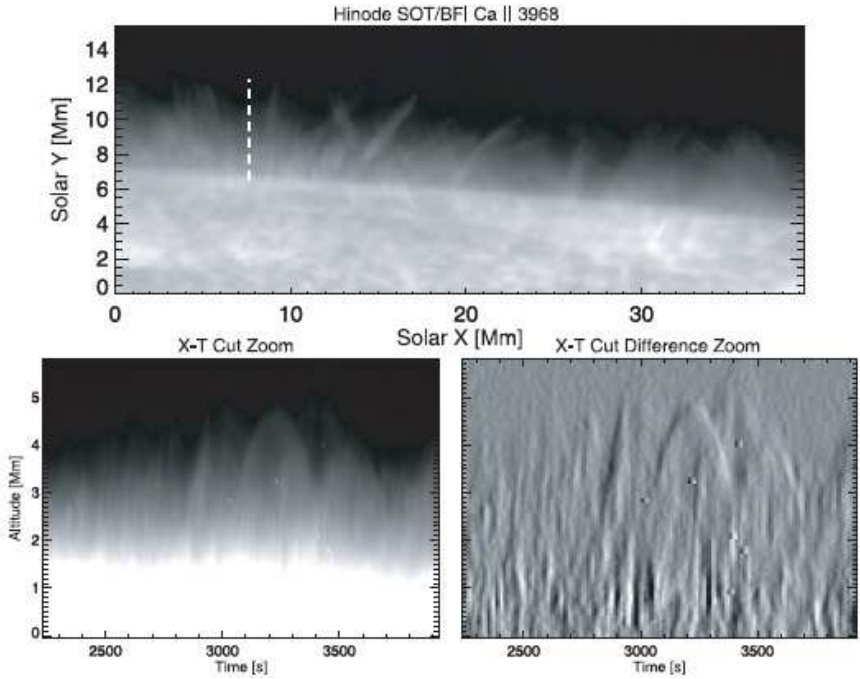}
\end{center}
\caption{The top panel shows spicules at the quiet Sun limb obtained
with the SOT/Hinode in the \caii\ H passband. The bottom panels show
space-time (X -- T) plots along the location indicated by a dashed
line in the top panels, for both the original data and time
difference data. The space-time plot is dominated by short-lived
vertical stripes (Type II spicules) and longer-lived parabolic paths
(Type I spicules) (from \citet{depontieu2007b})}\label{spic_i_ii}
\end{figure}

The morphology and properties of spicules have been under study for
the last sixty years although with rather limited interest after the
early years of their characterization. Only recently, the interest
of the solar physicists on them has been revived by the fact that
high resolution instruments are being used. High resolution movies
in the \caii\ H line obtained with the Broadband Filter Imager (BFI)
of SOT on the seeing-free Hinode spacecraft have revealed a
remarkably dynamic, spicule-dominated limb. Based on some physical
properties derived from SOT/Hinode \caii\ H observations,
\citet{depontieu2007b} suggested that there exist two types of
spicules: a) ``Type I'' spicules, similar to the traditional
spicules, which appear to rise up from the limb and then fall back,
to be long-lived (3 -- 7\,min) and to exhibit longitudinal motions
of the order of 20\kms\ and b) ``Type II'' spicules which seem to
have much shorter lifetimes (50 -- 100\,s), higher velocities
($\sim$\,100\kms), to be considerably taller and to exhibit only
upward motion and then rapidly disappear (Fig.~\ref{spic_i_ii}).

In that work the authors following the time evolution of the
intensity along a fixed cut almost perpendicular to the limb (see
Fig.~\ref{spic_i_ii}), they were able to show at least one clear
example of a ``Type I'' spicule which shows a distinctive different
profile than the steep linear streaks caused by ``Type II''
spicules. However, since spicules are so dynamic and may move in and
out of the fixed cut, such intensity profiles that cover the full
time series cannot be used to reliably measure the properties of
spicules. Thus the authors had to rely on spicules that did not move
transversely so that their full life time could be covered by a
fixed cut. Quite recently, \citet{zhang2012} re-analyzing the same
data sets adopted by \citet{depontieu2007b} identified and traced 105
and 102 spicules in quiet Sun and coronal hole (CH), respectively.
They claimed that they could not find a single convincing example of
``Type II'' spicules. Furthermore, they reported that more than 60\%
of the identified spicules in each region showed a complete cycle,
i.e., the majority are ``Type I'' spicules. Thirdly, the lifetimes
of the spicules in the quiet Sun and CH are 148 s and 112 s,
respectively, but there is no fundamental lifetime difference
between the spicules in the quiet Sun and CH reported earlier. In
the same paper of \citet{depontieu2007b}, the authors in order to
reveal the two different populations applied two different filters
to the time series. The first, low-frequency, filter was a Gaussian
centered on 3\,mHz with a width of 0.5\,mHz and brought out the
``Type I'' spicules, i.e. the relatively slowly evolving features
that typically move up and down during their lifetimes of order 3 --
7\,min. The second filter was designed as a combination of a
high-pass filter ($\geq$\,15\,mHz) with a low-pass filter placed at
a frequency which is 2\,mHz lower than the Nyquist frequency of the
time series studied. This filtering isolates the longer lived Type I
spicules from the more dynamic, Type II, spicules which appear and
disappear on very short time scales. It has not been checked,
however, if there exist spicules having lifetimes in-between the
above mentioned timescales.

\citet{tavabi2011} have recently shown that there exist four types of
spicules based on the spicule's diameter, ranging from 0.3\arcsec\
(220 km), 0.5\arcsec\ (360 km), 0.75\arcsec\ (550 km) to
1.15\arcsec\ (850 km). Unfortunately, detailed comparisons between
the ``classical'' spicules and those called Type I and Type II is
not easy to perform for several reasons among which the most
important is the lack of spatial and temporal resolution of
classical observations. Below we summarize properties of structures
seen at the limb.

\paragraph{Spicule abundances} In contrast to activity phenomena, spicules are omnipresent at
the solar limb. Their abundance has been measured, as a variation of
height and solar latitude. A point to emphasize is that the number
of counted spicules depends very much on the quality of the
observations. Thus their number, in total, is estimated to reach
$\sim10^{6}$ at zero height, while it is possible that this number
is larger \citep{Beckers1972}. Most of the observational studies
reviewed by Beckers (\citeyear{beckers1968}) show that the spicule
number increases with height reaching a maximum at a height and then
decreases again. This is a result of spicules overlapping one
another at lower heights and only few of them reaching large
heights. An estimation of the variation of spicules' abundance with
height gives around 56\,000 spicules on the full solar disk, around
5\,000\km\ from the surface. \citet{Athay1976} and \citet{zirin1988}
give 60\,000 and 70\,000 respectively, for the total number of
spicules on the whole Sun at any time. \citet{judge2010} assuming
Type II spicules to be randomly distributed along the boundaries of
circular supergranules into 200 small bushes with a common ``root'',
and each bush to have eight spicules, they obtained 1600 spicules
per supergranule. With each spicule having a diameter of 0.1\,Mm,
they obtained an area filling factor of 0.015 and a total of
2$\times$10$^7$ Type II spicules on the Sun at any time. This value
is some 20 times larger than the value given by \citet{Beckers1972},
while Type I spicules are not included. Type I classical spicules
being shorter are very difficult to count. They obtained this result
by comparing Monte Carlo radiative transfer calculations with
observations obtained by the SOT/Hinode in the \caii\ H line.
However, \citet{sekse2012} questioned this estimate and argue that
this number is too high and would imply a much more spicules than
observed. \citet{moore2011} surmised that at any given time there are
$\sim$ 50 Type II spicules present per supergranule, or $\sim$1 Type
II spicule per 2$\times$10$^{17}$~cm$^2$ of surface area in quiet
regions and coronal holes. All given values depend, of course, on
the threshold intensity adopted for the spicule visibility and in
which height from the solar limb spicules are counted. In some
studies, a variation of abundance as a function of solar latitude
has been found. More precisely, the number of spicules was found to
decrease from poles to equator \citep{lippincott1957, Athay1959}.

\paragraph{Spicule inclination} The inclination of spicules is undoubtedly
reflecting the local magnetic field topology. It should be mentioned
that what can be measured is their apparent tilt with respect to the
local vertical, since what we actually see is the projection of a
spicule into the plane normal to the LOS and this leads to an
underestimate of its true inclination. Measurements are also biased
towards smaller inclination values, since the most inclined spicules
are either not raised high enough to be visible or it is difficult
to discern the shorter ones due to superposition effects. Old
studies, reviewed by \citet{beckers1968} report a mean variation of
$\sim$ 20\degr\ of spicules' inclination from the local vertical.
\citet{Heristchi_mou1992} measured the apparent tilts to the vertical
of 843 spicules and from the distribution of the inclinations they
found an average tilt of 29\degr. It should be noted that their
distribution was referred to an axis with a tilt of 14\degr\ in the
yz plane as they found that spicules tend to lean towards the
equator. Recent calculations by \citet{pasachoff2009} did not find
such a large tilt toward the equator. They measured an average
absolute inclination to the vertical of 23\degr. Their statistics,
however, contained only 40 spicules. A hint on the inclination of
spicules has been achieved recently by the efforts to determine the
magnetic field and its inclination in spicules (see Section
\ref{mag_field}). \citet{LopezAriste_cas2005} found a good
correlation between the orientation of the magnetic field lines and
\ha\ spicules (see Fig.~\ref{fig:lop_ari}), while
\citet{trujillobueno2005} found that the best fit to the observed
Stokes profile is obtained for a magnetic field inclination of
$\theta = 37\degr$. \citet{pasachoff2009} pointed also out that it is
possible that spicules' tilt varies with the solar cycle. Since
spicules follow the pattern of the magnetic field, such a variation
is possible. Considerable differences in the inclinations of
spicules at different heliographic latitudes have also been
reported. At the polar regions, spicules seem to have less variable
inclinations and to be more perpendicular to the surface, whereas
they attain more and more larger inclinations towards the equator as
the latitude decreases. \citet{Bonnet1980} also found that in
Ly--$\alpha$ photographs of the solar limb, spicules appeared to be
symmetrically distributed around the polar axis. There are several
older studies \citep{vandehulst1953, lippincott1957} that associate
spicules' inclination with that of the coronal polar rays, but such
detailed correspondence should be re-examined with the current high
resolution observations.

\paragraph{Spicule lifetime and birthrate} A spicule's lifetime is defined as the average
duration of spicule visibility. There is a large dispersion in the
reported values of the lifetime of spicules. Most studies agree that
the lifetime of spicules is between 2 and 12\,min
\citep{Roberts1945, rush_rob1954, lippincott1957,
alissandrakis_mac1971, cook1984, georgakilas1999} with 5\,min
considered as a typical mean value. The same range of values has
been confirmed for \ha\ spicules by \citet{pasachoff2009} who found
lifetimes between 3 and 12\,min, with a mean value of
7.1$\pm$2.3\,min. The same authors find shorter lifetimes in TRACE
1600\,\AA\ counterparts of \ha\ spicules pointing out a possible
impact of the lower resolution of TRACE observations. The highest
resolution available (from SOT/Hinode in the \caii\ H passband) have
attributed lifetimes around 3 -- 7\,min for the classical spicules
and lifetimes of the order of 10 -- 150\,s to the extremely
short-lived ones termed Type II \citep{depontieu2007b}.
\citet{beckers1968} gives as a spicule birthrate the number
330\,s$^{-1}$.

\paragraph{Spicule heights and widths} The height of a spicule is conventionally
measured from its foot at the photospheric limb up to where the
spicule becomes invisible. It should be noted that what is finally
measured is the projected height of the structure on the plane of
the sky. The height of a spicule is not a well defined quantity,
since there is not a sharply defined upper boundary and its
foot-point, as well as its location in front or behind the limb
cannot be easily established, while its determination depends
amongst others on the seeing conditions, line of observation,
exposure time, etc. Especially, in the \ha\ line there is a
chromospheric limb, which inhibits the tracing of the spicule to the
photosphere. Apart from these difficulties, studies of the
distribution of the apparent heights of spicules shows good general
agreement. \citet{beckers1968, Beckers1972} reviewed older studies
based on \ha\ observations, which gave estimates of average heights
between 6\,500 and 9\,500\km. Since then, spicules have been
reported to reach up to 15\,000\km\ \citep{cook1984} in \ha, but
also in He\,II \citep{georgakilas1999}. A recent analysis by
\citet{pasachoff2009} based on high resolution \ha\ observations of
the solar North--Western limb combined with co-temporal TRACE images
in the 1600\,{\AA} channel gives apparent heights in the range
4\,200 -- 12\,200\km\ with a mean of 7\,200$\pm$2\,000\km\ and agree
very well with the old results given by \citet{lippincott1957}. They
also identified spicules common in both \ha\ and 1600\,{\AA} image
data sets and found that in 1600\,{\AA} images, they were taller
(5\,600 -- 14\,700\,km) and correlate very well with the height in
\ha. They concluded that the heights reached by spicules observed in
the TRACE UV continuum are $\sim$ 2\,800\km\ greater than that
reached in \ha. Similarly, spicular structure up to $\approx$ 18\,Mm
has been reported obtained by TRACE in Ly--$\alpha$ an TRACE
1600\,\AA\ at the north solar limb \citep{alissandrakis2005}.
According to \citet{depontieu2007b}, the maximum lengths of spicules
observed in \caii\ by Hinode vary from a few hundred kilometers to
10\,000\km with most below 5\,000\km. Type II spicules attain
heights between 1\,000 and 7\,000\km\ and they are tallest in
coronal holes.

The diameter of spicules is probably the property most influenced by
observing conditions, such as atmospheric seeing and the diffraction
limit of the instrument, as well as overlapping effects that make it
difficult to separate individual spicules. Older studies based on
\ha\ and \caii\ H and K lines estimate the spicules' diameters to
range between 400 and 2\,500\km\ \citep{beckers1968, Beckers1972},
with most spicules having diameters between 400 and 1\,500\km\
\citep{dun60, lynch1973} which were characterized, at the time, by
Beckers as the ``most reliable'' values. Recent high resolution \ha\
observations provide a more strict estimate, in the range
300--1\,100\km\ with a mean diameter of 660\km\
\citep{pasachoff2009} which is in very good agreement with results
by \citet{nishikawa1988} who found an average diameter
615$\pm$250\km. \citet{pasachoff2009} also found that the width in
the 1600\,{\AA} TRACE channel is greater than the width measured
with the Swedish 1-meter Solar Telescope (SST) at \ha\ by a factor
of over 1.5, ranging between 700 and 2\,500\km. They argued that
TRACE widths are impacted by the telescope's broad point-broad
point-spread function, while the resolution of SST is approximately
four times higher that that of TRACE. With the advent of high
resolution instruments, thinner structures are being observed ($<$
200\km) and some have been attributed to Type II spicules. They have
also revealed that some \caii\ spicules consist of finer threads
(double thread structure) \citep{suematsu2008}. \citet{tavabi2011}
give an extensive Table together with references (see their Table 1)
of the various values obtained so far for the diameters of spicules.
Furthermore, these authors based on \caii\ H SOT/Hinode observations
at the northern solar limb found that spicule diameters show a whole
range of widths from 200 to more than 1\,000\km, with distributions
at four distinct diameters.

\begin{figure}[t]
\centering
\includegraphics[angle=180, width=0.95\textwidth]{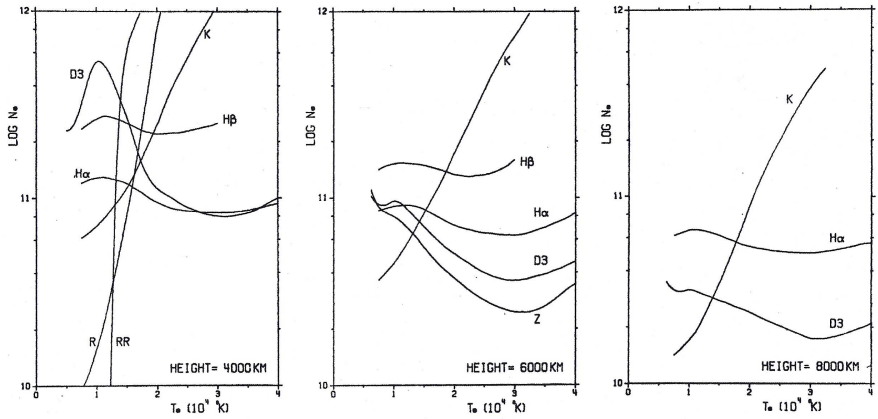}
\caption{Relation between spicule temperature and electron density
at different heights as derived for different lines. $Z$ stands for
the He 10830 line, $R$, for the ratio of 3888\,\AA\ helium to
hydrogen line intensity, $RR$ for the ratio of \caii\ and
He--D$_{3}$ line as derived from eclipse spectra (from
\citet{Beckers1972})} \label{temp_ne}
\end{figure}

\paragraph{Spicule temperatures and electron densities} Spicule temperatures and densities
can be estimated spectroscopically by using spectra taken
simultaneously in several different lines. It should be noted that
temperature (and electron density) diagnostics of spicules from line
intensities rely on data from constituents that are especially
difficult to model. In order to interpret line profiles, non-LTE
calculations have to be used which is not an easy task. Due to the
difficulty of such calculations there are very few reports on
temperature and density determinations in spicules and very large
discrepancies on the reported values. Unfortunately, recent
theoretical results do not exist. Compilation of older theoretical
results is given by \citet{Beckers1972}. Beckers derived these
parameters by assuming that a spicule is a cylinder situated
vertically on the Sun and irradiated by the solar radiation having a
diameter of 815\km. From the intensity of a spectral line he
obtained a $T_e$--$N_e$ curve. He then compared the observed line
intensities with the predictions of theoretical models. Curves
obtained by this way from different lines will intersect, giving a
set of temperature and density that is consistent with all
observations (Fig.~\ref{temp_ne}). \citet{Beckers1972} derived the
temperature from the observed intensities of \ha, \heii, and \caii\
lines from simple non-LTE calculations. He reports a temperature of
9\,000\,K and an electron density of 1.6$\times$10$^{11}$cm$^{-3}$
at a height of 2\,000\km\ and a temperature of $\sim$ 16\,000\,K at
4\,000 -- 8\,000\km, while for the electron densities he derived
values 1.5$\times$10$^{11}$cm$^{-3}$ at 4\,000\km\ and
4.3$\times$10$^{10}$cm$^{-3}$ at 8\,000\km.

\citet{krat_kra1971} obtained simultaneous spectrograms of spicules
in the lines \ha, \hb, D3 and \caii\ H and K at heights varying from
5\,000\km\ to 9\,000\km\ above the limb, the slit being set parallel
to the limb. They concluded that the emission in the different lines
originates in different parts of spicules with different radial and
turbulent velocities and that the electron density varies from
10$^{11}$cm$^{-3}$ to 10$^{12}$cm$^{-3}$.

\citet{alissandrakis1973} analyzed simultaneous spectra of spicules
at a height of 5\,400\km\ above the limb in the \ha, \hb\ and \caii\
lines. Since the intensity of the Balmer lines is a very good
indicator of the electron density, as they are almost insensitive to
temperature, while the K line is sensitive to both the $N_e$ and
$T_e$, he combined the observations to obtain the temperature and
density of spicules. He found that the values of the electron
density for 36 spicules at a height of 5\,400\,km range between
6$\times$10$^{10}$cm$^{-3}$ and 1.2$\times$10$^{11}$cm$^{-3}$, with
an average value of 6$\times$10$^{10}$cm$^{-3}$, while the range of
the electron temperature was between 12\,000\,K and 15\,000\,K with
an average of 13\,000\,K.

\citet{krall1976} analyzed time sequences of simultaneous spectra of
limb spicules in the \ha\ and \caii\ H and K lines. Using the
measured intensities in these lines they derived electron densities
averaged over the entire visible lifetime of spicules of $\sim$
6$\times$10$^{10}$cm$^{-3}$ at the height of 5\,000\km\ and minimum
and maximum values of $\sim$ 1.1$\times$10$^{11}$cm$^{-3}$ at
6\,000\km\ and 2.$\times$10$^{10}$cm$^{-3}$ at 10\,000\km,
respectively. They also found electron temperatures ranging between
12\,000\,K and 16\,000\,K, while from profile half-widths, they
concluded that turbulent velocities should be in the range between
12 and 22\kms.

\citet{braun_lindsey87} used brightness limb observations at 100 and
200\,$\mu$m and 2.6\,mm. Use of far-infrared continuum observations
are useful because the continuum emission is formed in LTE and the
source function is just the Planck function, which in the
Rayleigh--Jeans limit is proportional to the electron temperature.
They calculated limb intensity profiles for a variety of spicule
models and found that limb profiles are well fitted by spicules with
an electron temperature of the order of 7\,000\,K up to heights of
at least 7\,000\km\ above the photosphere. Particularly the 2.6\,mm
observations exclude spicule temperatures of 16\,000\,K below
7\,000\km, because this temperature should create a substantial limb
brightening, which is not observed.

\citet{matsuno1988} determined the height distribution of the kinetic
temperature of \ha\ spicules. The temperature was found to decrease
from 9\,000\,K at 2\,200\km\ to 5\,000\,K at 3\,250\km\ and to
increase up to 8\,200\,K at 6\,000\km. They suggested that the
decreasing temperature with height might be related to the lateral
expansion of rising spicular material along the expanding magnetic
field lines, while the increasing temperature above 3\,250\km\ may
be due to heating by the penetrating radiation in the Lyman
continuum originating in the EUV emission of the transition
region--corona.

\citet{socasnavarro_elm2005} used multiline limb observations of
spicules from different elements and, specifically, in the \caii\
8498~\AA\ and 8542~\AA\ lines using the Advanced Stokes Polarimeter
(ASP) at the Dunn Solar Telescope (DST) and the \hei\ multiplet at
10830\,\AA\ recorded with the Spectro-Polarimeter from INfrared and
Optical Regions (SPINOR), to derive some spicule properties. They
found that the Doppler widths of the observed spectral lines are
similar. Since the Ca atom is ten times heavier than the He atom, if
the lines were broadened by microscopic thermal velocities, one
would expect the He line to be more than three times broader than
the Ca lines, which is not the case. Assuming non-thermal broadening
they concluded that the electron temperature of the spicule should
be lower than 13\,000\,K to account for the observed line
broadening.

\begin{figure*}[t]
\begin{center}
\includegraphics[width=0.4\textwidth]{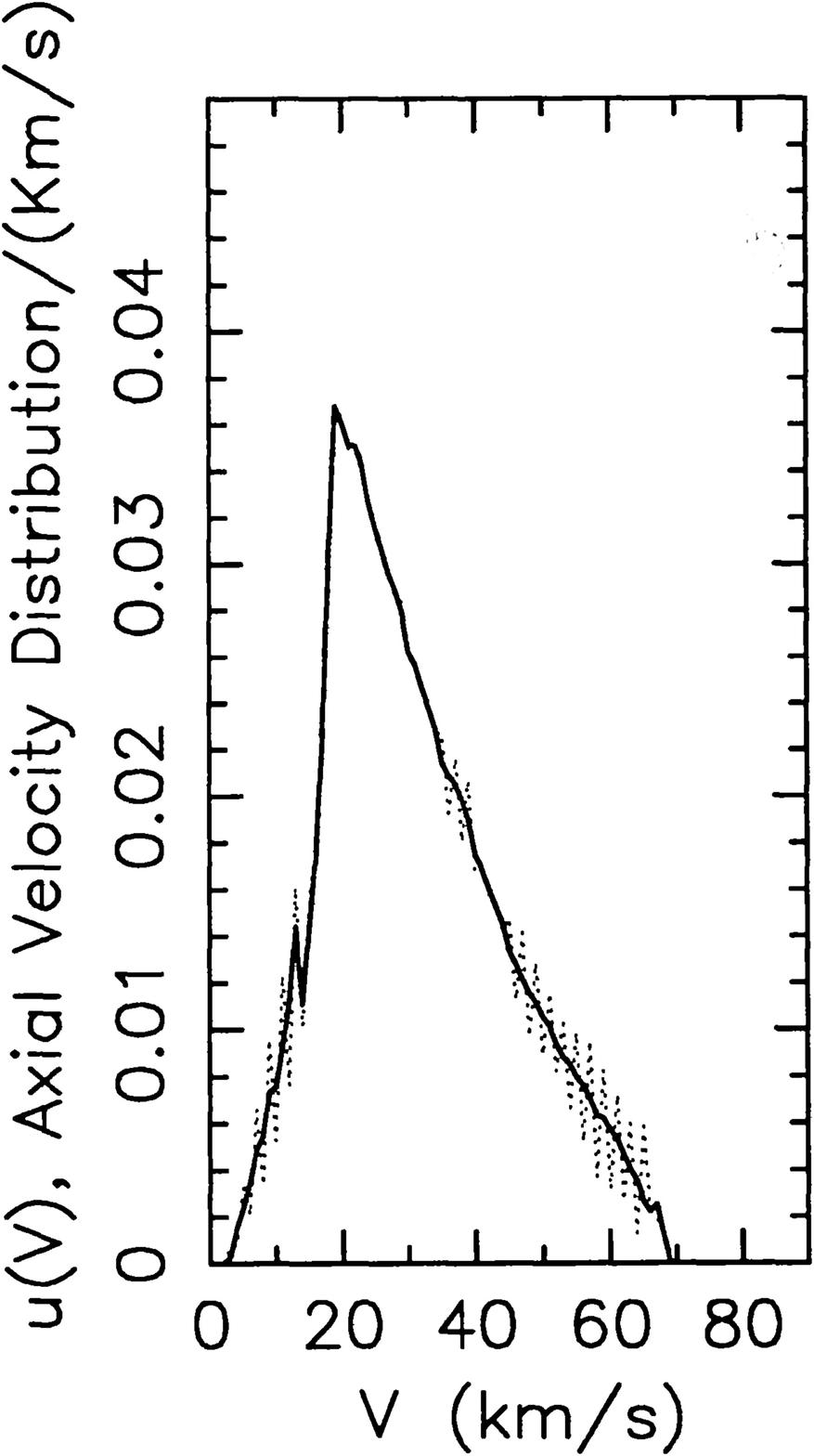}
\includegraphics[width=0.4\textwidth]{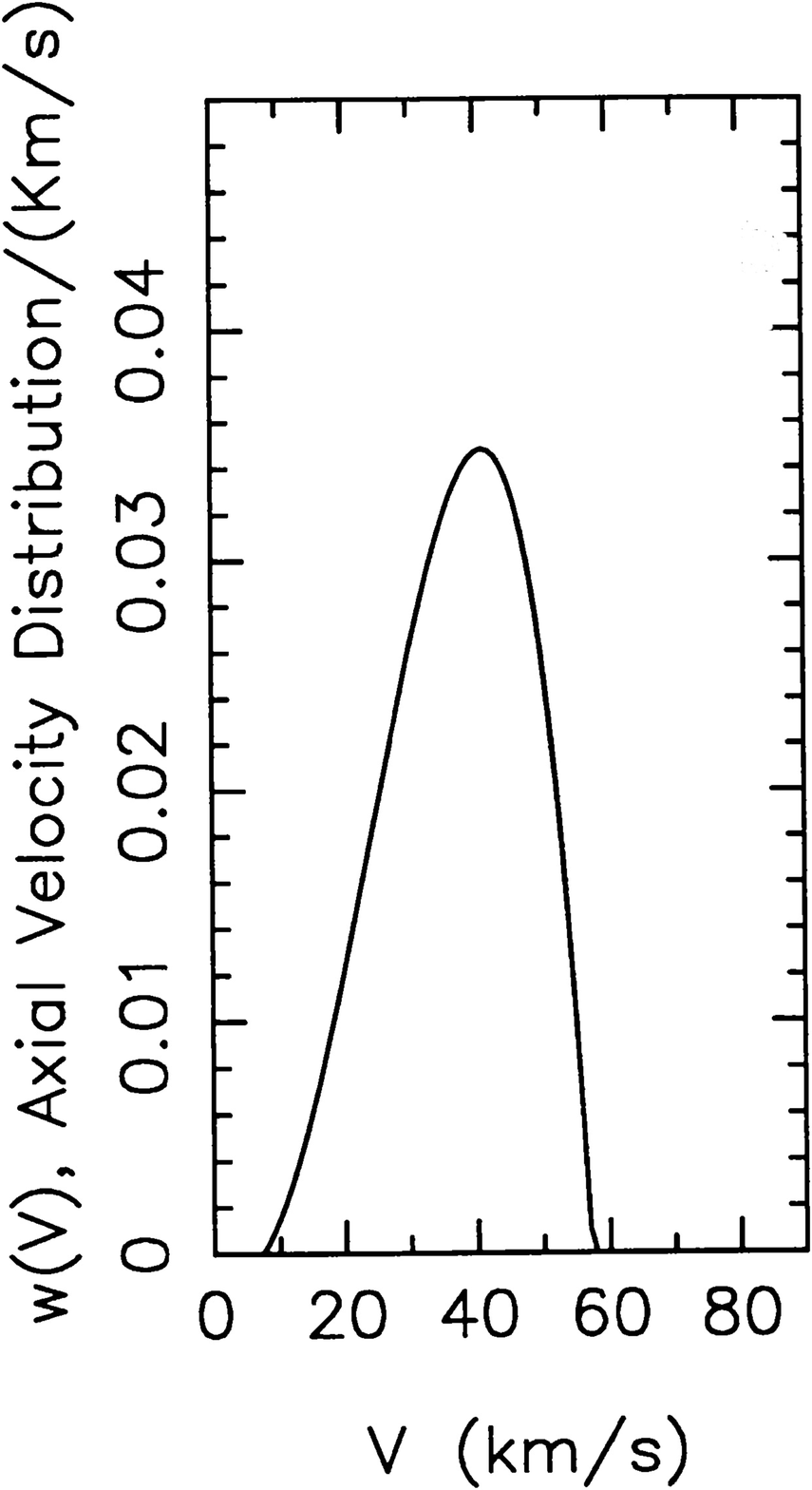}
\end{center}
\caption{{\it Left}. The apparent axial velocity distribution in spicules measured from
the time variation of their heights. {\it Right}. The axial mass velocity distribution
obtained from the Doppler velocities (from \citet{Heristchi_mou1992})} \label{axial_velo}
\end{figure*}

\begin{figure}[t]
\centering
\includegraphics[width=0.48\textwidth]{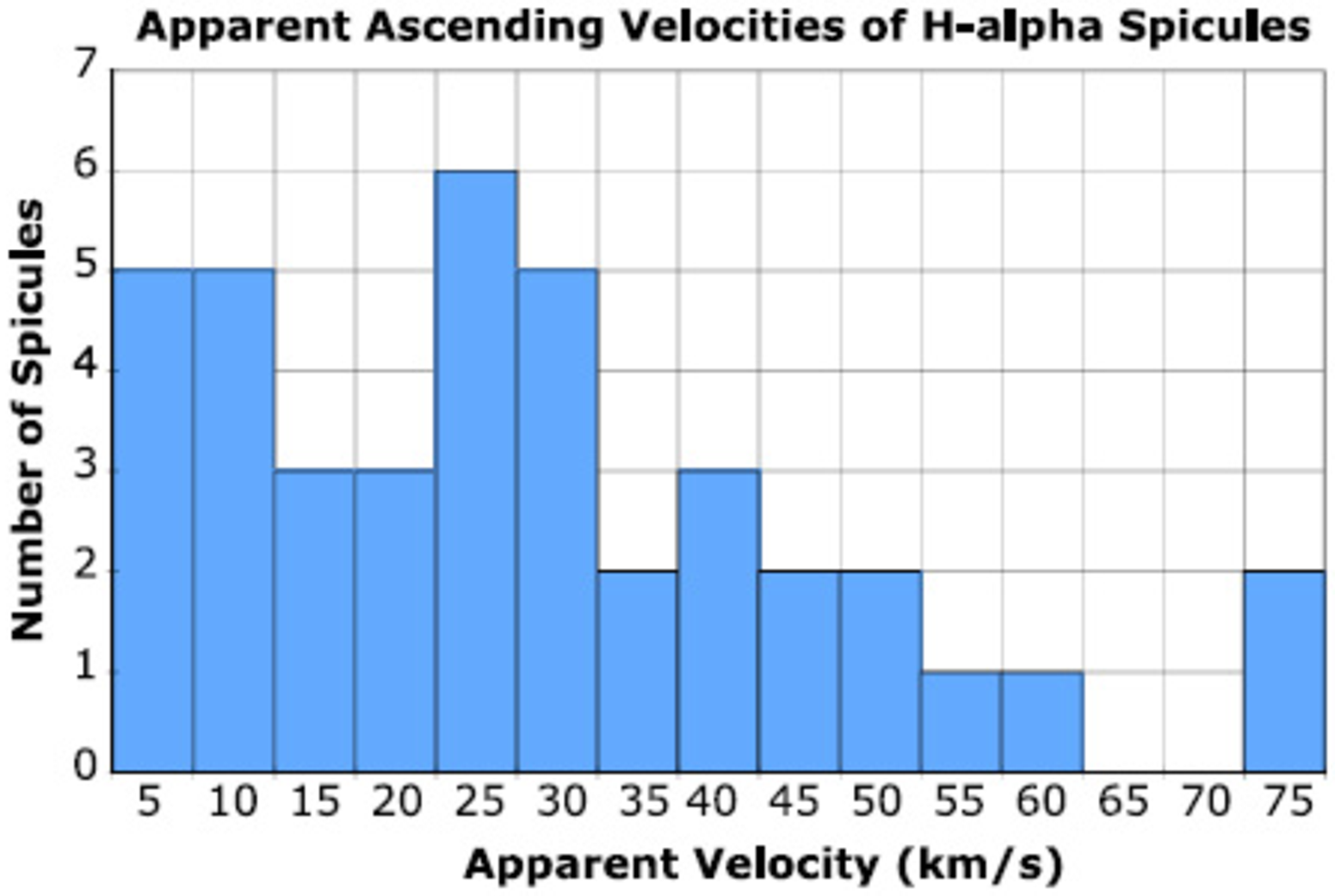}
\includegraphics[width=0.48\textwidth]{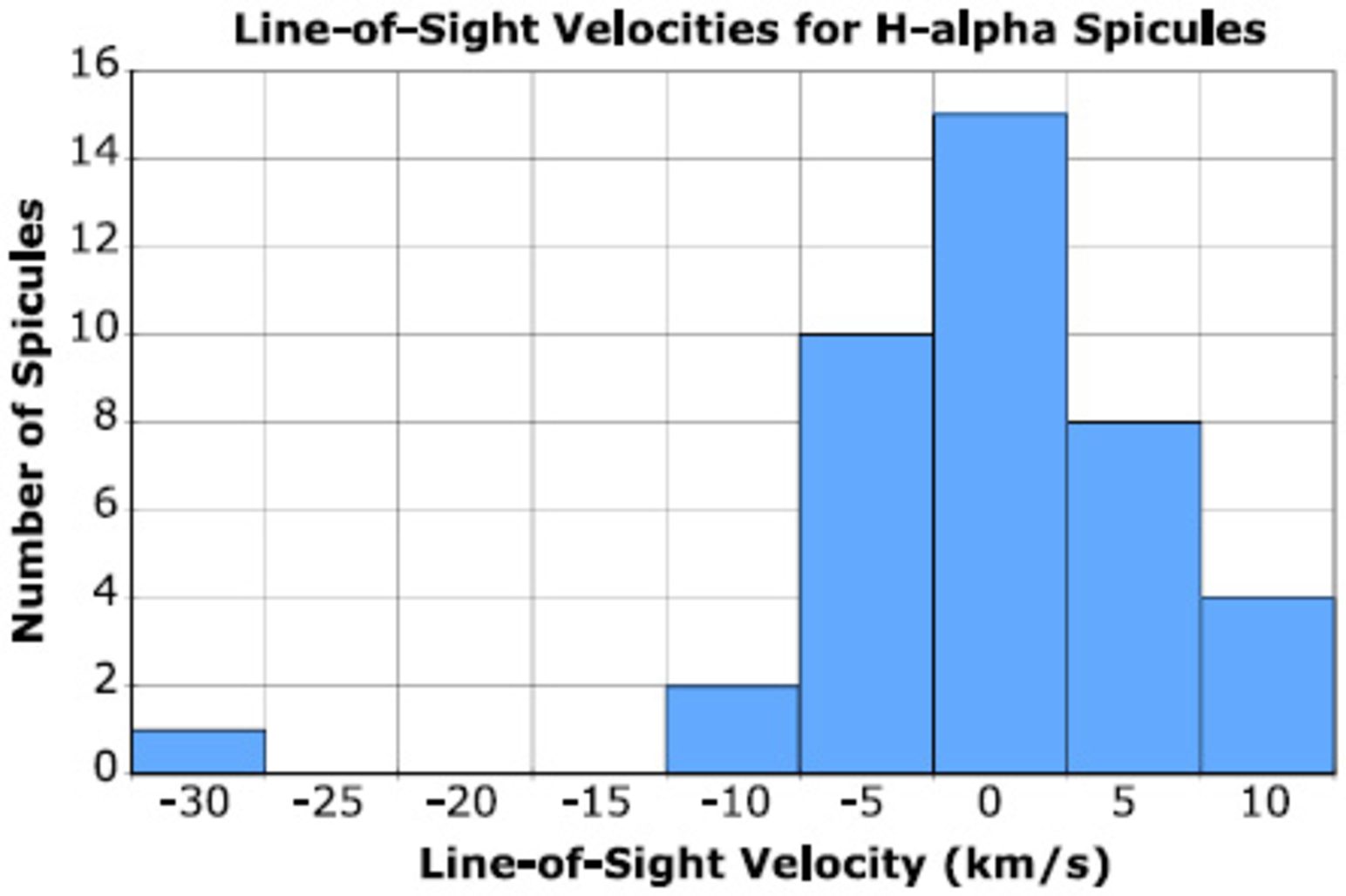}
\caption{{\it Left}. Histogram showing the distribution of apparent ascending
velocities. {\it Right}. Histogram showing the distribution of the LOS velocities of the
tops of the spicules when they were at their maximum heights (from
\citet{pasachoff2009})} \label{appar_velo}
\end{figure}

\paragraph{Evolution and velocities of limb structures} For limb spicules,
different and sometimes contradictory results on the velocities and
their variations have been reported. One of the reasons is due to
the different types of observations that have been used. Apparent
velocities correspond to motions in the plane perpendicular to the
LOS and could equally well result from the temporal variation of the
excitation or ionization of the spicular material, or the
propagation of a shock front. The general description based on
measurements of apparent velocities is consistent with a spicule
which rapidly elongates upwards with an average velocity of $\sim$
25\kms\ and reaches its maximum height within a minute or two after
its initial appearance. The upward motion is usually fairly regular
and continuous, the spicule rising and stopping abruptly at its
maximum height. Subsequently, the spicule may either fade from
visibility or else descend back to the low chromosphere with a
velocity comparable to that of its initial ascent. Measurements of
vertical proper motions were made in the 1950's and 1960's by
\citet{rush_rob1954} and \citet{lippincott1957}, and have been
reviewed by \citet{beckers1968, Beckers1972} and \citet{bray_lou1974}.
In an old work, \citet{lippincott1957}, after studying the apparent
motions of spicules, reported that in half of them the ascending
phase is followed by a descending one, while the other half is
fading away; she was also the first to report that some spicules
appear to rise from the same source several times. The majority of
descending spicules appear to retrace their original paths, but
according to \citet{Beckers1972} there are many which fall back along
different paths, often forming an arch. \citet{Heristchi_mou1992}
combining the observations of \citet{rush_rob1954} and
\citet{lippincott1957}, which gave the variation of height with time
of spicules, constructed the observed distribution of apparent
velocities relating them to the angular distribution of spicules
obtained from their axial velocity distribution
(Fig.~\ref{axial_velo}, left). They give as the most probable
velocity 20\kms\ and a range between 0\kms\ and 60\kms.
\citet{depontieu2007b} from the analysis of chromospheric \caii\ H
observations taken with the broadband filter on board the Hinode
mission found that spicules shoot upward at speeds between 20 and
150\kms. Recently, \citet{pasachoff2009} reported that in the
majority of the 40 spicules examined, they found only ascending
velocities because most of the spicules seemed to disappear at their
maximum heights. They measured ascending velocities with a mean of
27.0$\pm$18.1\kms, a median of 25\kms\ and a range of 3.0 to 75\kms\
(Fig.~\ref{appar_velo}, left).

Apart from the velocities deduced from direct observations of the changes of the height
of spicules with time, velocities can also be inferred from spectroscopic observations
which give information on Doppler shifts. The Doppler shifts of the emission features
can be interpreted as LOS mass motions of the spicular material. It should be noted that
for a spicule inclined to the plane of observation, the component of the rise velocity
corresponding to the sine of the angle of inclination would be observed.

\begin{figure}[t]
\centering
\includegraphics[angle=180, width=0.7\textwidth]{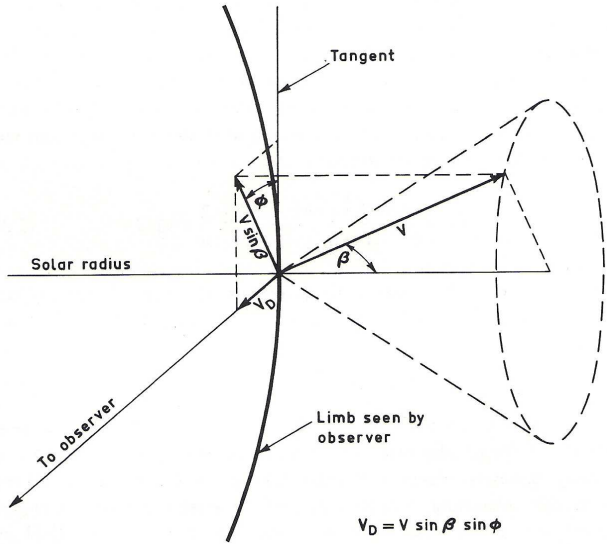}
\caption{Resolution of the velocity $V$ of a spicule along the LOS.
$V_{D}$ is the LOS component, $\beta$ is the angle made by the
spicule with the solar radius, and $\phi$ is the azimuthal angle of
the projection of the spicule velocity vector onto a plane tangent
to the limb (from \citet{bray_lou1974})} \label{geometry}
\end{figure}

\citet{alissandrakis1973} using simultaneous spectra in \ha, \hb\ and
\caii\ K at a height of 5\,400\km\ above the limb obtained the LOS
velocity distribution of 50 spicules, as inferred from the position
of the maximum of the line profile in the three lines. Thus he found
an absolute average value of the LOS velocity of 6.4\kms\ for \ha,
5.6\kms\ for \hb\ and 4.9\kms\ for \caii\ K. Using this technique
\citet{Heristchi_mou1992} measured the LOS velocity component of 90
well--isolated spicules at 5\,000--6\,000\km\ above the limb. To
improve the statistics they have also added some other available
distributions (see their Table I). Relating the measured Doppler
velocity distribution (which was approximately symmetric ranging
between -30 to 30\kms) with the angular distribution of inclinations
they obtained as a most probable axial velocity 40\kms\
(Fig.~\ref{axial_velo}, right), i.e. about 20\kms\ larger than the
most probable apparent upward velocity. Recently
\citet{pasachoff2009} measured LOS velocities from Doppler shifts of
the tops of spicules found at their maximum heights
(Fig.~\ref{appar_velo}, right). The Doppler shifts were calculated
by fitting Gaussians to 5--wavelength \ha\ spectra. They reported a
mean absolute value of 5.1$\pm$5.1\kms, a median of 3.8\kms\ and a
range of 0.3 to 30.2\kms. This value agrees with the mean value of
6\kms\ quoted by \citet{Beckers1972}. They also found that the
distribution of the LOS velocities at the bases of the 40 spicules
measured (where the base is defined as the location of the apparent
limb) has a mean of 3.1$\pm$11.9\kms\ and a range of $-$17.8 to
51.2\kms. As noted from their Doppler measurements, they regularly
found oppositely directed motions, i.e. bi--directional, which could
be an indication of magnetic reconnection. One has to bear in mind,
however, that whereas photographic observations give the vertical
velocity of the spicules in the plane of the limb, spectroscopic
observations give the velocity component along the LOS and, hence,
in the direction normal to the plane of the limb. This fact can
explain the discrepancy between the average apparent vertical
velocity value of 25\kms\ and an average value of $\sim$ 6\kms\
obtained for the LOS velocity of spicules. Following
\citet{athay_bes1964}, one can show that the magnitude of the average
Doppler velocity of spicules lying in a cone whose axis is normal to
the LOS and whose apex angle is 2$\theta$ (see Fig.~\ref{geometry})
is given by the formula:

\begin{equation}
\bar{V}_{D}=\frac{V\sin\theta\int_{0}^{\pi/2}\sin{\phi}d{\phi}}{\int_{0}^{\pi/2}d{\phi}}=\frac{2}{\pi}V\sin{\theta}
\end{equation}

\noindent where $\phi$ is the azimuthal angle of the projection of
the spicule velocity vector onto a plane tangent to the limb. At
greater heights, the observed Doppler shifts will be positive or
negative depending on whether the spicule is inclined away from or
towards the observer. Thus for $V$=25\kms\ and $\theta$=30\degr\ we
get 8\kms\ for the average LOS velocity of spicules, which is close
to the value reported from observations.

\citet{Heristchi_mou1992}, as stated above, found a large difference
($\sim$ 20\kms) between the maxima of the distributions of the axial
velocities deduced from the measurements of the apparent and Doppler
velocities of spicules. As they anticipated this difference is
presumably due to the ionization of the hydrogen atoms, when the
spicules, penetrating the corona, are heated. Thus if a partially
ionized plasma parcel is moving along the axis of a spicule with a
velocity of $\sim$ 40\kms\ and at the same time its front is being
ionized and an ionization wave is moving backwards with a velocity
of $\sim$ 20\kms\ then the apparent velocity would be equal to
20\kms. They also noted, however, that the differences in the
obtained values may also be due to the fact that the data they used
for their calculations were taken at different periods of the solar
cycle, at different locations on the Sun, with different instruments
and with different techniques.

The height dependence of the LOS velocity in spicules has also been
investigated by several authors. The results obtained differ
significantly from each other. For instance, \citet{mouradian1965}
has found that the Doppler shift decreases along the spicule,
\citet{beckers1966} found an increase of the Doppler shift with
height, while \citet{pasachoff1968} reported a weak or even no
observable change of the shift with height. The main reason for
these discrepancies are due to the fact that observations of
spicules at various heights simultaneously are related to
instrumental difficulties. \citet{kulidzanishvili1980} studied almost
simultaneously height sequences of 69 spicules in the \ha\ line. He
claimed that absolute values of LOS velocities increase linearly
with height and that no variation of the sign of the LOS velocity
along individual spicules is observed.

\begin{figure}[t]
\centering
\includegraphics[width=0.3\textwidth]{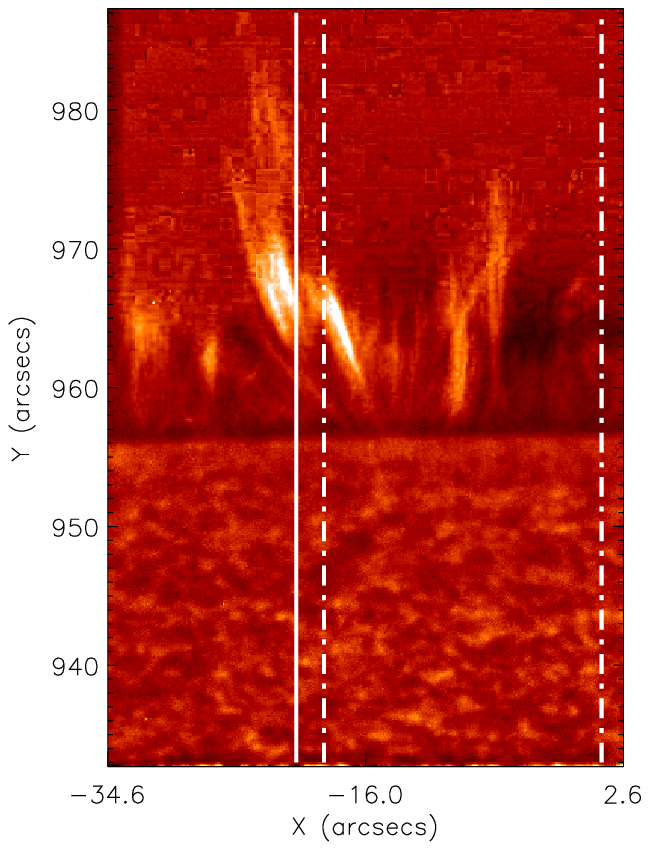}
\includegraphics[width=0.3\textwidth]{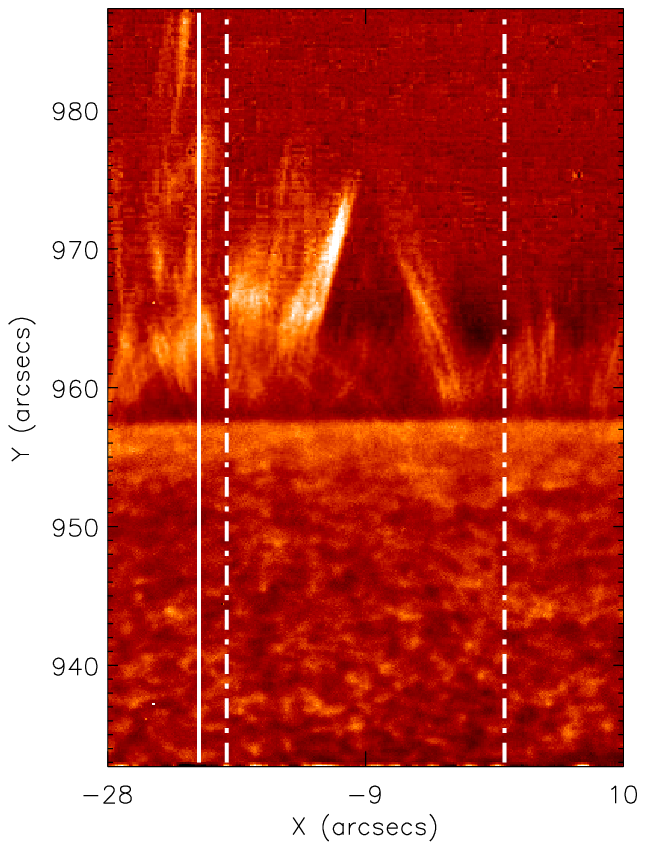}
\includegraphics[width=0.3\textwidth]{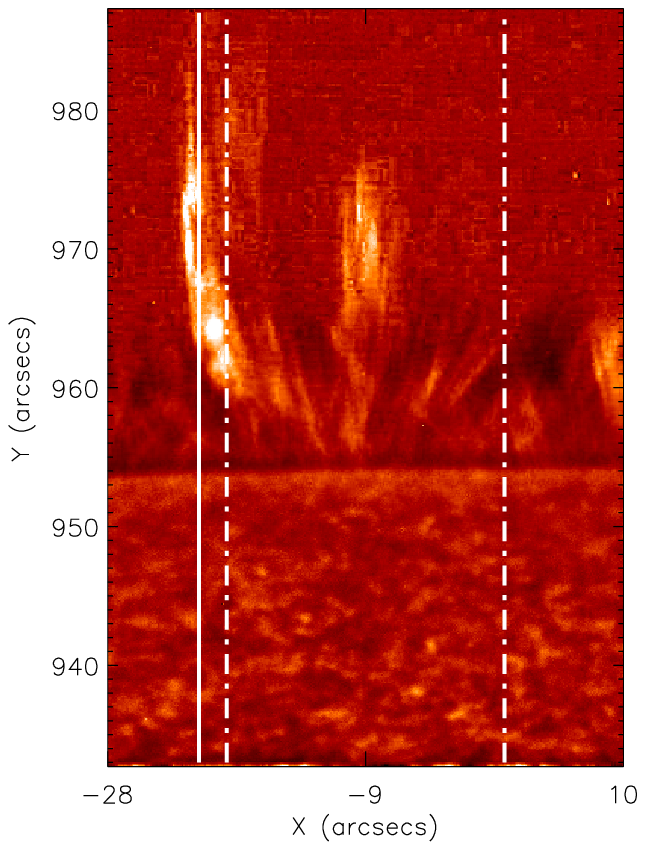}
\caption{SOT/Hinode \caii\ H images showing spicules taken from left
to right, on 2009, April 28 at 16:24 UT, on April 29 at 02:24 UT and
03:28 UT. The over-plotted solid line indicates the position of the
SUMER slit, while the dashed-dotted lines indicate the EIS raster
field-of-view (FOV) (from \citet{madjarska2011})}\label{spicules2011}
\end{figure}

It is quite evident that spicule motions and their time variations
are essential in clarifying the mechanism responsible for their
formation. However, reported time variations of Doppler shifts are
contradictory. \citet{mouradian1965} reported that the Doppler shift
of spicules increases progressively with time, goes through a
maximum and then decreases, but it does not change sign, being
always upwards. \citet{pasachoff1968} found that Doppler velocities
of a certain number of spicules do undergo quasi-periodic reversals.
Recently, \citet{wilhelm2000}, in a study of spicules observed in
several EUV lines by SUMER on SOHO, reported strong red and blue
shifts within a feature, which, furthermore, reverse sign.
\citet{madjarska2011} have recently analysed three large spicules
(Fig.~\ref{spicules2011}) and found them to be comprised of numerous
thin spicules which rise, rotate and descend simultaneously forming
a bush-like feature. Their rotation resembles the untwisting of a
large flux rope. They show velocities ranging from 50 to 250\kms.
These data will be further discussed in Part II of these review
series.

\citet{nishikawa1988} studied the proper motion of polar limb
spicules using time sequence filtergrams of single wavelength bands
centered on \ha--0.9\,\AA\ or \ha+0.9\,\AA\ (the passband was
0.5\,\AA) obtained with the Domeless Solar Telescope at the Hida
Observatory. After following successively from frame to frame four
individual spicules he examined whether spicule motions can be
interpreted as representing ballistic or constant velocity. By
fitting the height variation with a quadratic function of time (a
constant acceleration model) he found large initial velocities of 80
-- 100\kms\ and large acceleration of 0.45 -- 0.65\,km\,s$^{-2}$. No
disk features that can provide such initial velocities have been
observed. He pointed out, however, the possibility that such a
source could be small enough that it is not observed with the
current resolution and concluded that the ballistic model of spicule
motions cannot be rejected from the observational point of view. He
also pointed out the possibility that a rising and a falling motion
can be separately fitted with a constant velocity model. In this
case, the constant velocities ranged from 30 to 50\,\kms.

\citet{christopoulou2001} applied an image processing technique to
high-resolution observations obtained with the DST of the Sacramento
Peak Observatory. They estimated the proper motions of the apparent
tops of several spicules, by performing a least-square fit to the
height measurements as a function of time assuming a ballistic
motion, i.e. that the plasma is subject only to gravitational
forces. They obtained initial velocities larger than 40\kms\ in all
cases examined, suggesting that the magnetic field or other forces
should play an important role in the generation of spicules. It is
clear from the above  that the behaviour of the flow along spicules
remains an open question.

\begin{figure}[t]
\centering
\includegraphics[width=0.95\textwidth]{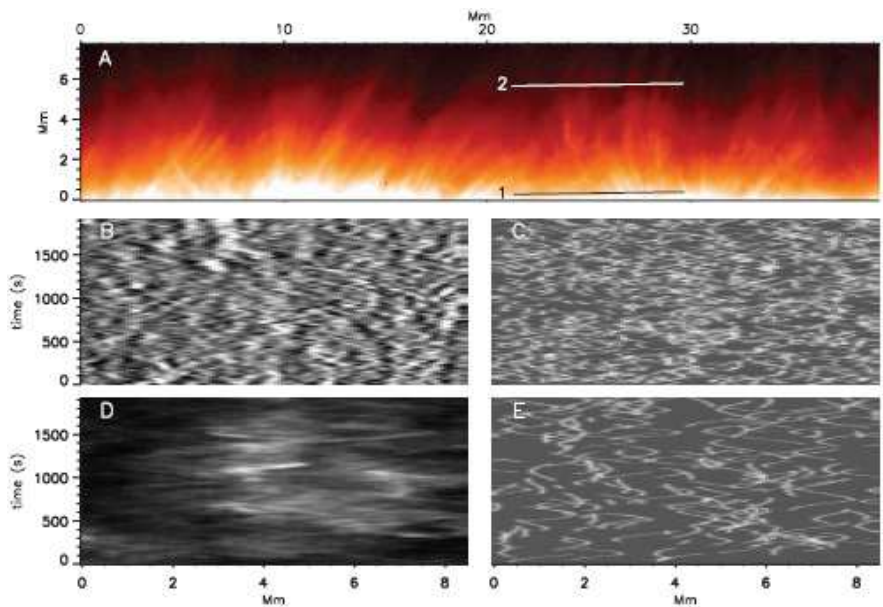}
\caption{(A) Spicules observed at the solar limb with SOT/Hinode in
\caii\ H. (B) A space-time plot (along the cut labeled 1 in (A)) of
the \caii\ H intensity. (D) A similar cut for the line labeled 2 in
(A). The plot is dominated by a multitude of criss-crossed short
linear tracks caused by spicular motion transverse to the magnetic
field direction. Similar linear tracks are visible, as well as
swings. The general characteristics (linear tracks and swings) of
(B) and (D) are well reproduced by cuts that are generated from
Monte Carlo simulations shown in (C) and (E) (from de
\citet{depontieu2007sci})} \label{swaying}
\end{figure}

Apart from the flows along spicules, motions of spicules themselves
parallel to the limb have been reported. From sequences of \ha\
spectra taken with the slit tangential to the limb,
\citet{pasachoff1968}, \citet{nikolsky1970}, \citet{weart1970},
\citet{nikolsky_pla1971} found evidence that some spicules change
their position along the slit relative to the average positions of
other features thus apparently showing a real horizontal component
of velocity. Nikolsky and Platova have concluded that the spicules
execute quasi--periodic oscillations parallel to the limb with a
period of $\sim$\,1\,min, characteristic amplitude of 1\arcsec\ and
velocities about 10 -- 15\kms. \citet{kukhianidze2006} performed
observations of solar limb spicules at 8 different heights, i.e.
$\sim$ 3\,800 -- 8\,700\km\ above the photosphere was covered. They
found that $\sim$ 20\% of the measured height series showed a
periodic spatial distributions in the Doppler velocities with a
wavelength of 3.5\,Mm and periods in the range of 35 -- 70\,s. They
suggested that the spatial distribution was caused by transverse
kink waves. Observations of such lateral or swaying motions have
again been revealed recently from the unprecedented high spatial and
temporal resolution images obtained with the SOT on-board the Hinode
spacecraft. \citet{depontieu2007sci} analyzed time series of \caii\ H
images taken with a SOT broadband filter. They reported that many of
the short-lived spicules undergo substantial transverse
displacements of the order of 500 to 1000\km\ during their short
lifetimes of 10 to 300\,s, while some longer-lived spicules undergo
a transverse motion with the displacement varying sinusoidally in
time (Fig.~\ref{swaying}). They suggested that this behavior is
strongly indicative of Alfv\'en waves, where the term Alfv\'en waves
is used to describe incompressible transverse MHD waves that
propagate along the magnetic field in an inhomogeneous medium. They
pointed out, however, that the observed waves could also be
interpreted as MHD kink-mode waves, should a stable waveguide exist
in the chromosphere. The fine structure and lateral motion of
spicules has also been observed and reported by \citet{suematsu2008}.
These authors suggested that spicules can be driven by magnetic
reconnection in unresolved spatial scale taking place at their
foot-points. They also suggested that since most spicules emanate
from a seemingly uni-polar magnetic region the relevant magnetic
reconnection must take place in unresolved spatial scale contrary to
the larger-scale jets associated with an emergence of a small
bipole.

Another unexplained feature that has been reported is that sometimes
a ``tilt'' of the spectral line is observed in spectrograms, i.e.,
an inclination at a small angle of the line profile of some spicules
to the direction of the spectrograph's dispersion
\citep{beckers1968, pasachoff1968, weart1970}. Such inclinations
have been observed in the \ha\ and Ca H and K spectra of spicules
and could be attributed to an actual difference in the LOS
velocities between the two sides of the spicule. One obvious
explanation would be that in these cases there is actually not one
but two separate unresolved spicules with different Doppler
velocities which are seen as one because of the low spatial
resolution. Another explanation which was given is that this
inclination may be interpreted as resulting from the rotation of the
spicule about its axis which would produce differential mass
motions. If, for example, the slit of the spectrograph is oriented
across the width of a rotating spicule, with the side rotating
towards the observer (towards the top of the slit) and the side
rotating away (towards the bottom of the slit), then the spectral
line will be inclined from the shorter wavelengths to longer
wavelengths going from top to bottom along the spectral line.
\citet{pasachoff1968} gave an upper velocity limit for this spicule
rotation at about 30\kms. By considering a simplified model of a
spicule as a column of gas 1000\km\ in diameter rotating with this
peripheral speed they pointed out that it would have a velocity
gradient of 60\ms\ per km across it which would produce a spectral
tilt of about 1\degr.9 in K and 2\degr.6 in \ha. As they noticed
this peripheral rotation velocity implies a centripetal acceleration
of $\sim$ 1.8\kms, i.e. about 6 times greater than the value of
solar gravity. \citet{alissandrakis1973}, on the other hand,
calculated a rotational velocity of $\sim$ 8\kms\ and concluded that
the rotation of spicules is almost negligible. Recently,
\citet{depontieu2012} using high-quality observations from SST
provided evidence that most, if not all, Type II spicules are
characterized by the simultaneous action of three different types of
motion: (1) field-aligned flows of order 50 -- 100\kms, (2) swaying
motions of order 15 -- 20\kms, and torsional motions of order 25 --
30\kms. These last motions confirm the existence of twisting
spicular motions.

\begin{figure}[t]
\centering
\includegraphics[width=0.95\textwidth]{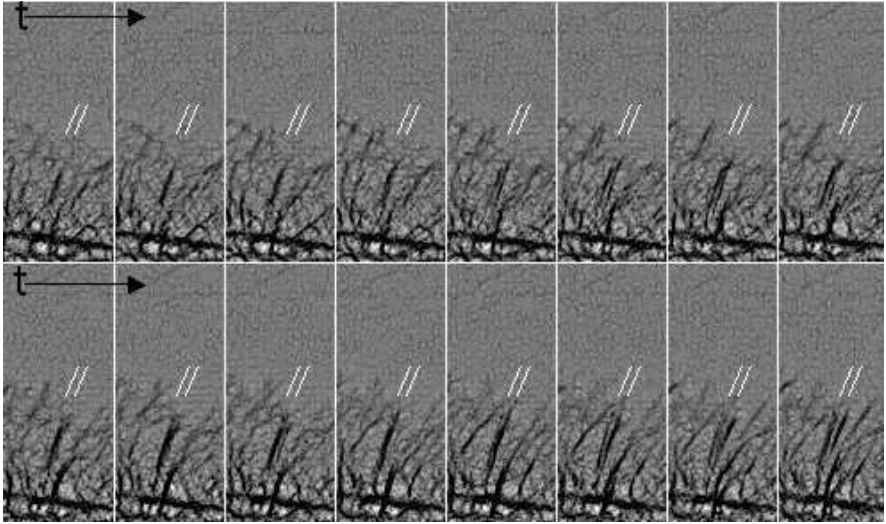}
\caption{Time series of sharpened images from Hinode \caii\ H
filtergrams. The cadence is 5\,s. The series shows that the spicule
consisting of double thread (indicated by white lines) is spinning
as a whole body (from \citet{suematsu2008})} \label{sue.fig2}
\end{figure}

\begin{figure}[t]
\centering
\includegraphics[width=0.8\textwidth]{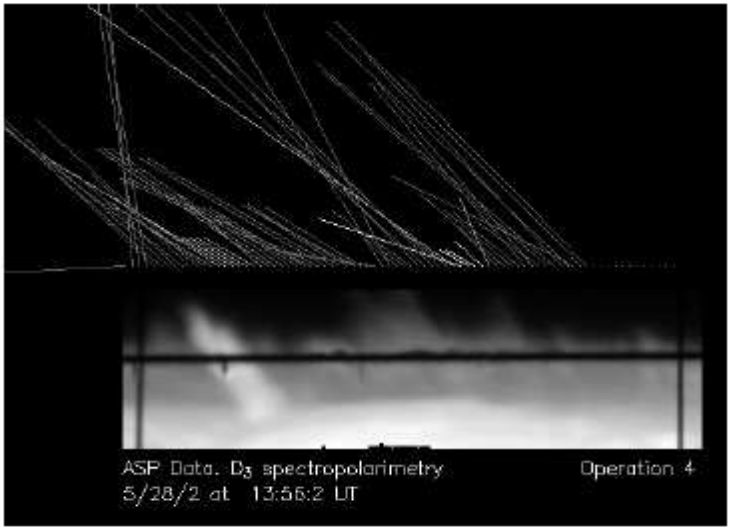}
\caption{In the background, is the slitjaw image in \ha\ showing the
position of the slit (horizontal) over the spicules. Superimposed on
the same column (with a vertical offset) the inverted solution for
the magnetic field vector for each point on the slit is plotted. The
direction of the lines is given by the azimuth in the LOS reference
frame, and their length by the field strength. The level of gray of
each line is scaled according to the intensity of the observed
feature in the slit-jaw image (from L\'{o}pez Ariste \& Casini
(2005))} {\label{fig:lop_ari}}
\end{figure}

Hinode high resolution observations revealed that most spicules show
up a double thread structure during their evolution
(\citet{suematsu2008}). They also revealed that the separation of
some of the double thread spicules changes with time, alternating
between a single-thread phase and double-thread one
(Fig.~\ref{sue.fig2}). This change in separation can be interpreted
by a spinning of spicules as a whole body (spin period of 1 -- 1.5
min and velocity $\sim$15\kms). \citet{sterling2010} again from
\caii\ H SOT/Hinode observations reported that they found many
spicules to expand laterally or split into two or more strands after
being ejected. A possible explanation could be that many of the
splitting or expanding spicules could be small-scale magnetic
eruptions, analogous to coronal mass ejections. Consistent with this
idea, the motion of the splitting spicules is similar to the
spreading of the legs of filament eruptions.

\paragraph{Magnetic fields in spicules} \label{mag_field} The determination of the magnetic
field vector in solar spicules is an important step to understand
them. This determination can be achieved through specto-polarimetric
measurements and theoretical modelling of the Zeeman and Hanle
effects in suitably chosen spectral lines. The He~{\sc i}
10830\,\AA\ triplet is a powerful diagnostic tool for chromospheric
magnetic fields \citep{trujillobueno2010}, because its spectral
signatures are sensitive to scattering polarization, while the
Zeeman, Hanle and Paschen--Back effects make it a very useful
magnetic indicator within a wide range of field strengths.

The Zeeman effect is the splitting of atomic energy levels (and of
the associated spectral line) into several components in the
presence of a magnetic field. In the linear regime this splitting is
proportional to the magnetic field strength and the Land{\'e} factor
of the energy level. The longitudinal Zeeman effect produces
measurable Stokes--{\it V} signals in the presence of weak magnetic
fields, while to observe the signature of the transverse Zeeman
effect on the Stokes {\it Q} and {\it U} profiles, field strengths
of more than 100\,G are needed. The Zeeman effect may produce
polarized radiation that can be analyzed to infer the properties of
the field. However, when the magnetic field is unresolved within the
resolution element or cancels out when mixed polarities are present
or too weak or absent then the splitting is negligible, there is no
polarization signal. In these cases, the Zeeman effect is of no
importance as a diagnostic tool of magnetic fields. There is,
however, another physical mechanism, the so called Hanle effect,
which allows us to ``see'' the ``Sun's hidden magnetism'' that the
Zeeman effect is impossible to diagnose. The Hanle effect, contrary
to the Zeeman effect, works in any topological complex weak-field
scenario (even if the net magnetic flux turns out to be exactly
zero). Atomic level polarization signals are produced by population
imbalances and quantum coherence among the magnetic sublevels of the
atom due to radiative transitions induced by the anisotropic
incident radiation. The Hanle effect can be defined as any
modification of the linear polarization signals due to the presence
of a magnetic field inclined with respect to the axis of symmetry of
the radiation field. We should note, however, that for magnetic
field strengths greater that $\approx$ 10\,G (i.e. the saturated
Hanle effect regime), the Hanle effect is sensitive only to the
orientation of the magnetic field vector, but not to its intensity.
In this context, the Hanle and Zeeman effects can be suitably
complemented for exploring magnetic fields in the solar atmosphere.

The first direct empirical demonstration of the existence of
magnetized, spicular material was achieved by
\citet{trujillobueno2005}. They applied a combined Hanle--Zeeman
diagnostic to the He I 10830\AA\ multiplet to spectropolarimetric
observations obtained with the Tenerife Infrared Polarimeter (TIP)
at the German Vacuum Tower Telescope (VTT) on Tenerife, Spain. The
spectrograph slit was placed off--limb parallel to the solar limb at
an atmospheric height of $\approx$ 2\,000\km. They detected non-zero
Stokes $U$ profiles which according to the Hanle effect theory is
the observational signature of the presence of a magnetic field
inclined with respect to the local vertical. To correctly model the
Stokes I profiles, the authors solved the radiative transfer
equation assuming an optically thick atmosphere. They found from the
analysis of quiet-Sun limb spicules at a height of 2\,000\km\ above
the photosphere that the best fit to the observed Stokes profiles is
obtained for a magnetic field strength of $B \approx 10\,G$ and an
inclination angle $\theta \approx 37\degr$. The authors noted that
the observed Stokes profiles are due to an averaging along the LOS
and thus the possibility of stronger fields occupying a small
fraction of the integration volume along the LOS cannot be excluded.

\citet{LopezAriste_cas2005} used spectropolarimetric observations of
spicules in the He I D$_{3}$ line done with the ASP at the DST of
the Sacramento Peak Observatory. The spectrograph slit was placed
parallel to the solar limb at $\sim$ 3\,500\km\ above the \ha\ limb.
The data consisted of full Stokes profiles showing significant
broadening due to non--thermal effects. They constructed a database
of synthetic profiles with different magnetic field parameters. They
then applied the Principal Component Analysis (PCA) inversion to a
series of spectropolarimetric profiles which is performed by
searching for the best match of the observed profiles within the
precomputed database of profiles. They found a good correlation
between the magnetic field orientation and the spicular structures
visible in \ha\ (see Fig.~\ref{fig:lop_ari}). However, due to a
90$\degr$ ambiguity which they attributed to the small significance
of Stokes $V$ in their data, a solution giving a magnetic field
approximately perpendicular to the axis of a spicule cannot be
excluded a priori. They concluded that weak fields (of the order of
10\,G) are more common in spicules, although stronger fields above
30\,G should not be excluded in a minority of cases. We must note,
however, that these spicules emanated from an active plage.

In the same year, \citet{socasnavarro_elm2005} obtained off-limb
spectropolarimetric observations in the \hei\ 10830\,\AA\ triplet
using SPINOR in the \caii\ 8498\AA\ and 8542\AA\ lines using the ASP
at the DST. They found that the ratio of the $Q$ and $U$ signals is
different for the Ca and He lines implying a different orientation
of the polarization plane. They concluded that this result provides
clear evidence of the presence of magnetic fields in spicules by the
following reasoning: if there was no magnetic field, one would
observe some degree of scattering polarization. The plane of
polarization of this signal is given by the geometry of the
scattering process and would be the same for all three lines
observed. If, however, the scattering atoms are embedded in a
magnetic field, then the Hanle effect would rotate the plane of
polarization by an amount that depends on some parameters of the
transition. In this case, the orientation of the linear polarization
would be different in different lines, as was observed.

Recently, \citet{centeno2010} carried out spectropolarimetric
observations of off-limb spicules at various distances from the
visible limb in the He I 10830\AA\ triplet with the TIP at the VTT
on Tenerife. With these particular observations, they detected clear
Zeeman-induced Stokes $V$ profiles. Such profiles are due to a net
LOS component of the magnetic field. They used the inversion code
HAZEL (HAnle and ZEeman Light) developed by \citet{asensioramos2008}
to invert the Stokes profiles caused by the joint action of atomic
level polarization and the Zeeman and Hanle effects in order to
infer a number of atmospheric properties together with the magnetic
field vector. They inverted the Stokes profiles for all the
positions along the slit for two data-sets taken one hour apart from
each other. In a small region of the slit they found magnetic field
strengths as large as 48\,G, where the Zeeman-induced Stokes $V$
signals were particularly large. They also pointed out that the
determination of the longitudinal component of the magnetic field
depends on the distribution of the orientation of the magnetic
fields which may produce a cancellation of the Zeeman-induced Stokes
$V$ signal when magnetic field vectors are not aligned, or a
significant non-negligible Stokes $V$ signature (and thus a
significant $B_{LOS}$ component) when the magnetic field lines team
up.

Another method that has been used for the determination of magnetic
fields in spicules is based on the {\it MHD coronal seismology}
\citep{nakariakov_ofm2001}. Properties of waves and oscillations
(e.g. amplitudes, periods), as well as physical parameters of the
medium (e.g. temperature, density) determined from observations can
be connected through MHD wave theory to the unknown magnetic field
strength and transport coefficients.

Magnetic fields in spicules can also be inferred from observations
of waves. Several observations have shown oscillatory transverse
displacement of spicule axes. This displacement can be caused by two
types of waves: kink or Alfv\'{e}n waves. In \citet{zaqar2009}, there
is a discussion on difficulties associated with the Alfv\'{e}n wave
scenario. Most authors have interpreted this kind of oscillation in
spicules as the observational signature of propagating kink waves.
It is well known that transverse kink waves in flux tubes anchored
in the photosphere can be generated by buffeting of granular motions
\citep{roberts1979, hasan_kalk1999}. The propagation of kink waves
can be traced either by direct observation of the tube displacement
along the limb or spectroscopically by the Doppler shift of spectral
lines when the velocity of the kink wave is polarized in the plane
of observation. If spicules are considered as flux tubes anchored in
the photosphere then the observed transverse displacements of their
axes can be interpreted by the propagation of kink waves.
\citet{kukhianidze2006} analyzed \ha\ spectra of limb spicules
obtained at different heights. They found that $\sim$ 20\% of the
measured heights showed a periodic spatial distribution in the
Doppler velocities which they attributed to kink wave propagation.
Wave periods were estimated as 35 -- 70\,s based on the expected
kink speed in the chromosphere (50--100\kms). Estimated wavelengths
at the photospheric level are comparable to the spatial dimensions
of granular cells, suggesting a granular origin of the waves.
\citet{singh_dwiv2007} used these observations and applied the method
of MHD seismology including the effects of gravitational
stratification. They estimated magnetic fields in spicules in the
range 8 -- 16\,G. \citet{zaqar2007} analyzed the same observational
data, estimating the kink speed in the range 90 -- 115\kms\ they
obtained magnetic field strengths in the range 12 -- 15\,G.
\citet{kim2008} used high resolution \caii\ H limb observations
obtained by the SOT/Hinode instrument to determine oscillation
parameters of spicules, such as periods, amplitudes, transverse
velocities, wavelengths and wave speeds. They interpreted the
observed oscillations as MHD kink waves and adopting spicule
densities in the range 2.2~10$^{-11}$~kg~m$^{-3}$ --
4.0~10$^{-10}$~kg~m$^{-3}$ they estimated magnetic field strengths
in the range 10 -- 76\,G. \citet{verth2011} used also SOT/Hinode
\caii\ H observations and magnetoseismology to determine the
vertical gradient in both magnetic field and plasma density in a
spicule by studying the change in velocity and phase speed with
height. They found that the magnetic field decreases in strength by
a factor of 245 between the photosphere and low corona, while the
plasma density decreases by a factor of $\sim$ 1000. However, in all
of these works, various assumptions have been used.

The measured values for magnetic field and inclination, therefore, are lower than the
values needed for efficient p-mode excitation in inclined flux tubes. A closer
investigation of the mechanism, the assumptions made in the analysis coupled with a larger
number of spicule magnetic field measurements are required in order to resolve this issue.

\subsubsection{Oscillations and waves in spicules}
Oscillations and waves in spicules have been detected by both imaging and spectroscopic
observations. Reported observations on oscillatory and wave phenomena in spicules, as
well as views and discussions about their interpretation can be found in the recent
review by \citet{zaqar2009}.

\subsection{On-disk chromospheric structures}
The chromosphere is the layer where the plasma-$\beta$, changes from above to below
unity, signaling a shift from hydrodynamic to magnetic forces as the dominant agent in
the structuring of the atmosphere. Due to the combined effects of the pivotal role
played by the magnetic field and the small--scale gas thermodynamics, this part of the
atmosphere is characterized by an impressive amount of fine-scale structures. In active
regions, long and relatively stable fibrils are observed, together with shorter dynamic
fibrils, while in quiet Sun regions dark mottles emanate from the magnetic network.
Whether or not these structures are similar and  are driven by the same mechanism, as
well as their relationship to the limb spicules, has been the subject of long-standing
discussions for several years.

The line which has been widely used for chromospheric observations,
especially following the development of the Lyot filter, is the
Balmer \ha\ line. Thus despite the complicated formation mechanism
of this line and the difficulties in the interpretation of \ha\
observations, much of what we know about the fine-scale on-disk
structures (e.g. mottles, fibrils) has been obtained from
observations in this line. Substantial work has also been done by
using the unique capabilities of the Multichannel Subtractive Double
Pass (MSDP) spectrograph which can acquire spectra over an extended
FOV \citep{mein2002}. The \caii\ H and K resonance lines, as well as
the \caii\ infrared triplet (IR) (8498\,\AA, 8542\,\AA\ and
8662\,\AA), have also been used for the solar chromosphere
diagnostics. The former are the broadest lines in the visible
spectrum and sample a wide height of the solar atmosphere. On-disk
filtergrams in the \caii\ H and K line are usually obtained with
broad-band filters and differ significantly in appearance from those
obtained in \ha. \citet{rutten2007} has remarked that dark structures
in \caii\ H and K filtergrams are barely visible. Observations in
these lines close to the limb show long thin emission features
called ``straws''. \cite{reardon2009}, however, have shown that the
lack of well defined chromospheric features in the \caii\ K
filtergrams, such as fibrils and mottles, is essentially due to
observational limitations. \citet{pietarila2009} found that bright
fibrils are ubiquitous in the \caii\ K line provided the spatial
resolution is sufficiently high and the wavelength band is
sufficiently narrow to avoid contamination by photospheric
radiation. The \caii\ IR and \ha\ lines differ substantially in
their sensitivity to the temperature. The former are, like \ha,
subordinate, but whereas the lower level of \ha\ is coupled to the
hydrogen ground level via the extremely strong Ly-$\alpha$ radiative
transition, the lower level of the \caii\ IR lines is metastable and
only coupled to the \caii\ ground level via electronic collisions.
For more extensive discussion on the formation characteristics of
these important chromospheric lines see \citet{cauzzi2008},
\citet{cauzzi2009} and \citet{leenaarts2012}. High resolution imaging
spectroscopy at fast cadence in \caii\ IR was initiated by
\citet{vecchio2007} with the Interferometric BIdimensional
Spectrometer (IBIS) installed at the DST of the National Solar
Observatory (NSO). \citet{cauzzi2008} using the same instrument
obtained monochromatic images at several wavelengths within the
\caii\ IR line, analyzing several structures. They found that the
appearance of the structures in this line is strongly reminiscent of
the structures seen in \ha. As in \ha, the structures observed in
\caii\ IR lines differ noticeably between active regions and the
quiet Sun.

\subsubsection{Mottles}
\label{parameters} The traditional term mottles refers to rapidly
changing hair-like jets observed in quiet Sun regions on the solar
disc, usually in the \ha\ or the \caii\ lines. They have long been
recognized to be one of the basic elements that constitute the
inhomogeneous chromosphere. Several authors tend to refer to these
features as spicules, although spicules do not always necessarily
represent only quiet Sun structures seen at the limb. Mottles are
organized in a complex geometric pattern over the solar disk
outlining the boundaries of the chromospheric network which is more
prominent in the Ca lines. They cluster into: a) either small groups
called chains, consisting of almost parallel structures emanating
along the common boundary of two supergranular cells or b) in larger
groups called rosettes, consisting of usually radially expanding
structures around the common boundary area of three or more
supergranular cells. \citet{tanaka1974} found about 30\% of all the
dark mottles are double or can be resolved into double structures
when viewed with high resolution \ha\ wing observations.

Mottles appear as dark (absorbing) against the disk when observed in
the wings of \ha\ and especially enhanced in contrast at
\ha$\pm$0.5\,\AA. Near the centre of the \ha\ line they are much
less distinct. Very often in the past it has been reported that at
the same locations dark as well as bright mottles are observed.
However, even such basic considerations as whether dark and bright
mottles are the same feature seen at different heights or not,
remains unanswered. Some authors suggested that a bright mottle is
the base of a dark elongated one \citep{banos_mac1970}, whereas
others claimed that bright and dark mottles are distinctly different
phenomena \citep{alissandrakis_mac1971}. Today with the achievement
of higher resolution observations one has the impression that what
was called bright mottles is simply the bright background below the
dark mottles. It is very important to note that the resolution of
observations affect drastically the observed morphology of the
chromospheric fine structures. Since mottles are usually grouped in
rosettes or chains one will fail to discriminate between different
structures when using low resolution observations, since each thread
and its separation are a few tenths of arcsec wide. With the
increasing capability of instruments such as SOUP, CRISP, IBIS or
ROSA, we are able to detect fine structure with dimensions at the
limit of the resolution.

\paragraph{Lengths and widths of mottles} \citet{bray_lou1974} in their monograph, summarize
their properties. Their horizontal dimension ranges between 725 --
10\,000\km\ and they exhibit a variety of shapes. \citet{sawyer1972}
categorized the shapes of mottles into oval, round, filament, arch
and lumpy and gave horizontal dimensions between
3\arcsec--10.6\arcsec\ (2\,000 -- 7\,500\km). Although generally
elongated, the larger mottles tend to have more irregular shapes.
Observations taken towards the limb, at large heliographic latitudes
show mottles reduced in size, with widths of the order of 1\arcsec\
and dominantly oriented towards the limb.

\paragraph{Lifetime of mottles} The lifetime of mottles is difficult to determine. Although the
general configuration of mottles around bright points may be preserved for more than
15\,min, features change within minutes. Old studies give values ranging between 3 and
15\,min, but differences in resolution may affect the measurements. A more recent
analysis by \citet{bratsolis1993} determines the mean lifetime of mottles between
13-14\,min.

\begin{figure}[t]
\includegraphics[width=0.85\textwidth]{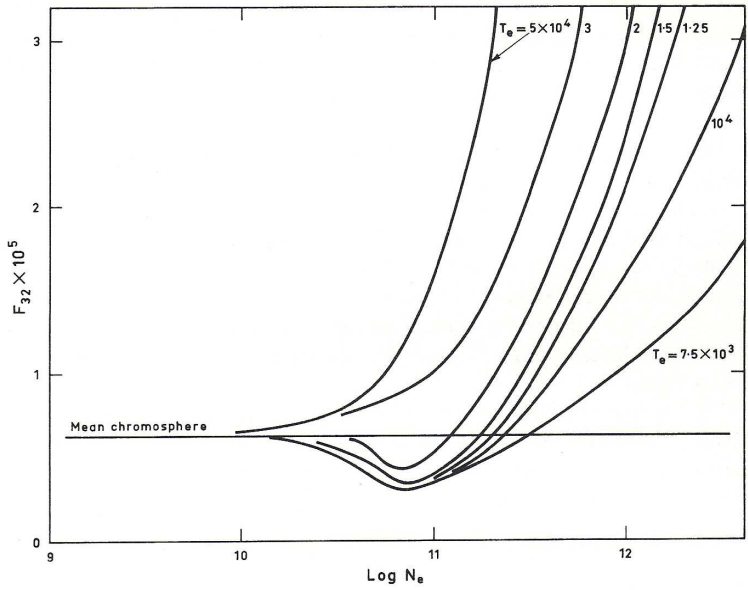}
\caption{\ha\ intensities of the mottles vs the logarithm of the
electron density $N_e$ as computed by \citet{giovanelli1967b} (in cgs
units). Different curves correspond to different temperatures.
Geometrical thickness of a mottle is $D$ = 2\,Mm} \label{nlte}
\end{figure}

\paragraph{Spectroscopic properties of mottles} \citet{giovanelli1967a, giovanelli1967b} did
pioneering extensive NLTE computations of the \ha\ line contrast
profiles relevant to chromospheric fine-scale modelling. In his
computations he assumed 1D slabs illuminated by the surrounding
atmosphere, having a geometrical thickness 2\,000\km\ and uniform
$T{_e}$ and $N{_e}$. For mottles' diagnostics, an important figure
is Fig.~\ref{nlte} given by \citet{giovanelli1967b}. In this figure,
individual curves represent the \ha\ line-center intensities as a
function of electron density for different temperatures. From this
figure one can conclude that portions of the curves below the
horizontal line correspond to dark mottles, while those above the
line are structures considered as bright mottles. An important
ambiguity one can notice is that for each negative contrast, for a
given temperature, two values of electron density are obtained.
Using the observed contrast measurements, \citet{giovanelli1967b} has
concluded that: a) for dark mottles $T$ $<$ 10\,000\,K, $N_e$
$\simeq$ 2$\times$10$^{11}$cm$^{-3}$, b) for less opaque mottles $T$
$>$ 20\,000\,K, $N_e$ $\simeq$ 10$^{11}$cm$^{-3}$ and c) for bright
mottles $T$ $<$ 25\,000\,K, $N_e$ $>$
5$\times$10$^{10}-$10$^{11}$cm$^{-3}$.

\citet{heinzel_sch1994} used \ha\ line profile observations of bright
and dark mottles together with a grid of prominence-like NLTE models
of \citet{gouttebroze1993} to derive the physical conditions in these
structures. In these models, the structures are considered as
vertically--standing 1D slabs irradiated from both sides by an
isotropic incident radiation. They have shown that higher-pressure
models ($p_g \approx$ 0.5\,dyn\,cm$^{-2}$) with temperatures around
10$^4$\,K can explain the profiles of both dark and bright
structures. However, it should be pointed out that the NLTE models
they used are more suited for prominences located at a height of
10\,000\km.

\begin{figure}[t]
\centering
\includegraphics[width=0.95\textwidth]{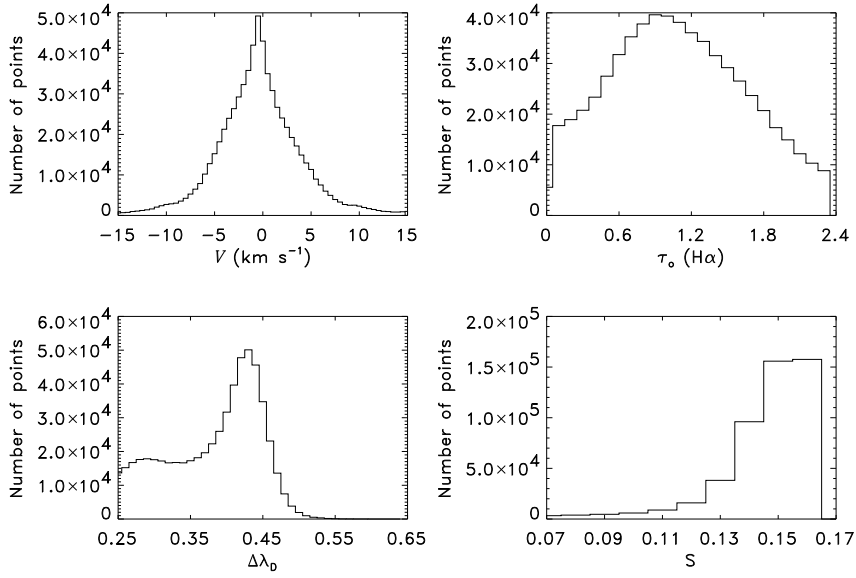}
\caption{Histograms of velocity $v$, optical thickness $\tau_{\rm 0}$, Doppler width
$\Delta\lambda_{\rm D}$ and source function $S$ derived from the cloud model for mottles
(from \citet{tziotziou2003})} \label{hist}
\end{figure}

\citet{alissandrakis1990}, \citet{tsiropoula1993}, and
\citet{tziotziou2003} used \ha\ time series observations of mottles
obtained by the MSDP and the cloud model (see Section
\ref{cloud_model}). They derived four adjustable parameters of the
model, i.e. the source function, the Doppler width
$\Delta\lambda_{\rm D}$, the optical thickness $\tau_{\rm 0}$, and
the LOS cloud velocity $v$. Although the \ha\ observations were
obtained in different times and regions the results they quoted are
very similar. According to the last paper these structures are
mostly optically thin with an optical thickness peaking at $\sim$
0.9. This is reflected on the source function distribution which
peaks at $\sim$ 0.156 (the line center quiet-Sun \ha\ intensity
relative to the continuum being 0.169), since optically thin
structures allow more of the background radiation to be transmitted
through them. Their Doppler width distribution shows a peak around
0.43\,\AA\ and has a mean value of 0.35\,\AA. The corresponding
histograms of the four cloud parameters (including the LOS velocity
which is described in the following sub-section) are given in
Fig.~\ref{hist}.

\citet{tsiropoula_sch1997} proposed a method which allows the
estimation of several physical parameters of dark structures once
the four parameters described above have been derived through the
cloud model for \ha\ observations. These parameters are: the
population densities at levels 1, 2, 3 ($N_1$, $N_2$, $N_3$) of the
hydrogen atom, the total particle density of hydrogen $N_H$,
electron density $N_e$, electron temperature $T_e$, gas pressure
$p$, total column mass $M$, mass density $\rho$, and degree of
hydrogen ionization $\chi_H$. Mean values of these parameters, as
well as their standard deviations, estimated from a set of several
mottles are given in Table 2. The temperature has been calculated
from the Doppler width obtained by the cloud model. They have
assumed two different values for the microturbulent velocity,
$\xi_t$ (i.e. 10\kms\ and 15\kms), showing its effect on the values
of the temperature and pressure. An assumed value of $\xi_t$ equal
to 10\kms\ lead to $\sim$ 1.7 times greater values of temperature
and pressure than the values obtained by assuming $\xi_t$ equal to
15\kms.

\begin{table}
\caption{Physical parameters of dark mottles from \citet{tsiropoula_sch1997}}
\begin{flushleft}
\begin{tabular}{lccc}
\hline \noalign{\smallskip}
   Parameter        &     Average          &        Standard\\
                    &      value           &       deviation\\
\noalign{\smallskip} \hline \noalign{\smallskip}
$N_1$, cm$^{-3}$    &     1.6 10$^{10}$      &    8.3 10$^{9}$\\
$N_2$, cm$^{-3}$    &     1.4 10$^{4}$       &    1.1 10$^{4}$\\
$N_3$, cm$^{-3}$    &     1.6 10$^{2}$       &    1.3 10$^{2}$\\
$N_H$  cm$^{-3}$    &     5.1 10$^{10}$      &    2.1 10$^{10}$\\
$N_e$  cm$^{-3}$    &     3.4 10$^{10}$      &    1.5 10$^{10}$\\
$M$, g cm$^{-2}$    &     2.2 10$^{-5}$      &    9.4 10$^{-6}$  \\
$\rho$, g cm$^{-3}$ &     1.1 10$^{-13}$     &    4.7 10$^{-14}$ \\
$T$, K (for $\xi_t$ = 10 \kms) &     1.4 10$^{4}$  &  9.2 10$^{3}$\\
$T$, K (for $\xi_t$ = 15 \kms) &     1.0 10$^{4}$  &  7.7 10$^{3}$\\
$p$, dyn cm$^{-2}$ (for $\xi_t$ = 10 \kms) &     0.20    &    0.1\\
$p$, dyn cm$^{-2}$ (for $\xi_t$ = 15 \kms) &     0.14    &    0.1\\
$\chi_H$             &     0.65            &     0.1\\
\noalign{\smallskip} \hline
\end{tabular}
\end{flushleft}
\end{table}

\paragraph{Spatial and temporal evolution and velocities of dark mottles}
Among other physical conditions, the study of the flows along these
structures is essential because it can add to the understanding of
the mechanism driving them. However, the problem is difficult and
the present situation is rather confusing because to determine a
mottle's motion and its variation with time is a very delicate and
difficult matter. This is not surprising in view of the difficulties
of recovering information on the velocities from line profile
observations formed from moving material embedded in an absorbing
medium at rest. Stable and good seeing conditions are required,
while wide wavelength coverage is necessary in order to follow the
changes of the LOS velocity. Furthermore, careful co-alignment of
images from different times, but also of different wavelengths is
required.

Another problem is that the apparent shape of mottles is not simple
and is changing with time. Some are very straight, while others are
curved. Some mottles are very thin and others are very thick. Some
are tapered and others are not. In high resolution \ha\ line wing
filtergrams, as described by \citet{tanaka1974} and \citet{dara1998},
they show a twined or multiple appearance. It is worth noting that
\citet{suematsu2008} found the similar twined structure for limb
spicules (see Fig.~\ref{sue.fig2}). Furthermore, the mottles in the
\ha\ blue-wing are thinner and closer to the network boundaries than
those observed in the \ha\ red-wing \citep{suematsu1995}.

A common practice to infer LOS velocities is to measure Doppler
velocities by determining the wavelength of a point midway between
two positions of equal intensity in the absorption line profile. In
this technique the moving structure is assumed to produce its own
profile which is simply shifted with respect to the background
profile by an amount $\Delta\lambda_{I}=\lambda_{0}\upsilon/c$,
where $\upsilon$ is the LOS velocity. The determination of the
velocities by this technique usually yields inconsistent results due
to the lack of knowledge of the effective height of formation of the
lines considered. Another technique used for the determination of
LOS velocities, when intensities on either side of a line profile
are known, is based on the red-blue wing subtraction

\begin{equation}
\label{1}
    DS=\frac{I(+\Delta\lambda)-I(-\Delta\lambda)}{I(+\Delta\lambda)+I(-\Delta\lambda)}
\end{equation}

\noindent This technique gives the so-called ``Doppler signal''. According to this
formula, positive Doppler signal denotes upward motion of absorbing material. Of course,
one must be aware that the derived values give only a parametric description of the
actual velocity field, they can give, however, a qualitative picture of upward and
downward moving material \citep{tsirop00}.

Although mottles show different appearance and are sometimes
displaced laterally among filtergrams of different wavelengths in
the \ha\ line \citep{suematsu1995}, we should note that no one has
analyzed their horizontal motions. \citet{beckers1963} reported that
dark mottles, after disappearing in the \ha-0.5\,\AA\ image, often
became visible in the \ha+0.5\,\AA, consistent with up-flows
followed by down-flows. \citet{bhavilai1965} has shown that some
mottles are visible in the blue wing, while others appear only in
the red wing. Upward and downward velocities, as inferred from \ha\
spectra, have been reported by \citet{suematsu1995}. These authors
after careful co-alignment and examination of a time series of high
resolution images at \ha-0.65\,\AA, \ha\ line center, and
\ha+0.65\,\AA\ and Doppler signals proposed the following
description for the evolution of disk mottles: at first they appear
as faint dark features at the \ha\ blue-wing filtergrams, elongate
and increase in contrast, and then fade out as a whole in this wing.
A few minutes later, they appear again as dark features at the \ha\
line center, and show increased contrast and length. Finally, they
appear at the \ha\ red-wing as faint, but long dark features, and
shrink towards the footpoint, increasing their contrast and then
fading out. They concluded that Doppler signals confirm this
pattern, i.e. upward radial velocity during the extending phase of
the structure and downward during the receding phase or of an
up--and--down moving jet as a whole. They also found that in some
cases, the red-wing mottles appear near the foot-points of the
blue-wing mottles in their extending phase. \citet{cauzzi2009}, on
the other hand, compared profile minimum intensities and Doppler
shifts of the minimum intensities of \ha\ profile observed with
IBIS. Although there are large scale spatial similarities between
the \ha\ profile-minimum map and the corresponding Doppler shift
map, they found two dissimilarities: a) there is a lack of
pixel-by-pixel similarity between the two maps and b) mottles in
Doppler-shift maps tend to be shorter and are dark or bright (moving
upward or downward) without obvious relation to the contrast in the
profile-minimum image, except for the brightest and darkest features
whose co-spatiality indicates exceptional motions in their lower
ends. In the corresponding \caii\ infrared line observations of
intensities and Doppler shifts, they found slightly higher coherence
but with negative correlation, i.e. higher intensity values at
locations with larger blue-shifts (dark in the maps).

\begin{figure*}[t]
\begin{center}
\includegraphics[width=0.45\textwidth, bb= 0 50 450 510]{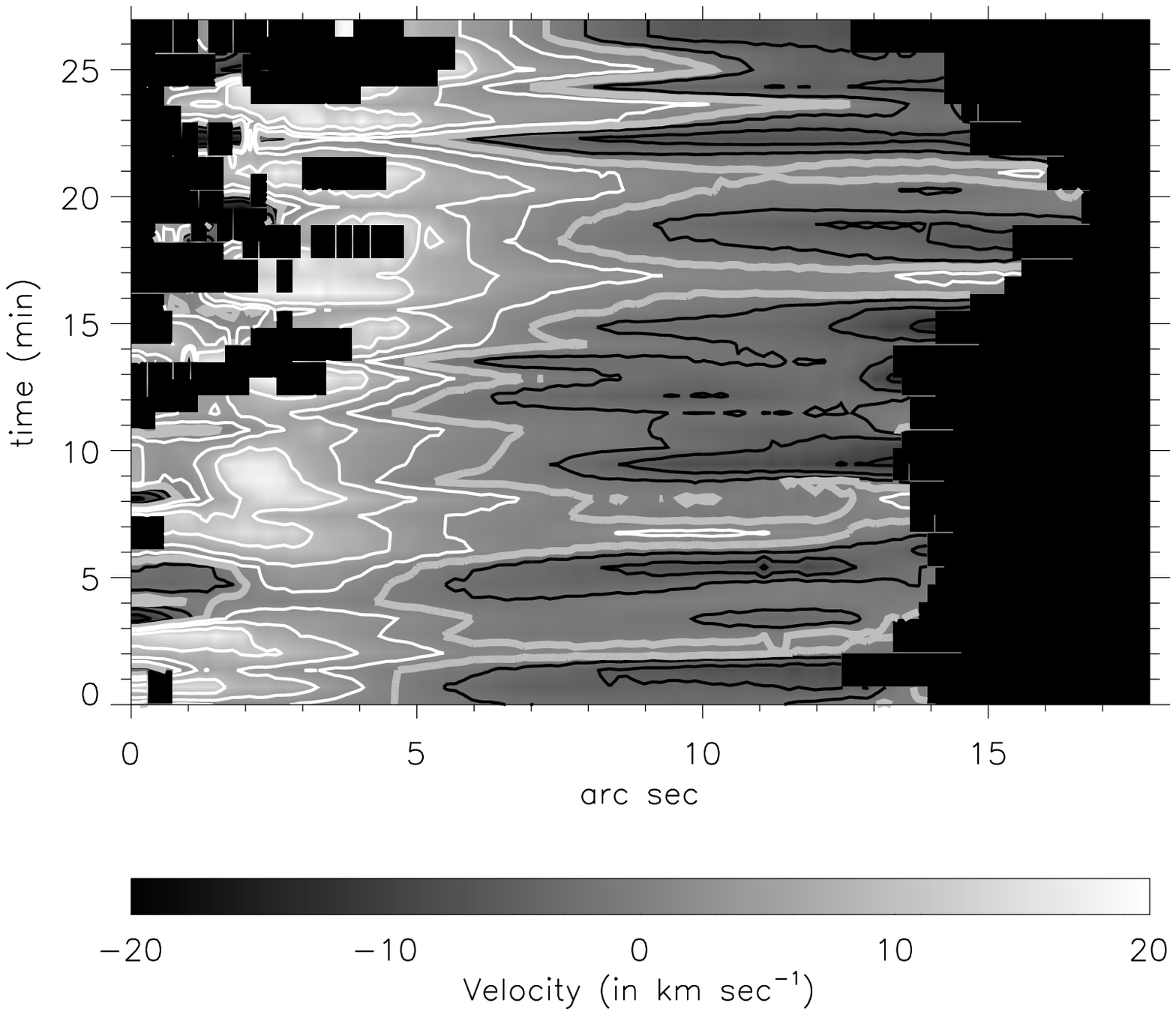}
\includegraphics[width=0.45\textwidth, bb= 0 50 450 510]{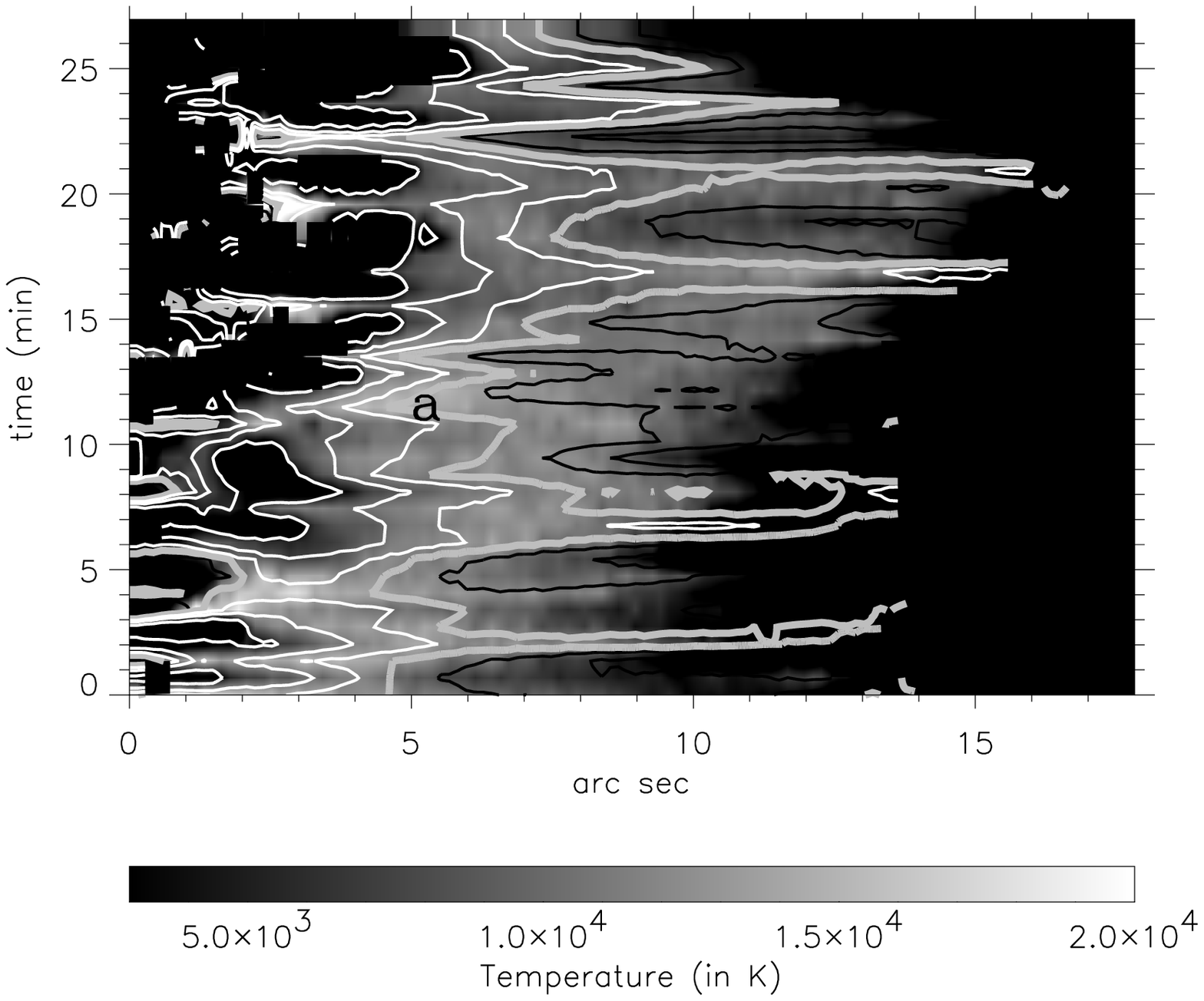}
\end{center}
\caption{Time slice images of cloud velocity $\upsilon$ (left) and temperature (right)
along a dark mottle. On the velocity images the black contours denote upward velocities,
white contours downward velocities, while the thick gray line represents the zero
velocity contour (from \citet{tziotziou2003})}\label{slicec}
\end{figure*}

\citet{suematsu1995} examined also the proper motion of the apparent
top of mottles. They showed that the tops of some mottles
approximately follow a parabolic trajectory, when they are traced
through the three wavelengths along the \ha\ line profile. Fitting
the trajectory with a quadratic function, they derived an initial
velocity and a deceleration of proper motion. Furthermore, they gave
a method to derive a tilt angle from the vertical, assuming that
mottles move on inclined straight flux tubes and are decelerated
only by solar gravity. They found initial velocities ranging from 10
to 50\kms\ with a mean of 28\kms\ and decelerations ranging from
$-$0.2 to $-$0.02\,km\,s$^{-2}$ with a mean of
$-$0.07\,km\,s$^{-2}$, which are much smaller than the solar
gravitational acceleration of 0.274\,km\,s$^{-2}$
\citep{suematsu1998}.

If mottles follow a ballistic trajectory (material motion under a
constant gravitational field), we can expect to find some
relationship between their lengths and lifetimes. For example, if
the mottle moves along a rigid magnetic flux tube of a given
inclination angle, with an initial velocity along the flux tube of
$v_{0}$, decelerated by solar gravity $g$, we have a relation
between the maximum length $L$ projected on a horizontal plane and a
half-lifetime $T$ of

\begin{equation}
L = \frac{v_{0}}{2} \sqrt{T^{2} - T_{0}^{2}}
\end{equation}

\noindent where $T_{0}=v_{0}/g$ is the lifetime of a vertical mottle
whose apparent length is zero. This equation implies that longer
lived mottles have larger apparent lengths, if every mottle has a
similar initial velocity. \citet{suematsu1995} studied this relation
and found that longer lived mottles are physically longer and that
the relation is consistent with the ballistic hypothesis. They
noted, however, that although some mottles are represented by
ballistic motions, a pure ballistic trajectory is unrealistic,
because the observed values of initial velocity and lifetime are too
small to fit the data, while the small decelerations would require
the mottles to be tilted typically 60\degr\ -- 70\degr\ from the
LOS. \citet{christopoulou2001}, from light curves of mottles, studied
their proper motions performing a least--square fit of the ballistic
motion to the height measurements as a function of time. Their best
fit was for an inclination angle from the local vertical of $\sim$
53\degr\ and an initial velocity of $\sim$ 56\kms. \cite{rouppe2007}
used high spatial and temporal resolution observations of a quiet
Sun region obtained with SST on La Palma. They reported that it was
very difficult to track individual dark mottles during their
lifetime for several reasons among which: a) many mottles lack a
sharp top end, b) they display significant fading which appears to
be induced by opacity changes with time and c) many mottles undergo
not only up and down motions, but also significant motion transverse
to the magnetic field. These factors contribute to LOS superposition
and render the identification and follow-up of mottles throughout
their lifetime quite problematic. They were able, however, to follow
some mottles and found that their tops undergo a parabolic path. In
addition, they found that the deceleration and maximum velocity are
linearly correlated and they suggested that leakage of global
oscillations from the photosphere (with dominant periods around
5\,min) plays an important role in the formation and dynamics of at
least a subset of quiet Sun mottles. This finding further suggests
that the driving mechanism for at least some of the mottles is
similar to the driving mechanism of DFs.

\begin{figure}[t]
\includegraphics[width=0.8\textwidth]{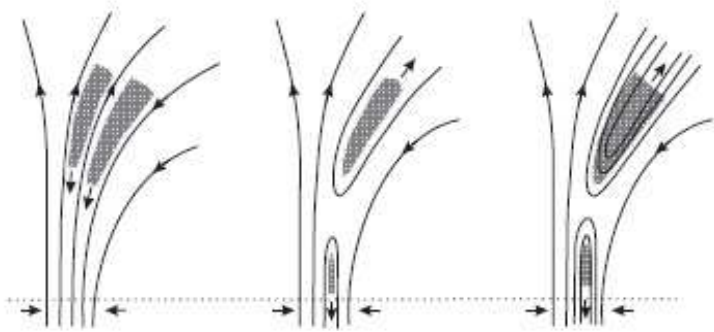}
\caption{A simple reconnection model (see text) explaining the
observed velocity behaviour in mottles, see \cite{tziotziou2003}}
\label{recon}
\end{figure}

Another technique to infer velocities of structures seen in
absorption on the solar disk has been introduced by
\cite{beckers1964phd}. It is based on the contrast profile and is
called the ``cloud'' model (see Section \ref{cloud_model}). It has
been used extensively in the analysis of observations of
identifiable fine structures in absorption on the solar disk
\citep{grossmann1971, grossmann1973, grossmann1977, bray1973,
loughhead1973, bray1974, bray_lou1983, tsiropoula1994,
tziotziou2003}. \cite{grossmann1971}, as well as \cite{bray1973},
applied this model to chromospheric mottles observed at disk centre
and derived values for the 4 parameters from the observed contrast
profiles. \cite{loughhead1973}, however, showed that this model
failed to explain contrast profiles near the limb, also
\cite{bray1974} could not reproduce some contrast profiles of
sunspot fibrils. \cite{alissandrakis1990} have described the range
of circumstances under which this model is applicable and were able
to map the 2D spatial variation of the obtained parameters. In
subsequent work, \cite{tsiropoula1994} using the ``cloud model''
showed that the predominant pattern of bulk motion in dark mottles
is that of down-flows at their foot-points and alternating up-flows
and down-flows at their tops. This result was re-confirmed by
\cite{tziotziou2003} who used another set of \ha\ observations
obtained by a different instrument. In these works the
bi-directional, as well as the recurrent character of the velocities
was revealed. This bi-directional flow followed by down-flowing
material along the whole structure is shown in Fig.~\ref{slicec}. As
can be revealed from this figure, the upper parts of the mottles
show flows in both directions, while the bottom parts show flows in
the downward direction only and the whole process is repeated with a
period of $\sim$ 5\,minutes. In order to explain this kind of flow
pattern they suggested that magnetic reconnection might be the
mechanism responsible for the formation and dynamics of these
structures. This suggestion is reasonable, since this process seems
to be the most suitable physical mechanism that can explain the
observed flow of material towards and away from the solar surface.
Furthermore, this process can occur at the network boundaries where
neighboring pairs of flux tubes of opposite polarity are driven
together by external plasma flows and reconnect. Based on these
observational findings, they proposed a simple reconnection model.
Such a model was also proposed by \cite{pikelner1969}. A schematic
representation of this model is given in Fig.~\ref{recon} (of
course, unipolar as well as bipolar fields of appropriate polarity
could be involved, see Figure 8 of \citet{wilhelm2000}). According
to this model, the squeezing of opposite polarity field lines at the
chromospheric level leads to cooling by radiation of the compressed
gas, which is trapped between them and consequently to a downflow
due to gravity (left part of Fig.~\ref{recon}). When opposite field
lines come close, reconnection occurs, then part of the material is
carried upward by the reconnected field lines, while the material
below the reconnection region moves downwards under the action of
both gravity and magnetic forces (middle and right part of
Fig.~\ref{recon}). This cycle is repeated until the field is
annihilated. Analyzing the same set of data, \cite{tziotziou2004}
showed that sometimes there is a temperature excess at the location
where reconnection seems to be occurring, indicating the presence of
local heating (Fig.~\ref{slicec}, right). Their study of the
intensity and Doppler velocity time sequence indicated also that
individual mottles appear in bursts, lasting typically for about
5\,min and usually reappear several times at the same location.
\cite{murawski2010} developed a 2D rebound shock model in which a
velocity pulse is steepened into shocks which produce
mottles/spicules and reported quasi-periodic behaviour of the rising
material due to consecutive shocks, as well as bi-directional flows
due to the superposition of falling off and rising plasma portions.

\cite{contarino2009} analyzed data acquired along the \caii\
8542\,\AA\ line with the IBIS instrument at the DST. They applied
the cloud model to four mottles and obtained values for the four
parameters given by this model. They found an almost symmetric
velocity distribution between $-$10 and $+$10\kms, which, they
suggested, is an indication of the presence of both upward and
downward velocities inside the mottles. As they reported, however,
the LOS velocities they found did not show a regular behavior. In
one of the mottles examined, they obtained alternating upward and
downward motion occurring in phase in all the examined segments. In
another mottle, they obtained up-flows in one endpoint and
down-flows in the other which reverse with a $\sim$ 4\,min period.
Finally, in the other two mottles, they found both positive and
negative velocities which change with an apparent irregularity.

\paragraph{Oscillations in dark mottles}
Different physical but mainly magnetic conditions in the
internetwork (IN) and network boundaries (NB) are clearly reflected
in the oscillatory behaviour of both regions,as well as in dark
mottles emanating from the network boundaries. Several issues
concerning the dynamics of network and internetwork have been
already addressed in the review by \cite{rutten1999}, while
intensity and velocity oscillations at both IN and NB have been
extensively investigated by several other authors.

Nowadays, the general belief is that NB regions show no power of
particular significance below 5\,min, in contrast to IN regions that
display a dominant period of 3\,min, especially in Doppler shifts
\citep{dame1984, deub_flec1990, lites1993, uexk_kneer1995,
cauzzi2000, krijger2001}. However, different results have also been
reported (see the Introduction of \cite{tsirop2009}), stemming
mostly from the use of different methods or from observations in
different lines formed at different atmospheric layers.
\cite{hansteen2000} and \cite{banerjee2001} for example, have
reported network oscillations probably attributed to waves, which
are produced in bursts with coherence time of about 10 -- 20\,min.
\cite{jess2009} used observations obtained by SST investigated a
group of network bright points. They reported full-width
half-maximum (FWHM) oscillations with periodicities ranging from 126
to 700\,s originating above a bright point. They also reported a
lack of cospatial intensity oscillations and transversal
displacements which, as they stated, rules out the presence of
magnetic-acoustic waves. They suggested that the FWHM oscillations
could be produced by a torsional Alfv\'{e}n perturbation.

\begin{figure}[t]
\centering
\includegraphics[width=0.45\textwidth]{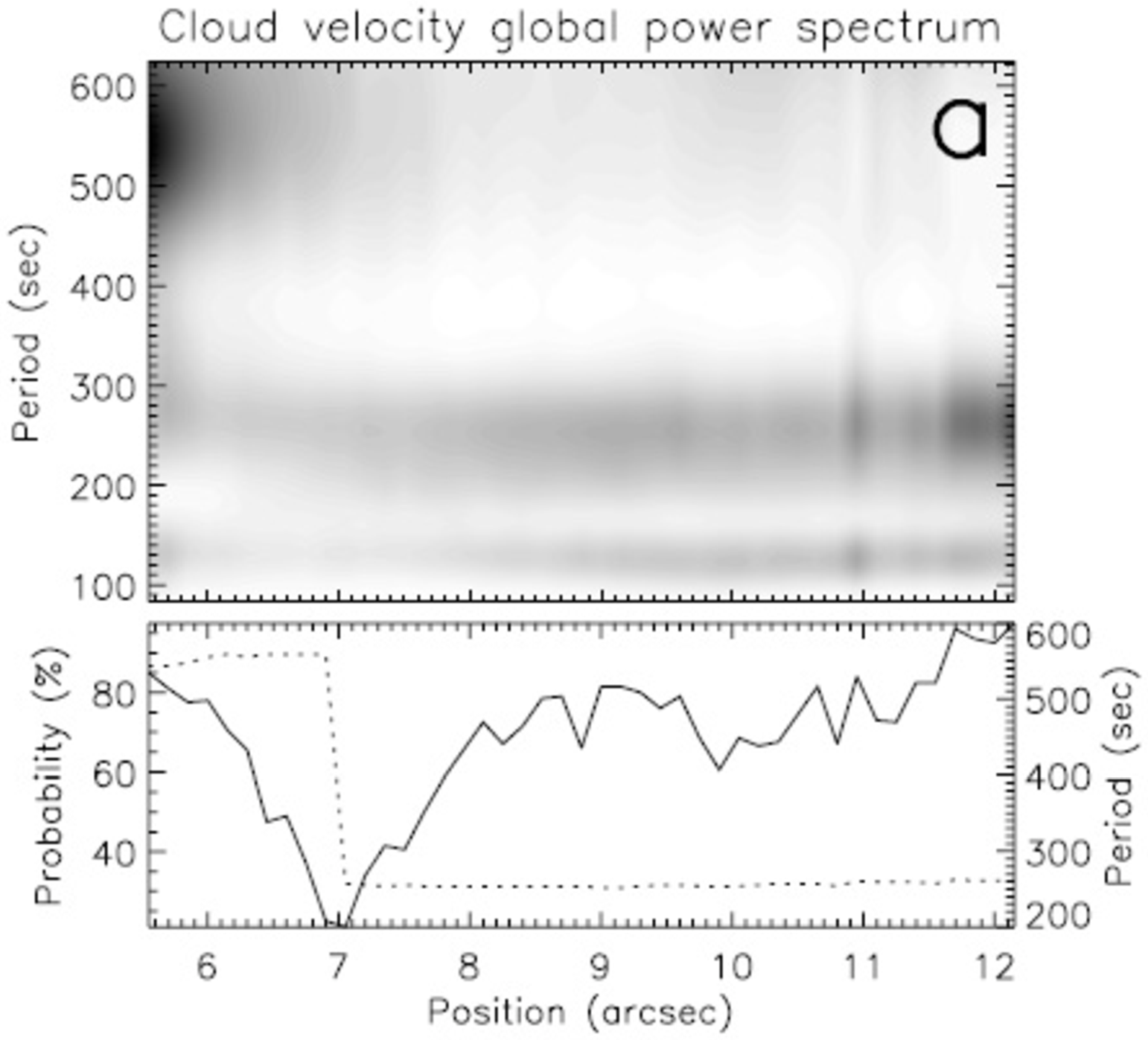}
\includegraphics[width=0.45\textwidth]{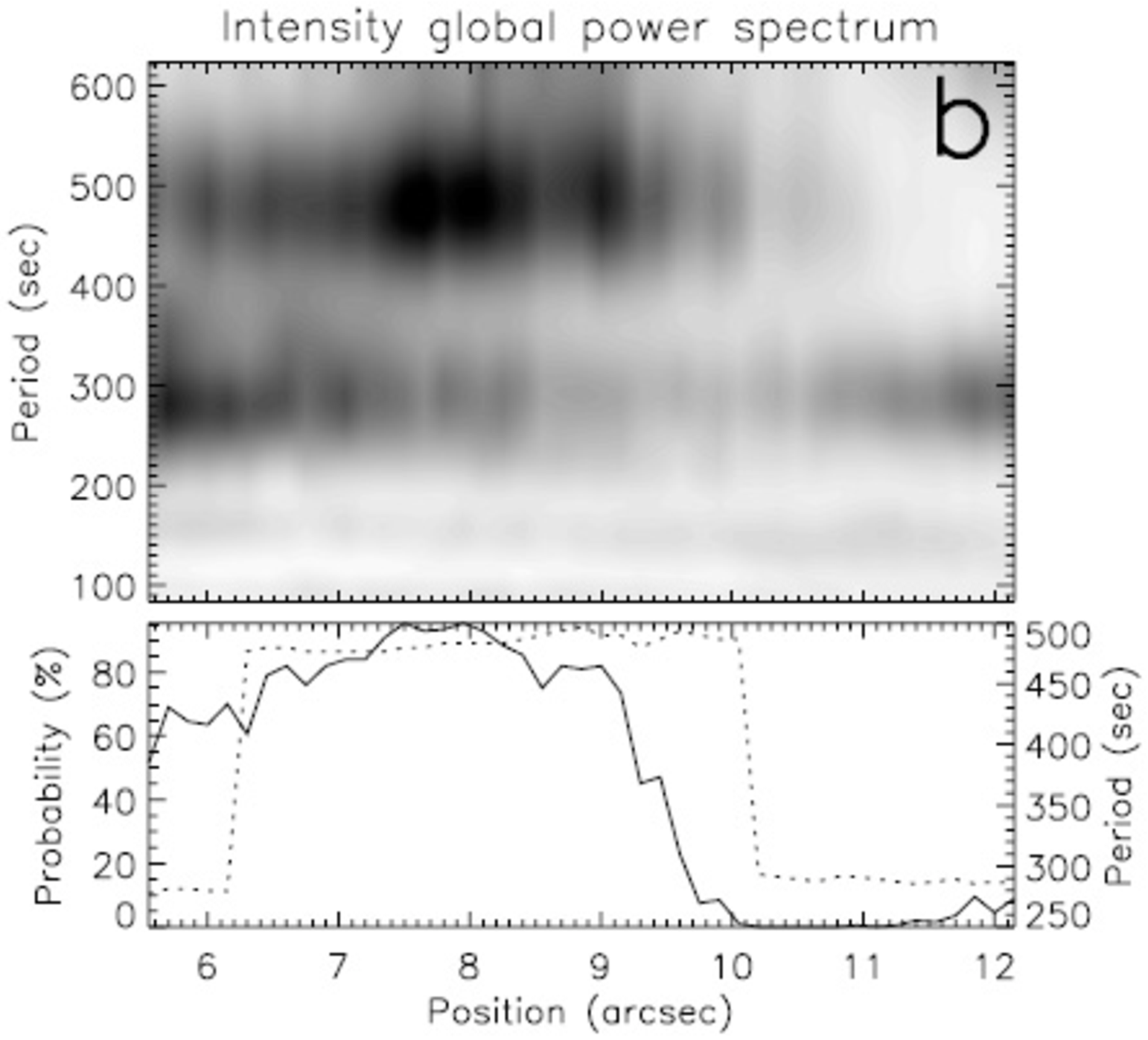}\\
\caption{Wavelet analysis results for a) the velocity and b) the
\ha\ line-center intensity variations along the axis of a
mottle, showing a dominant period close to 5\,min. Top panels show
the global power spectrum as a function of period and position along
the length of the mottle while bottom panels show the variations of
the global probability (solid line) obtained with a randomization
method and the period (dotted line) of the maximum global power peak
corresponding to this probability (from \cite{tziotziou2004})}
\label{period-mottles}
\end{figure}

The linkage between NB oscillations and mottles has been extensively
addressed by \cite{tziotziou2004} in a study concerning
periodicities of fine structures in a quiet Sun network region where
a chain of mottles was observed. Concerning quiet Sun mottles,
although differences do exist in the periodicities of intensity and
velocity variations, which are often bursty and intermittent, with
velocity variations resulting in a large range of periods from 100
to 500\,s, individual mottles or regions of mottles do exhibit a
most prominent period in the 5\,min range (see
Fig.\ref{period-mottles}). They also show a 3\,min signature which,
however, is never the dominant one. Furthermore, an intermittent
signature of 100\,s (10\,mHz), reported also by \cite{hansteen2000},
seems to exist; it is, however, unclear whether this short period
could be associated with the dynamics of recently reported Type II
spicules.

More recently, \cite{tsirop2009} in a multi-wavelength analysis of a
solar network region showed that both mottles and NB show a
periodicity of $\sim$ 5\,min in all considered lines. Phase
differences in the network boundary region indicated an upward
propagation of waves, while in the region of mottles the phase
difference was mainly negative for periods of 250 -- 400\,s
suggesting a downward propagation due probably to refraction of
waves from the inclined magnetic field of mottles. A dominant peak
around 1000\,s has also been reported which is, however, unclear if
it is related to the presumed in literature lifetime of mottles. The
relationship between the oscillatory properties of a network region
and the magnetic configuration of the chromosphere has been the
subject of study by \cite{kontogiannis2010b}. Based on a previous
paper \citep{kontogiannis2010a}, they showed the existence of power
enhancement and suppression (power halos and magnetic shadows
respectively) in 3, 5 and 7\,min \ha\ oscillations and they
concluded that p-mode leakage, mode conversion, as well as
reflection and refraction of waves on the magnetic canopy (see
Section \ref{canopy} for further details) can play an important role
to the observed properties of network oscillations.

The dominant 5-min periodicity present in both NB and dark mottles
is compatible with the idea of the leakage of photospheric 5-min
acoustic oscillations, which are the well-known p-modes, to higher
layers. Commonly, it was accepted that only waves with periods above
the acoustic cut--off period (which at the chromospheric level is of
the order of 3\,min) could travel freely in the solar atmosphere,
whereas waves with higher periods were considered as evanescent.
However, several theoretical studies \citep{michalitsanos1973,
bel_leroy1977, suematsu1990} have shown that the acoustic cut-off
period is increased when waves are propagating along inclined flux
tubes allowing the leakage of p-modes to higher layers. For the
first time p-mode leakage has been suggested by \cite{depontieu2004}
to be related to the formation of active region fibrils. p-mode
leakage, however, seems to be present also in quiet Sun mottles
which are also inclined flux tubes \citep{suematsu1990, tsirop2009,
kontogiannis2010a, kontogiannis2010b}. It should also be stressed
that measured periods of the order of 5\,min in IN regions close to
NB could well be attributed to the presence of inclined mottles that
cover these regions. Recent results reported by \cite{jefferies2006}
show that a sizable fraction of the photospheric acoustic power at
periods above the acoustic cut--off might propagate to higher layers
within and around the magnetic NBPs. They referred to these regions
as ``magneto-acoustic portals'' and showed that through these
regions a significant portion of energy is provided for heating the
solar chromosphere. \cite{khomenko2008a} has also suggested that
propagation of longer-period waves (i.e. equal or greater than
5\,min) is even possible in vertical magnetic flux tubes, resulting
from radiative losses in NBPs that may lead to the lowering of the
acoustic cut-off frequency. Recently, \cite{murawski2010} developed
a 2D rebound shock model which predicts quasi-periodic raising of
chromospheric jet-like structures with a period of $\sim$ 5\,min. In
their simulations this periodicity results from a non-linear wake
that is formed behind a leading pulse rather than from p-mode
leakage. They found that the periodicity strongly depends on the
amplitude of the initial pulse and can be longer for stronger
pulses.

\subsubsection{Fibrils} The term fibrils refers to mottle-like, dark (absorbing),
highly dynamic, jet-like, on-disc structures which are observed
above or in the direct vicinity of active region plages. The term
fibrils refers also to dark structures which are observed in the
penumbra of sunspots (penumbral fibrils) radially expanding around
the umbra and creating a filamentary structure and to structures in
active region plages which do not show jet-like behaviour; both are
probably associated with low-lying loop-like structures that connect
regions with opposite polarity magnetic flux. Jet--like fibrils
related to active region plages are relatively shorter than
penumbral fibrils or non jet-like fibrils or mottles and probably
constitute also a fraction of the observed on-limb spicules.

\begin{figure}[t]
\begin{center}
\includegraphics[width=0.95\textwidth]{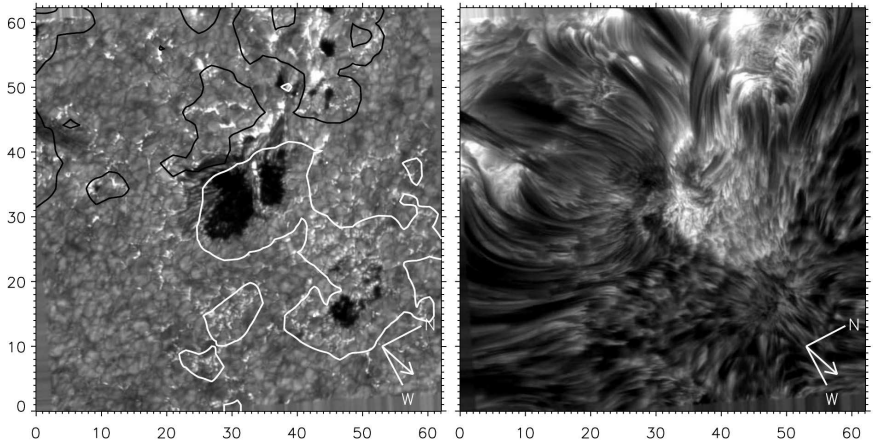}
\end{center}
\caption{\ha\ wideband ({\it left panel}) and \ha\ line center ({\it
right panel}) images of NOAA AR 10813 taken by SST. The white and
black contours in the left image outline the positive and negative
magnetic flux, respectively. The DFs are predominantly observed in
the mostly unipolar plage region between the two sunspots. The arrow
indicates the direction of the disk center (from
\cite{depontieu2007a})} \label{DFs}
\end{figure}

A close inspection of \ha\ filtergrams of plages reveals several
marked resemblances to the quiet Sun regions. In particular, the
basic components of the rosette structure, i.e. NBPs and dark
mottles are similar in several respects to the fine structure of
plages, i.e. facular granules and dark fibrils. The principal
differences of the two regions are: a) the number of facular
granules in a plage is much larger than the number of NBPs in the
centre of a rosette and b) the fibrils around a plage are generally
much more elongated than the mottles around a rosette.
\cite{Foukal1971a} proposed a continuous relationship between the
basic constituents of the quiet and active chromosphere. According
to this author there is a gradual progression between the mottles
oriented more close to the vertical and the fibrils which are rather
horizontally deflected because of the stronger field which
constrains the plasma to follow low-lying field lines.
\cite{Foukal1971b} gives some observed parameters of fibrils, e.g.
length, lifetime, density, temperature, velocity and magnetic field
and compared them to the corresponding parameters of spicules. He
concluded that the corresponding parameters are rather similar
except from the length (fibrils are longer) and the magnetic field
strength which is higher in fibrils. March (1976) performed a
systematic study of the lifetimes, lengths and evolution of fibrils
in an active region. He reported that all fibrils he examined
evolved by an upward extension followed by a downward retraction. He
also found that the lifetime of a fibril is a monotonically
increasing function of its maximum apparent length. Based on this
relationship, together with the variation of fibril lengths with
time he suggested that fibrils result from material being
impulsively injected into magnetic field lines at $\approx$ 30\kms,
and returning back under gravity.

\cite{pietarila2009} presented high-resolution observations of an
active region in the \caii\ K line obtained from the SST on La
Palma. They found that very thin, bright fibrils are a prevailing
feature over large portions of the observed field. Fibrils have not
been observed with such detail in this line before for several
reasons. One of them is that observations in this line are, usually,
obtained with broad-band filters. However, only the \caii\ K line
core is purely chromospheric, while the wings are formed in the
photosphere. With filters having a large passband, the intensity is
summed over a significant portion of the line profile and thus the
intensity of the line core originating from the chromosphere is
rather small relative to the intensity of the line wings originating
from the photosphere. If, however, observations are performed in a
sufficiently narrow wavelength band to avoid contamination by
photospheric radiation then fibrils can be observed. Another reason
is that these structures are very thin and thus very high-resolution
observations are required. The fibrils are structured by the
underlying magnetic field and the bright endpoints are clearly
co-spatial with the magnetic concentrations in the photosphere. This
is not as obvious in the strong plage where the density of the
fibrils is high enough to make identifying the endpoints difficult.
They reported that in a strong plage, where there is more magnetic
flux and due to unipolar crowding the field is more vertical, and so
are the fibrils, which are also short, and form a thick carpet
covering the underlying photosphere. At the plage edges, on the
other hand, the fibrils become longer and more organized, they
extend radially away from the magnetic concentrations, are nearly
parallel to one another, more inclined and extend over multiple
granules towards the quiet Sun.

While mottles have been the major constituent of the quiet solar
chromosphere, recent high resolution observations revealed that in
active regions there exist mottle--like structures called dynamic
fibrils (DF) which appear to form a subset of what have
traditionally been called active region fibrils. Another subset is
constituted from fibrils that do not show jet-like behaviour and
which are apparently low-lying, heavily inclined structures
connecting opposite polarity magnetic flux. This subset of much
longer fibrils is more stable in time than the DFs.

\paragraph{Lengths, widths and lifetimes of DFs}
DFs were extensively studied by \cite{hansteen2006} and
\cite{depontieu2007a} who have measured their morphological
properties and described their dynamic behavior using very high
spatial  (120\km) and temporal resolution (1\,s) \ha\ observations
obtained with the SST on La Palma. They found DFs are shorter than
mottles; their mean length is 1\,250\,km with values ranging between
400--5\,200\km. Their average width, which does not vary much with
time and height, is $\sim$340\km\ with most of them being
120--380\km\ wide, although much wider structures can be found
($\sim$ 1\,100\km). The lifetime of DFs is shorter than that of
mottles and lies between 120\,s and 650\,s, with an average of
290\,s. These properties are subject to regional differences, i.e.
they vary significantly with position in the FOV.
\cite{depontieu2007a} found that DFs (see Fig.~\ref{DFs}) were
shorter in length and lifetime in the denser plage.
\cite{pietarila2009} applied a method to isolate individual fibrils
and measured their widths and lengths. They found that most fibrils
have widths between 80\km\ and 100\km, while widths of fibrils in a
strong plage have a wide distribution with a max over 150\km.
\cite{anan2010} using SOT/Hinode \caii\ H observations of an active
plage identified 169 fibrils. They found that 80\% were
``parabolic'' showing a cycle of rise and retreat and 10\% ``faded
out'', i.e. did not show a complete retreat cycle. It should be
pointed out that these authors use the term ``parabolic'' (or
``ballistic'') motion in the context of constant deceleration during
the downward phase which is not necessarily a free-fall dominated by
gravity. The mean heights of the ``parabolic'' and the ``fade out''
fibrils were 1\,300\km\ and 820\km, while their lifetimes were
179\,s and 197\,s respectively. It has not been examined yet which
is the exact relationship between the DFs observed in \ha\ with the
SST and those observed in \caii\ H with the SOT/Hinode broadband
filter which for on-disk observations gives a predominantly
photospheric signal.

\begin{figure}[t]
\begin{center}
\includegraphics[width=0.85\textwidth]{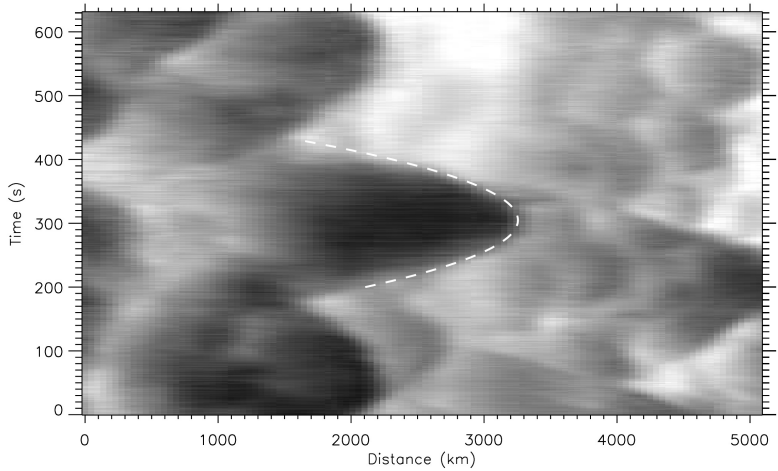}
\end{center}
\caption{An $x-t$ plot of the evolution of a DF (along the projected path). It follows a
near perfect parabolic path in its rise and descent. The white dashed line indicates the
best fit used to derive deceleration, maximum velocity, duration and maximum length
 \citep[from][]{depontieu2007a}}. \label{fig_para}
\end{figure}

\begin{figure}[t]
\begin{center}
\includegraphics[width=0.85\textwidth]{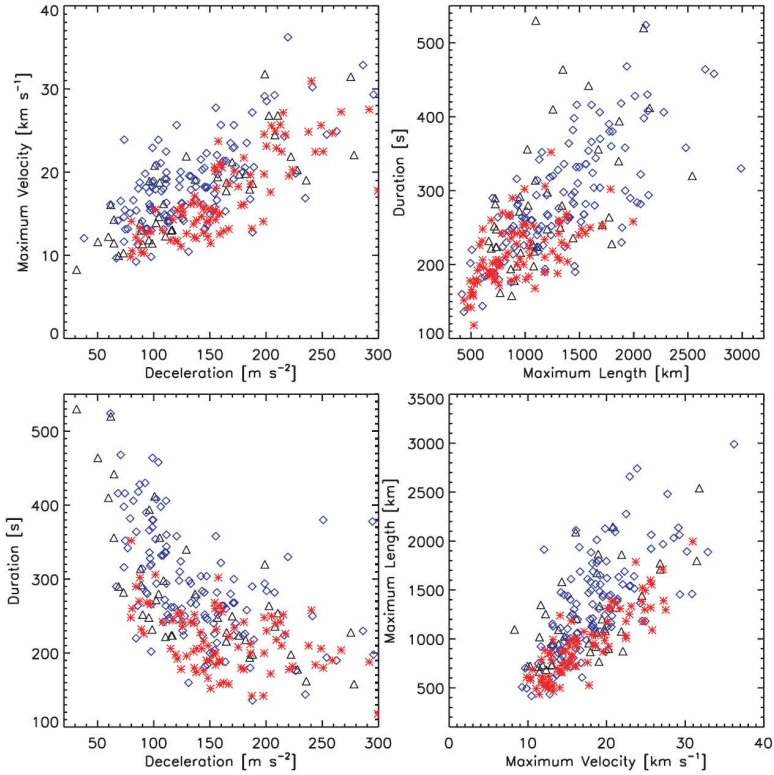}
\end{center}
\caption{Scatter plots of deceleration vs maximum velocity ({\it top
left}), maximum length vs duration ({\it top right}), deceleration
vs duration ({\it bottom left}), and maximum velocity vs maximum
length ({\it bottom right}). DFs in two different regions were
considered: one region contained dense plage ({\it red asterisks})
and the other was found at the edge of the plage ({\it blue
diamonds}). The scatter plots for the two regions often seem to have
different slopes (from \cite{depontieu2007a})}\label{vel_accel}
\end{figure}

\paragraph{Evolution and velocities of DFs} \cite{depontieu2007a} studied the evolution of
these structures using very high spatial and temporal resolution
\ha\ observations. In order to find the temporal behavior of
individual DFs they constructed $x-t$ plots showing the evolution of
the extent of a DF as a function of time. They analyzed 257 DFs and
concluded that the length of the DFs as a function of time is well
described by a parabola (see, e.g. Fig.~\ref{fig_para}). Fitting the
time evolution of the length of the DFs with a parabola they derived
the maximum velocity (on ascent or descent), maximum extent,
deceleration and duration. The parabolic paths are characterized by
a large initial velocity, usually of order of 15 -- 20\kms, that
decrease linearly with time. The deceleration is, usually, between
120 and 280\mss. The deceleration is typically only a fraction of
the solar gravity and incompatible with a ballistic path at solar
gravity. For the particular DF shown in Fig.~\ref{fig_para} they
found a deceleration of 216\mss, a maximum velocity of 27\kms, a
maximum extent of 1\,800\km, and a lifetime of 240\,s. They also
found that DFs exhibit lower velocities and higher deceleration in
the denser plage and that these differences cannot be attributed to
projection effects. They also constructed scatter plots of various
DF properties found in two different regions within the FOV. One of
the regions contained a dense plage, i.e. strong magnetic field
concentrations in which the magnetic field is generally more
vertical, while the other was located at the edge of the plage
region containing more inclined magnetic fields. They reported
significant differences of fibril properties between those occurring
above the dense plage and those occurring at the edge of the plage.
From the scatter plots they derived some intriguing correlations.
One interesting correlation is that between the deceleration and the
maximum velocity of the DFs which shows a clear linear relationship
(Fig.~\ref{vel_accel}, {\it top left}). Another interesting
correlation is that between the maximum length and the duration of
the DFs (Fig.~\ref{vel_accel}, {\it top right}). The longer DFs tend
to have longer lifetimes. They also found that the maximum velocity
and maximum length of DFs are well correlated: DFs with higher
maximum velocity tend to be longer (Fig.~\ref{vel_accel}, {\it
bottom right}). The deceleration of DFs shows a somewhat less clear
correlation with the DF duration (Fig.~\ref{vel_accel}, {\it bottom
left}). Although it seems that longer lived DFs typically experience
lower decelerations, there is a large spread in the values. There
are clear differences between the two regions, while the
correlations are not quite linear. \cite{anan2010} found that the
DFs they examined follow a ballistic motion with constant
deceleration. They found mean maximum velocities of the parabolic
and fade out fibrils of 34\kms\ and 16\kms, respectively, and mean
accelerations of $-$510\mss\ and $-$130\mss, respectively. They also
found that the deceleration was proportional to the maximum
velocity, i.e. the initial velocity of the ejection. The range of
the maximum velocity was between 6\kms -- 110\kms for the parabolic
fibrils, whereas it was between 5\kms\ -- 40\kms\ for the fade out
spicules.

Apart from the above mentioned analyses of fibrils based on
filtergrams, only a few quantitative spectroscopic studies have been
undertaken. \cite{langangen2008a} obtained spectrograms in the
\caii\ 8662\,\AA\ line using the SST spectrograph. They identified
26 DFs and infer Doppler shifts from the spectra. They also derived
mean values $\sim$ 89\mss\ for the deceleration, 11.3\kms\ for the
maximum velocity and 217\,s for the lifetimes. These values are
lower than those obtained by \cite{depontieu2007a} who measured
proper motions in narrow-band images. \cite{langangen2008a}
performed also numerical simulations and explained the discrepancies
in the lower maximal velocities derived from Doppler measurements
compared to the proper-motion velocities as due to both the low
formation height and the extensive width of the contribution
function of the \caii\ 8662\,\AA\ line. They also concluded that
their observations support the result that DFs are driven by
magnetoacoustic shocks excited by convective flows and p--modes.
\cite{langangen2008b} used a time series of filtergrams obtained in
the red and blue wings of the \ha\ line near the solar disk center
and measured proper motions and Doppler signals in 124 DFs. They
obtained mean lifetimes of $\sim$ 258\,s. Again they found Doppler
velocities to be a factor of 2 -- 3 smaller than velocities derived
from proper motions in the image plane and the corresponding
decelerations to have a difference of a factor of 5. They attributed
the difference to the radiative processes involved, since the
Doppler velocity originates from a wide range of heights in the
atmosphere, in contrast to the proper-motion velocity, which is a
very local quantity, because it is measured from the sharply defined
bright tops of the DFs.

A combination of high-resolution observations and advanced numerical
simulations have shown that DFs are most likely driven by
magneto-acoustic shocks that form when photospheric oscillations
leak into the chromosphere along inclined flux tubes
(\cite{hansteen2006}; \cite{depontieu2007a}). The inclination of the
magnetic field lowers the acoustic cut-off frequency sufficiently to
allow p-modes with the dominant low frequency to propagate along
flux tubes \citep{michalitsanos1973, bel_leroy1977, suematsu1990}.
The results of the simulations spanning from the upper convection
zone to the corona, lead to the conclusion that DFs are formed by
chromospheric shocks driven by global p-modes and convective flows.

\begin{figure}[t]
\centering
\includegraphics[width=0.8\linewidth]{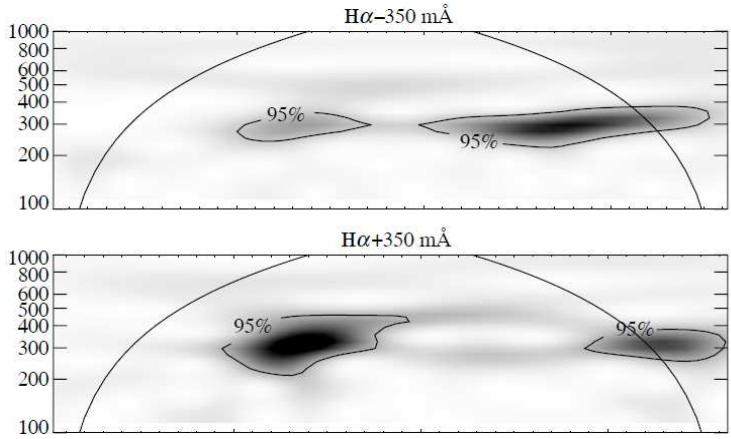}\\
\caption{Wavelet power spectra for \ha-350 m\AA\ and \ha+350
m\AA\ chromospheric oscillations in fibrils showing intermittent
5-min periods (from \cite{depontieu2006})} \label{period-fibrils}
\end{figure}

\paragraph{Oscillations in fibrils}

Although quiet and active regions are considered to be made up by
the same elementary magnetic flux tubes, there are important
differences in the observed magnetic flux densities mainly due to
the different filling factors, which are manifested also in the
respective oscillatory behavior. However, far less work exists in
the literature for oscillations in plages and fibrils compared to
work for oscillations in IN, NB and mottles.

\cite{bhatnagar1972} examining \ha\ filtergrams have reported 5\,min
oscillations in plages, while \cite{muglach2003} using UV
observations from the Transition Region and Coronal Explorer (TRACE)
found 5\,min oscillations in both plage and network regions and
3\,min oscillations in IN regions. \cite{depontieu2003} examining
high-resolution SST observations of dynamic fibrils in a plage
region reported the presence of significant oscillatory power,
although only a small fraction of dynamic fibrils showed
oscillations or recurrence. Oscillatory periods in dynamic fibrils
range between 4 and 6\,min, similarly to dark mottles in quiet Sun
regions, and seem to be concentrated mostly in the more inclined
structures \citep{depontieu2004, depontieu2006}. This similarity in
the oscillatory behavior of fibrils and dark mottles has been
clearly shown by \cite{tziotziou2004}, who analyzed a plage region
containing almost vertical fibrils, named as dark grains, which
exhibited a unique, high-probability period of $\sim$ 5\,min. This
period has been explained as the period signature of either flows
along vertical structures anchored in the photosphere or as a
possible signature of a maximum amplitude p-mode interference
wavetrain, compatible with the idea of p-mode leakage reported by
\cite{depontieu2004} or as due to consecutive shocks as predicted by
the 2D rebound shock of \cite{murawski2010}. Furthermore,
\cite{tziotziou2004} showed that the fibril-free area, although in
close spatial proximity to the fibril area, had IN-like
characteristics with periods closer to 3\,min in contrast to the
5-min period of fibrils which is similar to the period obtained for
dark mottles.

The oscillatory characteristics of DFs, as \cite{depontieu2004} have
shown, exhibit similarities to photospheric p-mode oscillations,
such as variable periods between 200 -- 600\,s and intermittency
(existence of wave packets). Moreover, observations have indicated
that DFs tend to occur near the peripheries of plages where magnetic
field lines are often more inclined. This has led
\cite{depontieu2004} to suggest that DFs are driven by leakage of
normally evanescent p-modes in inclined magnetic fields, as
\cite{suematsu1990} had also previously suggested, something that
can explain the observed 5-min periodicities.

\begin{figure}[t]
\centering
\includegraphics[width=0.8\linewidth]{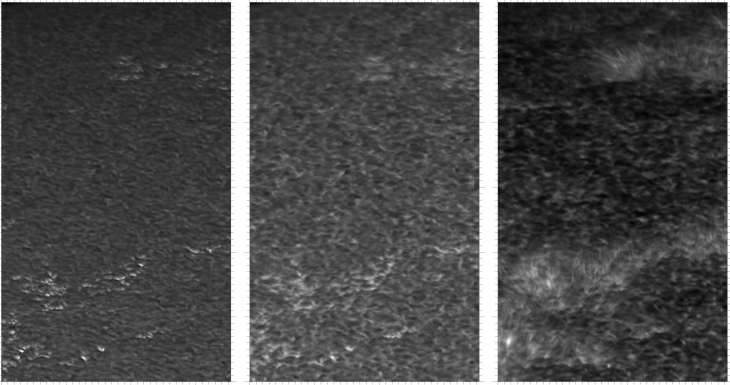}\\
\caption{DOT picture of the sun taken on June 18, 2003 in the G
band, the \caii\ H wing, and the \caii\ H core. The region is close
to the solar limb. Scale: the tick marks are separated by 1\arcsec.
The magnetic network, shows up as bright patches in the \caii\ H
wing and the G band. The long thin structures in the \caii\ H core
sticking out of the network are termed ``straws'' (from
\cite{rutten2006})} \label{straws}
\end{figure}

\subsubsection{Straws} \label{strawssub}
\cite{rutten2006, rutten2007} using high-resolution ground-based
observations obtained with DOT in the wide-passband (of 1.4\,\AA)
\caii\ H filter observed some new features, which are finer than the
traditional mottles and were termed ``straws'' (Fig.~\ref{straws}).
These structures are long, highly dynamic, bright against dark
internetwork background, extend radially outward from network bright
points, and occur in ``hedge rows''. The hedge rows of upright
straws are bright and thin in \caii\ H, bright, optically thick,
twice as high and much thicker than in \caii\ H in Ly-$\alpha$, much
less distinct and dark in \ha\ line center, and more prominent, but
less upright and very dark in the \ha\ wings. \cite{rutten2006}
concluded that straws reflect transition region conditions and the
differences in the different lines are due to radiative transfer
effects. Such features were also recently observed with the Solar
Optical Telescope at the solar limb in \caii\ H images.
\cite{depontieu2007b} refers to these straw-like features as ``Type
II'' spicules.

\subsubsection{Rapid blueshifted excursions (RBEs)} 
\begin{figure}[t]
\centering
\includegraphics[width=0.95\textwidth]{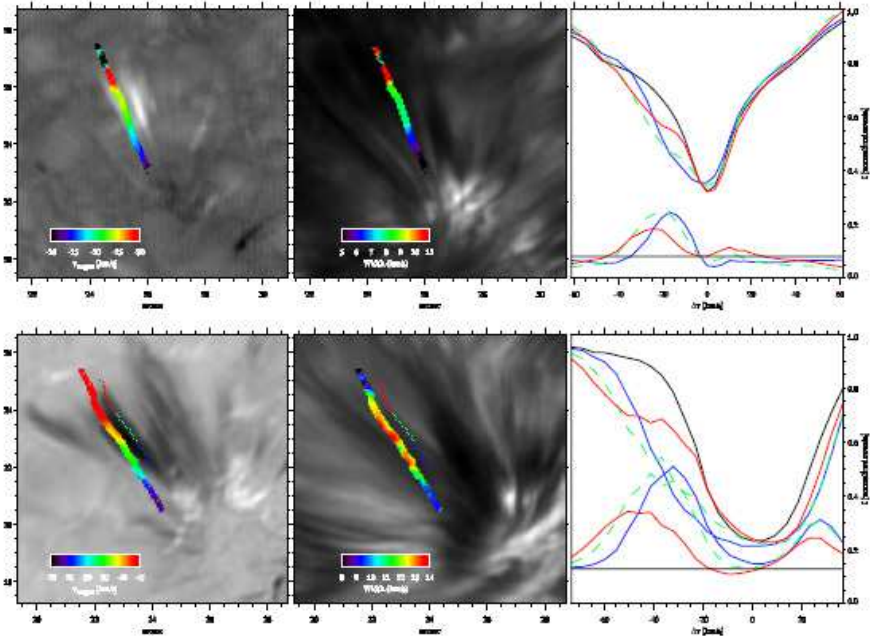}
\caption{Properties as a function of position of an RBE in the
\caii\ 8542\,\AA\ data set (top panels) and the \ha\ data set
(bottom panels). Color-coded-measured Doppler velocity (left
panels), width (middle panels), and mean spectra over three portions
of the RBE (right panels: closest to foot-point in blue, middle part
as dashed green, part furthest away in red and mean spectrum in
black. Upper curves show the spectral profile, the lower curves show
the subtraction of the average spectrum and the spectral profile).
The extracted RBE is shown with a thin colored line in the left and
middle panels with blue for the third of the length closest to the
foot-point, green for the middle part, and red for the part furthest
away from the foot-point. The measured parameter is shown displaced
to the left of the RBE. The background image is at line center
(middle panels), at a blue position of 36\kms\ (bottom left panel),
and a Dopplergram at 30\kms\ (top left panel). At both the bottom
and top end of the RBEs the amount of blue-ward absorption drops
significantly. When this happens, the Doppler velocity and width is
set to zero. This is why the black color coding at both ends of the
RBEs is not indicative of lower velocities or widths, but rather
signifies the spatial extent of the blue absorbing feature. From
\cite{rouppe2009}} {\label{rbes_ha_ca}}
\end{figure}

A class of short-lived events
observed on the solar disk with IBIS/DST and characterized by large
Doppler shifts that appear only in the blue wing of the \caii\ IR
line have been recently reported by \cite{langangen2008c}. They were
found at the edges of the rosettes as sudden broadening of the line
profile on the blue side of the line without an associated redshift
and were denoted as rapid blue-shifted excursions (RBEs) by the same
authors. For the measured RBEs they derived an average length of
1.2\,Mm, an average width of 0.5\,Mm, a mean lifetime of
45$\pm$13\,s, and velocities of the order of 15 -- 20\kms. Using
Monte Carlo simulations, they showed that the observed blue-shifts
can be explained by a wide range of spicule orientations combined
with a lack of opacity in the upper chromosphere. Similar events,
but longer lived and spatially more extended, have been reported
earlier \citep{wang_etal98}. These events were also revealed using
spectral imaging data in the \caii\ 854.2 nm and \ha\ lines obtained
with the CRisP Imaging Spectropolarimeter at the 1-m aperture SST on
La Palma \citep{rouppe2009}. These authors used a manual detection
scheme together with an automated algorithm to detect RBEs. They
identified 413 features in the \caii\ 8542\,\AA\ data set and 608 in
the \ha\ data set and claimed that the number of RBEs if they were
observed at the limb would be 1.9 per linear arcsec.
\cite{judge2011} exrapolating the number of RBE detections of
\cite{rouppe2009} estimated that there should be $\sim$ 10$^5$ RBEs
on the Sun at any given moment. They found rapid blue-ward
excursions in the line profiles of both chromospheric lines
(Fig.~\ref{rbes_ha_ca}) and suggested that these structures may be
the on-disk counterparts of Type II spicules. They measured lengths,
Doppler velocities and widths of RBEs and found that the values of
these quantities are consistently higher for the \ha\ RBEs compared
to the \caii\ RBEs. Thus \ha\ RBEs are on average longer (of order
3\,Mm vs 2\,Mm), higher widths (13\kms\ vs 7\kms) and have higher
average Doppler shifts toward the blue (35\kms\ vs 15\kms).

\section{Relation between mottles/fibrils and photospheric bright points}
\label{pbp} Multi-wavelength photospheric and chromospheric
observations of quiet solar regions, such as the ones presented in
Fig.~\ref{mag_elem}, show a clear spatial relationship between the
dark elongated structures, called mottles, emanating from the
network boundaries and the photospheric bright points. Hence, it is
quite natural to think that these structures are somehow related.
Bright points, which can be regarded as proxies of small-scale
magnetic elements in the photosphere, are prominent at network
boundaries in the \ha\ line-wing images at high spatial resolution
\citep[e.g.][]{Dunn_Zir1973, suematsu1995, Leenaarts2006}. As
\cite{suematsu1995} showed, by comparing \ha--0.65\,\AA\ and
\ha+0.65\,\AA\ filtergrams, bright points are very prominent at
network boundaries in the blue-wing, while only a few appear in the
red-wing or line center. Recently, \cite{Leenaarts2006} concluded
that in the blue \ha\ wing bright points represent a promising proxy
magnetometer to locate isolated magnetic elements.

It is of great interest to investigate whether the appearance of
bright points, which represent intense magnetic concentrations, is
related to fine structure formation at the network boundaries.
However, it is extremely difficult to directly associate bright
points to specific dark mottles. \cite{dunn1974}, used their
excellent \ha\ filtergrams, but they were not able to find a
definite relationship between them. \cite{suematsu1995}, using time
series of \ha\ observations were able to connect a bright point with
a mottle through its whole life. They reported, however, that there
were many mottles which were not associated with bright points. They
also noted that bright points often appear at the foot-points of
mottles at \ha--0.65\,\AA, but during their peak extension or
falling phase rather than at their beginning. In \ha$+$0.65\,\AA\,
bright points seem to be obscured by falling material. In the blue
wing, however, and in the falling phase of mottles one can see a
much deeper layer due to an effective red shift of the line.
Enhancement of the brightness in $-$0.65\,\AA\ could occur due to
compression of the atmosphere by in-falling spicule material that
penetrates deeper the solar atmosphere, delivering its potential
energy. The infalling material may even distort and kink the flux
tube inward; this is inferred from the inward offset of the spicule
base in the red-wing compared to its blue-wing position
\citep[see][]{suematsu1998, dara1998}. \cite{cauzzi2009}, in
simultaneous \ha\ and \caii\ infrared observations obtained from
IBIS noticed that in \ha\ at minimum intensity dark structures show
sometimes marked brightenings towards one end. Comparison with
magnetograms showed that these bright endings correspond to the
stronger magnetic concentrations, although the magnetic network,
which is easier to see in the respective \caii\ infrared images, is
not easily recognized in this line. \cite{sterling2010} used \caii\
H observations of a north pole coronal hole obtained with SOT/Hinode
and investigated the roots of some spicule-like features seen on the
disk near the limb. They found that these features seem to emanate
from fast moving \caii\ brightenings. Frequently the ejected
spicules, which have speeds $\sim$ 100\kms, seem to occur when these
brightenings, which have horizontal velocities of a few 10\kms\,
appear to collide and disappear. These spicules could be associated
to the so-called Type II spicules. Based on their findings, they
suggested that \caii\ brightenings could represent acoustic shocks
or fast-moving interacting magnetic elements. As they stated it is
hard to understand how the energy from acoustic shocks could be
converted to spicules of such high velocities. If, on the other
hand, \caii\ brightenings represent different magnetic elements,
their merging could lead to magnetic cancellation and reconnection,
resulting to the deposition of thermal energy in the low atmosphere.

The existence of a relationship between bright points and the
corresponding structures found in active regions, called fibrils,
has been examined by \cite{pietarila2009} in \caii\ K line
observations. They reported that bright foot-points of fibrils are
clearly co-spatial with the photospheric magnetic concentrations. As
they noticed, however, in a strong plage it is not easy to identify
such a correlation, because the density of the fibrils is very high,
the fibrils are more vertical and form a carpet covering the
surface.

\section{Mottles and fibrils and the formation of the magnetic canopy}
\label{canopy}

The highly ordered photospheric magnetic fields of the plages and
magnetic network spread out and occupy the whole atmosphere as they
extend upwards. Thus, instead of the bright points that constitute
the network and the facular granules that constitute the plages,
which more or less represent the cross-sections of rather vertical
flux tubes, thin, elongated, and usually inclined structures
dominate in the chromosphere and provide evidence of its complex
structure. \cite{gabriel1976} proposed that the magnetic field of
the upper solar atmosphere has a wine glass geometry as a result of
magnetic field lines fanning out of the network. Such a general
configuration of the chromospheric magnetic field was already
assumed before \citep[see e.g.][]{kopp1968} and is physically
anticipated. Due to the almost exponential decrease of the gas
pressure with height, the magnetic flux tubes expand and form a
funnel-like magnetic geometry with almost horizontal fields. One
interesting quantity describing the relative importance of gas and
magnetic field is the plasma $\beta$ parameter (where $\beta =
8\pi\,p/B^{2}$). The $\beta$ = 1 layer, i.e. the layer where the gas
and magnetic pressures are equal (or, equivalently, the sound speed
equals the Alfv\'{e}n speed) is called {\it magnetic canopy}. This
layer is of critical importance because it partitions the atmosphere
into contiguous volumes of gas pressure (high-$\beta$) and magnetic
pressure (low-$\beta$) dominated plasma. The location of the
magnetic canopy depends on the photospheric magnetic flux and on the
pressure profile of the atmosphere in which the flux tubes are
embedded. Therefore, in principle, one may expect that it should be
lower near active regions and higher around the chromospheric
network in quiet Sun. In the quiet Sun the canopy is situated
somewhere between $\sim$ 1 and 2\,Mm above $\tau_{5000}=1$
\citep{solanki_steiner1990}. Thus it is within the chromosphere that
$\beta$ falls below 1 and magnetic forces start to gain control on
the dynamics of the solar plasma. This results not only in mode
conversion, refraction and reflection of waves, but also, together
with the local thermodynamic properties, in the plethora of fine
structures that characterize this part of the atmosphere.

\begin{figure}[t]
\includegraphics[width=0.95\linewidth]{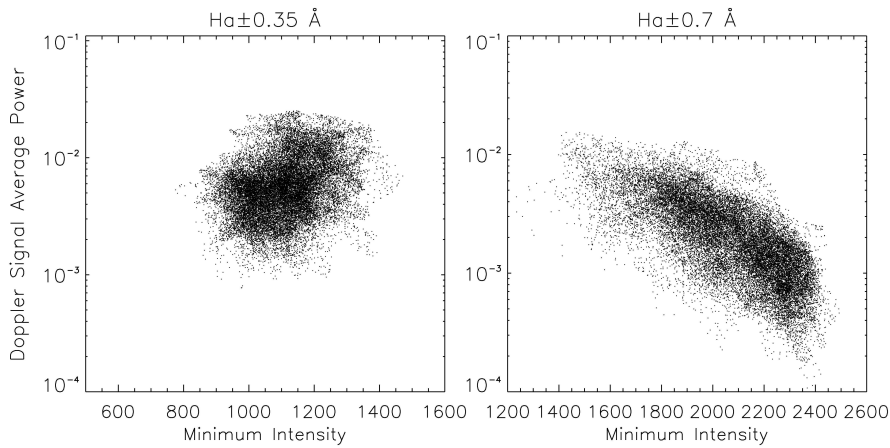}\\
\caption{Scatter plots of the average power of Doppler signal
variations in a logarithmic scale vs the \ha$\pm$0.35\,\AA\ minimum
intensity ({\it left column}) and of the \ha$\pm$0.7\,\AA\ minimum
intensity ({\it right column}) in a rosette region consisting of
mottles from \citet{kontogiannis2010a})} \label{power}
\end{figure}

Concerning wave propagation, it has been shown by numerical
simulations and theoretical studies that the canopy acts as a
boundary. Waves that propagate from the photosphere into the
chromosphere, undergo mode conversion, refraction and reflection
when they reach the canopy \citep[see e.g.][]{rosenthal2002,
bogdan2003, schun_cal2006, cally2007, khomenko2008b, kuridze2009,
nutto2012}. Through the interaction between the various MHD modes at
the canopy, energy is transferred at higher atmospheric layers. This
interaction has also been revealed by observational studies which
show the distribution of the oscillatory power on the FOV. It was
found that high frequency oscillatory power at the photosphere is
increased around intense magnetic concentrations, in active regions,
forming the so-called {\it power halos} \citep{braun1992, brown1992,
hindman1998, thomas2000, jain2002, muglach2003}. It was also noted
that the power is reduced above strong magnetic concentrations at
the chromospheric level. \cite{muglach2005} combined power maps and
magnetic field extrapolation in an active region and showed that
closed lines are associated with a further increase in power, due to
reflection of acoustic waves. The same areas of enhanced power were
less extended around nearly vertical open field lines. It is,
generally, accepted that this enhancement is due to the interaction
of acoustic waves with the canopy. On the other hand, over the
network regions the presence of reduced power has been revealed by
\cite{judge2001} through TRACE and SUMER observations. These authors
used the term {\it magnetic shadow} to describe this power deficit
observed in areas adjacent to the NBPs and not only above them.
Magnetic shadows were also detected in chromospheric lines, such as
\caii\ IR or \ha\ \citep{vecchio2007, kontogiannis2010a}. It was
explained that the interaction of waves with the magnetic canopy is
also responsible for the formation of the magnetic shadows
\citep{mcintosh2003}. In \cite{kontogiannis2010a} it was further
found that around NBPs and over rosettes magnetic shadows were
detected at \ha$\pm$0.35\,\AA, while at \ha$\pm$0.7\,\AA\ acoustic
halos were detected at the same positions. They also noticed that,
interestingly, the power maps show a fibrilar structure which
correlated very well with the positions of dark mottles.

\begin{figure}[t]
\includegraphics[width=0.85\linewidth]{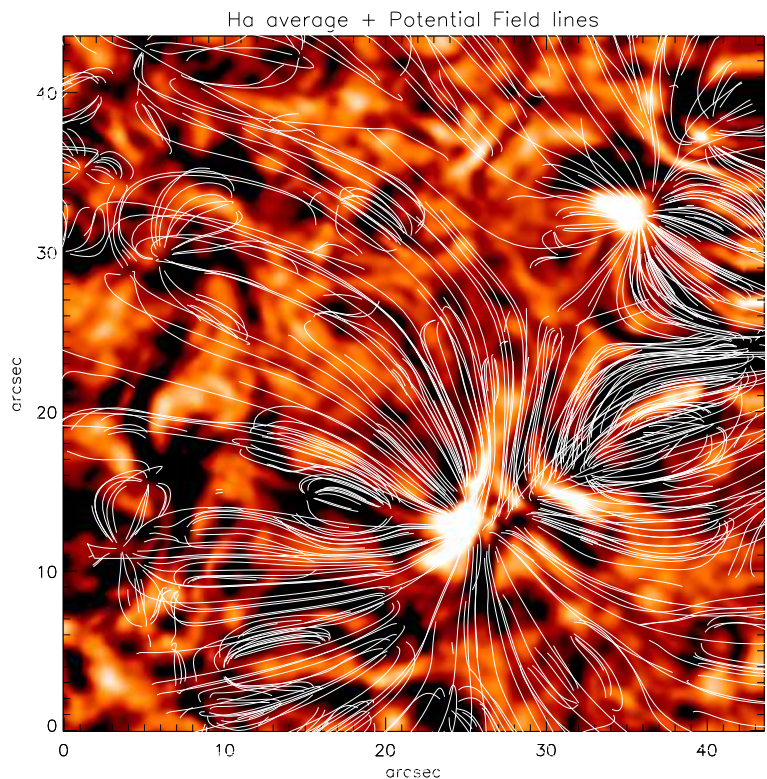}\\
\caption{\ha\ line centre 30-minute average intensity image.
Overlayed are the field lines of the extrapolated magnetic field
using the potential field approximation (from
\cite{kontogiannis2010b})} \label{flines}
\end{figure}

The different types of fibrilar structures seen in chromospheric
lines (especially in the \ha\ line) chart inclined magnetic flux
tubes. These flux tubes contain more atoms and ions in the right
state for absorption (or emission) in the corresponding lines than
their surroundings. It is reasonable to assume that the
chromospheric structures are somehow connected with the formation of
the magnetic canopy and should play an important role in wave
propagation. \cite{kontogiannis2010a} examined the relation between
power enhancement or suppression within a rosette region and the
positions of dark mottles. They gave scatter plots (shown in
Fig.~\ref{power}) between the average power of the oscillations of
the Doppler signal and the minimum intensity in \ha$\pm$0.35\,\AA\
and \ha$\pm$0.7\,\AA\ (which are obviously related to dark
structures). In the left scatter plot lower power values correspond
to lower minimum \ha$\pm$0.35\,\AA\ intensity values. The inverse is
observed for the \ha$\pm$0.7\,\AA\ intensities (right scatter plot).
The authors arrived to the conclusion that power
enhancement/suppression is directly related to mottles. Potential
magnetic field extrapolation using photospheric magnetograms has
shown that mottles follow, in most cases, the magnetic field lines
of the chromospheric field (see ~\ref{flines}),
\citep{kontogiannis2010b}.

In this same study the authors have assumed a VAL 3C model for the
solar atmosphere and used SOT/SP magnetograms. They obtained the LOS
and transversal components of the magnetic field, its inclination to
the vertical, as well as the plasma-$\beta$. From these parameters
they calculated the height of formation of the magnetic canopy. They
showed that the magnetic canopy is directly related to the dark
mottles, which seem to provide the loci where wave mode conversion,
refraction and reflection occur. They also showed that the
modulation of the oscillatory power is defined by the inclination of
the magnetic field and the relative positions of the magnetic canopy
and the height-of-formation of the line used for the observations.
In a subsequent paper \cite{kontogiannis2011} used high-precision
measurements of the magnetic field in a network region provided by
SOT/SP on-board the Hinode mission and MDI/SoHO measurements
together with potential magnetic field extrapolation. They showed
that the data of the latter instrument lead to higher formed and
less horizontally extended canopies. The authors concluded that the
sensitivity of the instrument used to measure the photospheric
magnetic field is very important in the derivation of the height of
formation of the magnetic canopy.

\section{Trends in observation}
The study of the fine and highly dynamic solar structures requires
high angular, temporal and spectral resolution observations. Solar
observations are moving into a new era of high-resolution
observations. Ground-based observations, traditionally restricted by
seeing conditions, are now profiting from adaptive optics and
post-detection image restoration techniques to achieve a much better
angular resolution than was possible before, while new
large-aperture telescope projects have already been initiated.
Space-based observations are providing large volumes of data in the
visible, ultraviolet and X-ray wavelengths, covering extended
temporal durations, and have revolutionized solar science during the
past years with successful missions such as YOHKOH, SoHO, TRACE,
STEREO, HINODE and more recently SDO, while new exciting projects
are under way.

Ground-based solar telescopes are equipped with state-of-the art
instruments and CCD cameras. The T\'{e}lescope H\'{e}liographique
pour l'\'{E}tude du Magn\'{e}tisme et des Instabilit\'{e}s Solaires
(THEMIS), the 90-cm aperture French-Italian solar physics facility
in Tenerife, is equipped with a Fabry-P\'{e}rot filter magnetograph
(MTR) and a large double-pass spectrograph (MSDP) for multi-line
spectropolarimetry and 2-D spectrometry. The 45-cm Dutch Open
Telescope (DOT) provides solar images at nearly 0.2\arcsec\
resolution while the SST has been recently operating a new CRisp
Imaging SpectroPolarimeter (CRISP). The Interferometric
BIdimensional Spectrometer (IBIS), a bi-dimensional spectrometry
instrument based on a dual Fabry-P\'{e}rot interferometric system,
plus ROSA (Rapid Oscillations in the Solar Atmosphere camera) are
already operating at Dunn Solar Telescope of the Sacramento Peak
Observatory. The German VTT telescope operates with
spectropolarimeter instruments such as POLIS (for the iron line pair
at 630.2 nm) and TIP (for the near infrared) and TESOS which is a 2D
spectrometer. The 1.6-meter New Solar Telescope (NST) at Big Bear
Solar Observatory (BBSO), uses a large set of new instruments such
as broad-band filter imagers (e.g. a TiO filter centered at a
wavelength of 705.7\,nm, with a bandpass of 1\,nm) to deliver high
resolution images of small-scale photospheric features with high
cadence exposures. Knowledge acquired from past and present
instrumentation will be used in new solar telescopes like the
1.5-meter German Gregory Coud\'{e} Telescope in Tenerife and future
large-telescopes like the US Advanced Technology Solar Telescope
(ATST) and the European Solar Telescope (EST). Post-detection image
restoration, although it requires a large amount of processing,
enhances the spatial resolution close to the theoretical telescope
diffraction limit through techniques, such as the phase-diverse
image registration \citep[e.g.,][]{lofdahl1998} which is used in SST
and speckle image reconstruction \citep[e.g.,][]{sutterlin2001}. An
extension of the first image restoration technique called
Multi-Object Multi-Frame Blind Deconvolution
\citep[MOMFBD,][]{noort2005} operates now at SST. Speckle image
reconstruction works by dividing the observed field in isoplanatic
subfields which are reconstructed independently by describing a
seeing-sampling sequence of perturbed images called ``speckle burst"
with a model of atmospheric turbulence, and finally rejoining them
in one single image. This technique has been extensively used for
post-restoring of images from several telescopes, such as the DOT,
VTT, DST and SST. However, although these techniques produce high
resolution solar images, usage of such images for the quantitative
determination of different physical parameters with inversion
techniques (e.g. cloud modelling) can sometimes lead to spurious
results and hence possible effects and artifacts of image
restoration on the data should always be examined before such a
scientific analysis.

Real-time image correction in ground-observations is nowadays
achieved with adaptive optics. Adaptive optics try to correct the
incoming wavefront, similarly to the speckle reconstruction
principles, by dividing the aperture plane into sub-apertures
(ideally smaller than the seeing) and applying a wavefront
correction to compensate for deformations caused by seeing
conditions and imperfections in the optical system. Adaptive optics
makes it feasible to build very large meter(s)-class innovative
solar telescopes such as the NST at Big Bear Solar Observatory and
the GREGOR in Tenerife and the even larger future telescopes like
ATST and EST.

\begin{figure}[t]
%\centering
\includegraphics[width=0.95\linewidth]{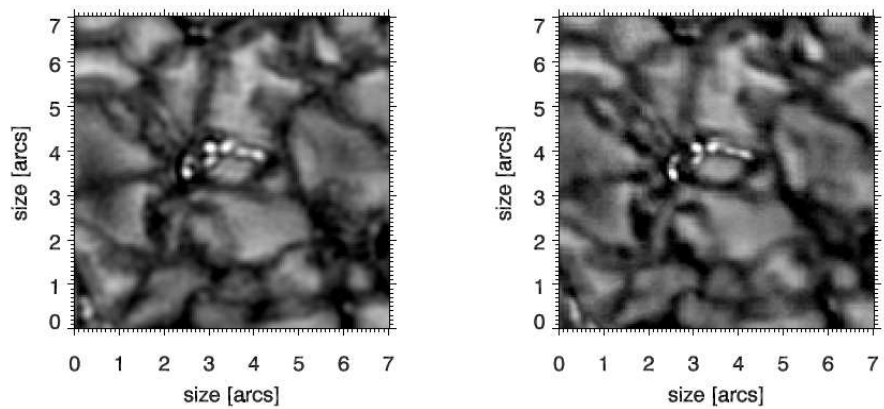}\\
\caption{{\it Left}: Deconvolved Hinode SOT image, {\it Right}:
Reconstructed DST image (\cite{woger2008})} \label{Woger}
\end{figure}

Adaptive optics systems are now commonly used on several advanced
solar telescopes to enhance the spatial resolution of the recorded
data. Post image reconstruction techniques such as speckle
interferometry can be used to achieve near-diffraction limited
resolution over a large FOV rather than the limited area where the
wavefront sensor of the adaptive optics system measures the
wavefront distortions. \cite{woger2008} investigated the
reconstruction properties of the Kiepenheuer-Institut Speckle
Interferometry Package (KISIP) code. Comparisons were made between
simultaneous observations of the Sun using the ground-based Dunn
Solar Telescope and the space-based SOT/Hinode telescope. The
analysis showed that near real-time image reconstruction with high
photometric accuracy of ground-based solar observations is possible,
even for observations in which an adaptive optics system was
utilized to obtain the speckle data (Fig.~\ref{Woger}).

\section{Trends in interpretation of line profiles}
The on-disk solar chromosphere can be observed in the cores of
strong absorption lines at visible or near-UV wavelengths, while
many of the emission lines in the EUV part of the spectrum arise in
the upper chromosphere. Fine-scale chromospheric features are
usually observed in the \ha\, Ly\,$\alpha$, \caii\ H \& K and the
\caii\ infrared triplet. In principle the solar chromosphere is
optically thick at these wavelengths and, consequently, their
computation from a model atmosphere requires solving the well known
radiative transfer equation together with the statistical
equilibrium equations self-consistently, a problem which is both
highly non-linear and non-local. However, a simple method called the
cloud model, which does not involve such elaborate computations and
assumes optically thin structures and a constant source function
along the observed structure has been successfully applied over the
years for several chromospheric structures observed in \ha\
\citep[see review by][]{tzio:07} for the determination of physical
parameters. Below we discuss in more detail the cloud model, other
radiative transfer techniques and recent and future trends in
interpretation of line profiles.

\subsection{The cloud model}
\label{cloud_model}

For the deduction of different physical parameters of chromospheric
structures from observations in the \ha\ line a method is used which
is based in the so called ``cloud" model, introduced by
\citet{beckers1964phd}, \citep[see also][]{alissandrakis1990}. The
model is valid for optically thin structures located well above the
background. It considers for an observed intensity profile
$I(\Delta\lambda)$ the contrast profile:

\begin{equation}
C(\Delta\lambda) = \:\frac{I(\Delta\lambda)- I_{\rm 0}(\Delta\lambda)}{I_{\rm 0}(\Delta\lambda)} = \\
\:\left(\frac{S}{I_{\rm 0}(\Delta\lambda)} - 1 \right) (1-e^{-\tau(\Delta\lambda)})
\end{equation}

\noindent with a Gaussian wavelength dependence for the optical thickness:

\begin{equation}
\tau(\Delta\lambda) = \tau_{\rm 0}\,e^{-\:({\frac{\Delta\lambda-\Delta\lambda_{\rm
I}}{\Delta\lambda_{\rm D}}})^2}
\end{equation}

\noindent where $I_{\rm 0}(\Delta\lambda)$ is the reference profile emitted by the
background and $\Delta\lambda_{\rm I}  = \lambda_{\rm 0}{\upsilon}/c$ is the Doppler
shift with $\lambda_{\rm 0}$ being the line center wavelength and $c$ the speed of
light. The four adjustable parameters of the model are the source function $S$, the
Doppler width $\Delta\lambda_{\rm D}$, the optical depth $\tau_{\rm 0}$ and the
line-of-sight (LOS) velocity $\upsilon$. All these parameters are assumed to be constant
through the structure.

Once the four parameters are determined from the cloud model, the
estimation of several other parameters can be made. From the
calculated Doppler width values and if we assume a value for the
micro-turbulent velocity, $\xi_t$, we can deduce the temperature,
$T_e$. From the optical depth at line center, which may be written
as:

\begin{equation}
\tau_0 = \:{\frac{\pi^\frac{1}{2} e^{2}}{m_e c}}\:{\frac{f\lambda^{2}}{c}}
\:{\frac{N_2}{\Delta\lambda_D}} d
\end{equation}

\noindent and  $\Delta\lambda_D$, $N_2$, e.g. the number density in the second hydrogen
level, can be obtained. Then the electron density, $N{_e}$, the total particle density
of hydrogen (i.e., neutral plus ionized), $N{_H}$, the gas pressure, $p$, the total
column mass, $m$, the mass density, $\rho$, can be determined from the following
relations:

\begin{equation}
N_{e} = 3.2\:10^{8}\:\sqrt{N_2}\:cm^{-3}\\
\end{equation}

\begin{equation}
N_{H} = 5\:10^{8}\:10^{0.5 \log{N_2}}\\
\end{equation}

\begin{equation}
p = k (N_{e} + 1.0851 N_{H}) T_{e}\\
\end{equation}

\begin{equation}
m = (N_{H} m_{H} + 0.0851 N_{H} \times 3.97 m_{H})\:d\\
\end{equation}

\begin{equation}
\rho = \:{\frac{m}{d}}
\end{equation}

\noindent where $d$ is the path length along the LOS (for details about the derivation
of these parameters \citet{tsiropoula_sch1997}).

\subsubsection{Optically thick radiative transfer}

For any given species the radiation field is determined from the
source function. The source function is derived from the atomic
level populations that are in turn determined by the local radiation
field. This inter-dependence defines the highly non-linear and
non-local aspect of optically thick radiative transfer.

As already mentioned (see Section ~\ref{parameters})
\cite{giovanelli1967a, giovanelli1967b} did pioneering extensive
NLTE computations of the \ha\ line contrast profiles relevant to
chromospheric fine-scale modelling assuming 1D slabs illuminated by
the surrounding atmosphere. \cite{heinzel_sch1994} used \ha\ line
profile observations of bright and dark mottles together with a grid
of NLTE models of prominence-like structures considered as
vertically-standing 1D slabs irradiated from both sides by an
isotropic incident radiation and derived the physical conditions in
these structures.

In general, most radiative transfer codes used so far for the
diagnostics of isolated solar structures, combine the accelerated
(or approximate) Lambda iteration (ALI) or multilevel ALI (MALI)
schemes
\citep[e.g.][]{cannon1973,scharmer1985,rybicki1991,rybicki1992,olson1986}
with efficient algorithms for computing the radiation field
\citep[Feautrier's method,][]{feautrier1964,auer1967} and direct
integration techniques \citep[e.g.][]{mihalas1978, olson1987} in
order to solve the intricate radiative transfer problem.

There already exist several 1D-geometry radiative codes, such as
MULTI \citep{carlsson1986}, MALI \citep{heinzel1995} and RH
\citep[][the 1D version]{uitenbroek2001} that compute the radiative
transfer problem in plane parallel atmospheres, using atoms
containing several or even hundreds of levels and/or thousands of
lines. However, the solar chromosphere has a complex 3D structure
which should be taken into account when evaluating the radiation
field with non-Local Thermodynamic Equilibrium (NLTE) codes. Such
codes are usually required to use local ALI operator, since any
non-local operator would require computationally expensive
inversions of very large sparse matrices. Since the typical size of
3D radiative transfer problems is quite large, some form of
parallelization (either over frequency and angle, or using domain
decomposition, or a not-yet implemented hybrid parallelization
strategy) is required to speed up convergence to a physical
solution. Nowadays, there exist three 3D NLTE radiative transfer
codes, the MUGA code \citep[see, e.g][]{fabiani1997,botnen1999} the
RH code by \citet{uitenbroek2001} and the Multi3D
\citep{fabiani1999, leenaarts2009} code, for calculating
chromospheric models.

The situation that prevails in the atmosphere of the Sun poses
several different, problems in the computation of the radiation
field with NLTE codes. For example, the hydrogen line radiative
transfer problem, requires to take into account non-equilibrium
hydrogen ionization in the chromosphere and hydrogen level
populations out of statistical equilibrium. This stems from the fact
that the assumption of statistical equilibrium for hydrogen fails in
the chromosphere, because the timescale of hydrogen ionization and
recombination is of the same order as the hydrodynamic timescale.
The existing codes fail to reproduce the \ha\ line center intensity,
as well as the fibrilar structure of the chromosphere. This is
because apart from taking into account non-equilibrium ionization
processes, 3D NLTE time-dependent radiative transfer codes are
required in order to compute the detailed hydrogen line profiles and
get reasonably realistic electron density and temperature which are
very important parameters for the modelling of the chromospheric
structures. On the other hand, the \caii\ lines suffer less than
\ha\ from non-equilibrium ionization processes and thus can be
modeled assuming statistical equilibrium. The cores of \caii\ H \&
K, however, form in partial redistribution, which is computationally
far less trivial than complete redistribution and still has not been
fully implemented in 3D. To make the problem of simulating these
lines even worse we mention that the \caii\ H \& K lines are usually
observed with wide filters which deliver intensity contributions
from a wide range of formation heights. Partial redistribution
effects are less important for the \caii\ infrared triplet and,
moreover, these lines can be computed in 3D. Current
state-of-the-art instruments have delivered many high-quality
chromospheric observations in the \caii\ 8542\,\AA\ line, in
particular, although this line has a lower line-core opacity than
\ha\ and \caii\ H \& K. The NLTE \caii\ 8542\,\AA\ computation using
a 3D simulation \citep{leenaarts2009, leenaarts_car2009}, does not
show, however, the fibrilar elongated structures delineating
magnetic field lines around the network. This lack may be do to
several reasons, such as the low resolution of the grid used for the
simulation, need for larger scale magnetic field configuration, etc.
Another possibility is that \caii\ 8542\,\AA\ opacity in the fibrils
is caused by physical processes currently included in the existing
simulation codes. On the other hand, quite recently
\cite{leenaarts2012} using 3D NLTE radiative transfer computations
and MHD simulations investigated the \ha\ line formation in the
solar chromosphere. They were able to reproduce the fibril-like dark
structures seen in \ha\ line-core observations. Furthermore, their
simulations support the commonly held notion that fibril-like
structures indeed trace the magnetic field.

Knowledge of the physical conditions in isolated structures is very
important in order to elucidate the physical processes which are
occurring in them. A state-of-the-art method to improve this
knowledge is the forward modelling by which data from a combination
of radiation-MHD simulation codes with 3D NLTE radiative transfer
codes are converted into observed quantities and after being
compared with observations can give more reliable initial values for
the codes. Although existing codes still suffer from severe
limitations, they are producing promising results, e.g. the
computation of the emerging \caii\ 8542\,\AA\ and \ha\ line profiles
with a 3D NLTE model using an RMHD simulation \citep{leenaarts2009,
leenaarts2012}, the spectral line response to reconnection events in
the photosphere and transition region \citep{heggland2009},
Monte-Carlo simulations of the line profile response to upward
propagating material, such as in the case of rapid blue-shifted
events, \citep{langangen2008c}, and toy models of the solar
atmosphere to qualitatively explain the appearance of the solar limb
in the Hinode \caii\ H passband caused by large line-of-sight
velocities or wide spicule line profiles \citep{judge2010}.

\section{Inter-relations between the different fine-scale chromospheric structures
seen on the disk and at the limb} One of the outstanding questions
in spicule research concerns the relationship between spicules at
the limb and structures observed on the disk. Although spicules,
mottles and fibrils are all nowadays treated as jet-like features of
the solar chromosphere and the similarity of many of their
properties strongly suggests that some of these phenomena are
related, if not the same, there has been a strong debate in the
solar community about the exact relationship between them. In
particular, it has long been suspected and even assumed that the
{\it dark mottles} are the direct manifestations of {\it spicules},
and considerable effort has been expended to prove this
relationship. The mass velocity is one of the properties that could
help establishing a kinship between mottles and spicules.
\cite{grossmann1992} have examined two distributions of axial
velocities, one based on limb the other on disk observations. They
arrived to the conclusion that the two distributions disagree, that
the velocity in spicules is significantly greater than in mottles
and thus that mottles are not the disk counterparts of spicules.
They discussed several possibilities for this discrepancy, but the
most important we think -- not discussed in that paper -- is that
the velocities from which these distributions were derived have been
obtained by different methods: for the limb velocities distribution
the Doppler shift method was used, while for the disk velocities
distribution they used the ``cloud'' model. These two techniques
yield very different results, as it is stated by
\cite{tsiropoula1994}. The discrepancies on the velocities between
structures seen on the disk and at the limb, apart from the problem
in associating with spicules some easily identified class of disk
structures, must, furthermore, be sought as due to several other
reasons. These include various selection effects arising from
different visibilities of the features, different wavelengths used
for their observations, different phases in their evolution and also
eventual bias in the data arising from the methods of analysis. They
are also arising from the fact that the structures are observed
under very different conditions: in one case (limb) at great heights
and along the horizontal LOS; in the other case (disk) at low
heights and along the vertical. Direct observational evidence for
the identification of fine dark mottles with spicules would be that
of a dark mottle crossing the solar limb. \cite{christopoulou2001}
found several examples of individual mottles crossing the solar limb
and provide further support to the association between spicules and
mottles. On the other hand, some authors suggested that similarities
in dimensions, lifetime, and physical conditions, such as density
and temperature, provide indirect evidence to the identity of
spicules and dark mottles \citep{tsiropoula_sch1997}.

It is also not clear which the relationship is between quiet Sun
{\it dark mottles} and active region {\it long fibrils} and {\it
short dynamic fibrils} (DFs). Long fibrils are found at the edges of
plages, are longer, more stable, longer lived and more horizontally
oriented than the most dynamic DFs, which are found towards the
center of the plages, are more vertically oriented and have shorter
lifetimes (although their lifetime depends on their inclination
being longer for the more inclined ones). The more complex topology
of the quiet Sun with its mixed polarity magnetic fields are to be
opposed to plage regions which are considered as unipolar with
magnetic flux tubes packed close together. Due to these differences
the conditions between quiet Sun and plage regions should differ
considerably and this must be reflected in the dynamics, as well as
the mechanisms responsible for the formation of mottles and fibrils.
Thus for mottles magnetic reconnection has been suggested as the
more likely formation mechanism \citep{tsiropoula1994,
tziotziou2003}, while for fibrils magnetoacoustic shock waves driven
by leakage of convective motions and oscillations from the
photosphere into the chromosphere along magnetic fields have been
suggested as the formation mechanism. \cite{tsiropoula1994}, and
\cite{tziotziou2003} derived LOS velocities along mottles using the
cloud model. Examining the temporal evolution of the LOS velocity
they found that the predominant motion is downward at their
foot-points and alternatively upward and downward at their tops.
Mottles are found at the network boundaries where bipolar elements
are swept by the supergranular flow. Interactions of the magnetic
fields have as a result the enhancement of the flux concentration in
the case of same polarities or its cancellation in the case of
opposite polarities. The bipolar nature of the flux cancellation
process most likely involves magnetic reconnection and thus it is
reasonable to consider this process as the driver of mottles.
\cite{murawski2010} studying the formation of chromospheric jet-like
structures in the framework of the rebound shock model which was
based on a localized velocity pulse launched from the lower
atmosphere which quickly steepens into shocks. They were able to
explain the observed speed, width, heights and periods of type I
spicules, as well as the observed multi-structural and bidirectional
flows. Fibrils, on the other hand, are found at active region plages
which are considered as unipolar, with high filling factor of the
magnetic field, while the field lines are only moderately inclined
and the spatial distribution of magnetic flux in their core evolves
slowly. It is possible that different mechanisms have to be
considered as their driver. For dynamic fibrils examination of their
proper motions have shown that their tops follow parabolic paths.
Recent observations, as well as several new, usually 2D, simulations
have suggested that dynamic fibrils are driven by magnetoacoustic
shock waves, which originate in the convection zone and photosphere.
It has long been recognized that along inclined magnetic flux tubes
p-modes leak sufficient energy from the gloabal resonant cavity into
the chromosphere \citep{hansteen2006, depontieu2007a, heggland2007}.
\cite{heggland2009} performed 2D simulations that included the
effects of radiation and heat conduction and that involved
reconnection events induced by waves propagating upward from the
photosphere/convection zone and were able to produce a number of jet
phenomena among them spicules. \cite{martinez2009} performed 3D
simulations and showed that some structures having the properties of
mottles and DFs can be formed by upwardly propagating chromospheric
waves through several driving mechanisms: collapsing granules,
magnetic energy release in the photosphere or low chromosphere,
p-modes, etc. They suggested that magnetic energy release events
might be related to magnetic field line reconnection and thus both
mechanisms, i.e. magnetic reconnection and shock waves could
co-exist in the jets. Their simulations, however, do not reproduce
the observed values for the duration and maximum height of the jets.
Thus, they found durations of 2 -- 3\,min (instead of 5 -- 10\,min)
and maximum heights less than 2\,Mm (instead of $\sim$ 10\,Mm).
There is a bias also towards rather vertical structures which cannot
account for the heavily inclined structures that, as it is known,
are responsible for the formation of the magnetic canopy. Their
simulations, on the other hand, showed that half of the events they
found were driven by magnetic energy release, most likely related to
the emergence of new field into a pre-existing ambient field. Such
conditions, however, do not prevail in plage regions and thus these
simulations are still far from what works in the real Sun. From the
above it seems more plausible to consider that both drivers, i.e.
waves and magnetic reconnection should be considered as responsible
for the formation of both mottles and fibrils.

It is also unclear what the relationship is between the recently
named {\it Type I} and {\it Type II spicules} and the {\it
traditional spicules}. Traditional spicules show mean velocities of
order 20 -- 30\kms, lifetimes of order 5 -- 10\,min, and heights of
5\,000 -- 10\,000\km. Type I spicules appear to rise up from the
limb and fall back again and show similar dynamical evolution as
active region DFs \citep{hansteen2006, depontieu2007a}, as well as a
subset of quiet--Sun mottles \citep{rouppe2007}. For these
structures magnetoacoustic shocks have also been suggested as their
driven mechanism (\citep{hansteen2006, depontieu2007b, heggland2007,
martinez2009}). The so-called Type II spicules, on the other hand,
appear to exhibit upward motion followed by rapid fading without a
downward moving phase. They have shorter lifetimes (10 -- 100\,s),
high apparent velocities (50 -- 150\kms), and lower widths (150 and
700\km). They undergo a swaying motion which has been suggested to
be caused by the upward propagation of Alfv\'{e}n waves
\citep{depontieu2007sci}. Thus it seems that Type I and Type II
spicules, as seen by SOT on Hinode, look totally different from what
are considered as traditional spicules described in the literature.
It has to be noticed, however, that the properties of the newly
discovered Type I and Type II spicules have been derived from
observations in the \caii\ H intensity images obtained with the
broad-band filter onboard the Hinode spacecraft, while most of the
earlier works were based on filtergrams and/or spectra obtained in
the \ha\ line. Furthermore, since spicules are so dynamic and many
undergo transverse motions during their lifetime, in order to infer
a reliable measure of their properties if a fixed cut almost
perpendicular to the limb is used then one has to rely only on
spicules that did not move transversely so that their full lifetime
can be covered. In a recent work \cite{zhang2012} re-analyzing the
same data sets used by \cite{depontieu2007b} traced the intensity
along individual structures in quiet Sun and CH. They claimed that
they could not find a single convincing example of Type II spicules.
Furthermore, they claimed that more than 60\% of the identified
spicules in each region showed a complete cycle, i.e., the majority
are Type I spicules. Due to these discrepancies, it cannot be clear
how the traditional spicules are related to the Type I and Type II
SOT/Hinode spicules. Furthermore, what is measured in \caii\ H
images is apparent motions and it is unclear whether they are
associated with bulk mass motions derived sometimes in the \ha\
line. The differences in the heights, lifetimes, velocities and
driving mechanisms that pointed that there are two types of spicules
should also be questioned. Are there really two different
distributions of the different parameters or one broad distribution
with a wide range of values? \cite{bray_lou1974} gave histograms of
upward and downward velocities in traditional spicules in the range
$\pm$60 -- 70\kms, lifetimes in the range 2 -- 10\,min, heights in
the range 6 -- 16\,Mm. \cite{Heristchi_mou1992} give also a
distribution of the apparent velocities in the so-called traditional
spicules from which it can be seen that they are in the range 0 --
70\kms\ (see Fig.\ref{axial_velo}, left). Regarding the formation
mechanisms, \cite{shibata1982a} found that: if the explosion
(reconnection) occurs below the middle chromosphere, the jets are
formed as a result of shocks; if the explosion (reconnection) occurs
above the middle chromosphere the jets are directly driven by
reconnection. \cite{shibata1982b} using the shock model explained
why spicules are taller in coronal holes and shorter in active
regions. Thus in our opinion the question whether there really exist
two types of spicules or one type of spicules with very broad
distributions of heights, lifetimes and velocities, remains open.

Recently an important effort has been undertaken to establish a
relationship between {\it Type I} or {\it Type II  spicules} and
{\it structures on the disk}, such as straws and RBEs. It is
suggested that straws seen in DOT \caii\ H near-limb images
appearing bright against the very dark background and fast-waving
might be identical to the fast-waving Type II spicules.
\cite{langangen2008c}, suggested that \caii\ 8542\,\AA\ RBEs might
be the on-disk counterparts of type II spicules observed at the limb
by the Hinode \caii\ H broad filter. This suggestion was based on
the similarity in evolution, lifetime, spatial extent and location.
RBEs, however, show low blue-shifts of 15 -- 20\kms, as opposed to
the high apparent velocities reported for Type II spicules. In order
to explain this discrepancy they used Monte Carlo simulations and
showed that the low observed blue-shifts can be explained by a wide
range of spicule orientations combined with a lack of opacity
\citep{langangen2008c}. Of course, type II spicules have very short
lifetimes, small spatial dimensions and change very fast and thus
they are on the limit of what can be resolved with present-day
telescopes. \cite{rouppe2009} from spectral observations in the \ha\
and \caii\ H 8542\,\AA\ lines reported also that the disk
counterparts of Type II spicules might be the RBEs based on the
similarities on their appearance, lifetimes, longitudinal and
transverse velocities and occurrence rate. Recently,
\citet{depontieu2011} used on-disk \ha\ observations of RBEs
obtained with SOT/Hinode and reported that they found direct
evidence of a strong correlation between the RBEs and short-lived
brightenings in a wide range of transition region (TR) and coronal
passbands observed with the Atmospheric Imaging Assembly (AIA)
on-board the Solar Dynamics Observatory (SDO) space mission. They,
also, found evidence that chromospheric spicules observed in \caii\
H with SOT/Hinode at the limb in coronal holes are intimately linked
to the formation of features at (TR) and coronal temperatures and
concluded that on-disk RBEs and \caii\ H limb spicules are
connected.

It should also be mentioned that in a recent work, \cite{judge2011}
questioned whether spicules (and fibrils) are straw-like fine
structures (as assumed by most authors) or whether they are best
described as warps in a sheet. Readron (2012, private comm.)
suggests that some features may appear within $\approx$2 sec,
clearly questioning the plasma ejection flux-tube model as this
implies Mach speeds well in excess of 100. Higher cadence data from
present-day imaging spectrometers will allow us to access whether
(or what fraction) of these features are jet/tube features or as
suggested by \cite{judge2011} as current sheets. Spicules produced
by these current sheets could also be a natural explanation for the
\caii\ H Type II spicules which are very fast and seem to disappear
rather than falling back to the lower atmosphere.

Furthermore, it is not known if spicule-like structures seen in UV
and EUV are some aspects of traditional spicules or totally
independent features. Several authors suggest that UV spicules may
be the result of heating of chromospheric spicules to UV
temperatures. This would provide an explanation for the reports that
spicules often ``fade from view'' when being observed in
chromospheric spectral lines (e.g., \cite{beckers1968}. Another
possibility is that UV spicules are a transition region ``sheath''
that surrounds the chromospheric spicules.

\section{Summary}
In this review we present observations and physical parameters of
fine-scale structures observed in the solar chromosphere. The solar
chromosphere is dominated by a variety of thin, jet-like features
omnipresent in quiet, as well as in active regions, and on the disk,
as well as at the limb. Because of their small size and the
limitations of ground-based observations, for many decades it was
difficult to characterize completely their properties, and much less
to determine with confidence their interrelationship and the
mechanism(s) responsible for their generation. In recent years
because of the improved quality of observations acquired by ground
and space-based instrumentation and, especially, with the current
ability to observe the chromosphere with high resolution and high
time cadence exciting jet-like features have been detected that were
not recognized earlier. Their properties are not yet fully
unraveled, while new names have been attributed to distinguish them
from those previously observed. The nomenclature is nowadays rather
confusing, but there are strong indications that, at least some if
not all of the several observed fine-scale structures, are
physically closely related, if not the same.

It seems also that all these small-scale structures are related to
the fine-scale structure of the magnetic field and its evolution.
Magnetic field lines are the most likely channels for transporting
convective energy to the solar upper atmospheric layers. Detailed
spatial and temporal high-resolution observations of the magnetic
fields are crucial for understanding the small-scale structures and
investigating their role in powering the Sun's outer atmosphere.
Magnetic field extrapolations are also very important tools in
reconstructing the 3-D topology of the upper solar layers. Different
methods have been proposed so far based on different assumptions. Of
particular interest is the reconstruction of the 3-D magnetic
topology in inclined fine-scale structures, such as e.g. spicules.
These structures, being magnetic flux tubes following the local
inclination of the magnetic field lines, define the layer of the
magnetic canopy, i.e. the layer where the plasma $\beta$ (ratio of
the gas to the magnetic pressure) is equal to 1. This layer is very
important, since acoustic waves generated by the convective motions
upon meeting this layer undergo mode conversion, refraction,
reflection and transmission with important effects to the upper
solar atmosphere.

The state-of-the-art in the understanding of small-scale events and
their interrelationship requires simultaneous, co-spatial,
multi-wavelength, high spatial and temporal resolution ground-based
observations combined with observations from forefront observing
space facilities of the same solar region. Such observations will
permit their follow-up from the photospheric to the lower coronal
level of the solar atmosphere. They will also permit a better
correlation between manifestations of the same phenomenon in
different atmospheric heights and, hence, are a powerful tool for
understanding the physics that governs the generation and dynamics
of fine-scale structures.

Apart from different types of studies based on analyses of
observations, there is an on-going development and refinement of
non-LTE and cloud model inversion codes of spectral line profiles
based on better understood physics of spectral line formation and
radiation transfer. These codes together with advances in theory are
an indispensable tool for a more accurate determination of several
physical parameters that describe the observed fine structures.
Quantitative determinations of such parameters is an essential
ingredient not only for their modelling, but also for the modelling
of the solar atmosphere as a whole. By their accurate determination
it will also be possible to define the basic physics of, at least,
some of the coronal heating processes and the origin and generation
of the solar wind.

These developments go hand in hand with improvements in numerical
simulations and improvement of observational technology. Numerical
simulations, using 3D MHD codes together with forward modelling that
produce precise synthetic data which are able to mimic accurately
improved observations are important to correctly interpret them. It
is hoped that, in the near future, the combination of state-of-the
art multi-wavelength observations, along with improved NLTE codes,
realistic MHD simulations and forward modelling, will provide a
breakthrough in our understanding of the fine-scale chromospheric
structures.

\begin{acknowledgements}
The authors would like to thank the International Space Science
Institute (ISSI) in Bern, Switzerland, for the hospitality provided
to the members of the team on ``Solar small-scale transient
phenomena and their role in coronal heating'', as well as the
members of the team for fruitful discussions. We also acknowledge
the remarks and suggestions made by the two referees which helped to
improve the paper.
\end{acknowledgements}

\bibliographystyle{aps-nameyear}

%\bibliography{review_references}

\end{document}